\documentclass[amsfonts,amsmath,prd,preprint,nofootinbib]{revtex4}
\newcommand{\beq}{\begin{equation}}
\newcommand{\eeq}{\end{equation}}

\newcommand{\bsp}{\begin{split}}

\usepackage{epsfig,bbm,cancel,ulem}
\usepackage[breaklinks=true]{hyperref}
\usepackage{xcolor}
\usepackage{wasysym}
\usepackage[mathscr]{eucal}
\usepackage{multirow}
\usepackage{amsmath}
\usepackage{enumitem}

\begin{document}

\title{Bound orbits around charged black holes with exponential and logarithmic electrodynamics}

\author{A S. Habibina}
\email{a.sayyidina@sci.ui.ac.id}
\author{B.~N. Jayawiguna}
\email{byon.nugraha@ui.ac.id}
\author{H.~S.~Ramadhan}
\email{hramad@sci.ui.ac.id}
\affiliation{Departemen Fisika, FMIPA, Universitas Indonesia, Depok, 16424, Indonesia. }
\def\changenote#1{\footnote{\bf #1}}

\begin{abstract}
We present exact black hole solutions endowed with magnetic charge coming from exponential and logarithmic nonlinear electrodynamics (NLED). Classically, we analyze the null and timelike geodesics, all of which contain both the bound and the scattering orbits. Using the effective geometry formalism, we found that photon can have nontrivial stable (both circular and non-circular) bound orbits. The noncircular bound orbits for the one-horizon case mostly take the form of precessed ellipse. For the extremal and three-horizon cases we find many-world orbits where photon crosses the outer horizon but bounces back without hitting the true (or second, respectively) horizon, producing the epicycloid and epitrochoid paths.Semiclassically, we investigate their Hawking temperature, stability, and phase transition. The nonlinearity enables black hole stability with smaller radius than its RN counterpart. However, for very-strong nonlinear regime, the thermodynamic behavior tends to be Schwarzschild-like.  
\end{abstract}

\maketitle
\thispagestyle{empty}
\setcounter{page}{1}
\section{Introduction}
\label{intro}


The behaviour of light around black hole (BH) is one of the most captivating phenomenon in the realm of astrophysics. One example is the notion of \textit{photon sphere}, a particular condition where null rays orbit in constant radius. Recent observation has captured ring-like image around a supermassive BH~\cite{Akiyama:2019cqa}. While this discovery is the result of realistic rotating BH (which have been investigated extensively, for example in~\cite{Teo}), it is still possible to gain physical insights from photon sphere in static cases. An even more intriguing phenomenon is the existence non-circular photon orbit. It is widely known that Schwarzchild and Reissner-Nordstr\"{o}m BHs do not possess a condition where stable photon orbit outside their corresponding horizons is possible. It is important to point out that the extremal case of RN can possess stable photon orbit exactly {\it on} its horizon \cite{Pradhan:2010ws}. Some findings around this issue shows that test particles can escape the horizons and emerge into another universe due to the potential barrier \cite{Grunau:2010gd} and that the presence of net charge of the test particle can be interpreted as the increase or decrease of the charge of the black hole \cite{Hong:2017dnf}. One of many ways to solve the mentioned problem is by replacing the linear Maxwell electrodynamics with a different model of Nonlinear Electrodynamics (NLED).

There has been many comprehensive studies surrounding the idea of NLED since the beginning of the twentieth century, with  Born and Infeld (BI) proposed the first nonlinear extension of Maxwell's electrodynamics to cure the singularity of electron's self-energy~\cite{Born:1934gh}. Euler and Heisenberg predicted the existence of vacuum magnetic birefringence in quantum electrodynamics (QED)~\cite{Heisenberg:1935qt}.
Although the classical NLED had been abandoned in favor of QED, in recent years interests in it resurrected, interestingly also due to the success of QED~\cite{Bamber:1999zt}. Arguably, nowhere does this NLED phenomena get more exciting than in gravitational physics. Almost as soon as the BI electrodynamics was proposed, their extensions (as well as other NLED's) to black holes or compact stars have been explored, both in general relativity (GR) and modified gravity theories~\cite{Hoffmann:1937noa,Hassaine:2008pw,Hendi:2014kha,Bronnikov:2000vy,Kruglov:2009he}. The development of NLED-coupled black hole is also followed with its phenomenological study. It is found that, in the presence of NLED, photon propagates along the null geodesic of its effective geometry~\cite{Novello:1999pg,Breton:2001yk}. The consequence of corresponding theory has been studied in various applications, such as particles trajectories \cite{Breton:2007bza}, gravitational lensing \cite{MosqueraCuesta:2004wh} and deflection angle \cite{Eiroa:2005ag}. Other studies about particle motion in NLED-charged black hole can be found in \cite{Linares:2014nda}.

Among the various NLED models, the so-called ``{\it logarithmic NLED}" (hereafter referred to as the LNE) is shown to possess interesting features. Its corresponding self-energy is finite~\cite{Soleng:1995kn}. 
A study by Gaete and Neto claimed that LNE also exhibits the vacuum birefringence phenomenon \cite{Gaete:2013dta}. Thus, it seems phenomenologically plausible to assume that at high energy regime Maxwell's electrodynamics exhibits a logarithmic correction. Soleng showed that when coupled to gravity, LNE contains in its spectrum static BH solutions with nonlinear $U(1)$ charge.  In s series of papers, Hendi further elaborated the BH spectrum of LNE along with another interesting NLED model, the ``{\it exponential NLED}"  (referred to as the ENE)~\cite{Hendi:2012zz}. Although the ENE does not remove the divergency of electric field at the origin, its singularity is much weaker than in Einstein-Maxwell gravity. The BH solutions in both models reduce to RN in weak-coupling limit. They have also been studied from various perspectives, both classically as well as thermodynamically, in $4d$ as well as in higher dimensions. To the best of our knowledge, none of them deal with the null geodesics and corresponding photon orbits.

Intrigued by the properties of the mentioned models (LNE and ENE), we decide to focus our study to their effects as charged black hole on test particles and the possibility of bound orbits around them. The summary of this work is given by the following. In Section \ref{nled} we give a brief description of the models and NLED in general. The black hole solution is evaluated in Section \ref{bh}. Sections \ref{tlg} and \ref{ng} are reserved for the discussion on the possible timelike and null orbit scenarios, respectively. The detailed discussion of two and three horizons case if provided in Section \ref{two} and \ref{three}. We then present a comprehensive discussion of the thermodynamics in Section \ref{t}. Finally, we summarize our result in Section \ref{con}.

\section{Logarithmic and Exponential Electrodynamics}
\label{nled}

In this paper, we employ the ENE and LNE lagrangians as exhibited in~\cite{Hendi:2013mka}, 
\begin{eqnarray}
\label{L}
\mathcal{L}(\mathcal{F})=
\left\{\begin{array}{cl}
\beta ^2 \left(\exp (-\frac{\mathcal{F}}{\beta ^2})-1\right) &, \textrm{ENE}\\
-8\beta ^2 \ln \left(\frac{\mathcal{F}}{8\beta ^2}+1\right) &, \textrm{LNE}
\end{array} \right\},
\end{eqnarray}
where $\beta$ is called the nonlinearity parameter and $\mathcal{F}\equiv 1/4\ F_{\mu\nu}F^{\mu\nu}$, the Maxwell Lagrangian. They both satisfy
\begin{equation}
\lim_{\beta\to\infty} \mathcal{L}(\mathcal{F}) = -\mathcal{F}.
\end{equation}
Here we use the following notations
\begin{equation}
\label{L_}
\mathcal{L}=\mathcal{L}(\mathcal{F}) \,\,\,\, , \,\,\,\,
\mathcal{L}_F\equiv{\partial\mathcal{L}\over \partial\mathcal{F}} \,\,\,\, , \,\,\,\,
\mathcal{L}_{FF}\equiv{\partial^2\mathcal{L}\over\partial\mathcal{F}^2}.
\end{equation}
The general field equation of nonlinear electrodynamics is given as
\begin{equation}
\label{feq}
\nabla_{\mu} (\mathcal{L}_F F^{\mu\nu})=0 \,\,\,\, \Rightarrow \,\,\,\, \frac{1}{\sqrt{-g}}\partial_{\mu} (\sqrt{-g}\mathcal{L}_F F^{\mu\nu})=0
\end{equation}
and the energy-momentum tensor in general is in the form of
\begin{equation}
\label{emt}
T_{\mu\nu} = \mathcal{L}_F F_{\mu\gamma} F_{\nu}^{~\gamma} -g_{\mu\nu} \mathcal{L}.
\end{equation}

\section{Magnetically Charged Black Holes Solution}
\label{bh}

We define an action with a nonlinear electrodynamics as
\begin{equation}
\label{action}
S = \int d^4 x \sqrt{-g}\bigg[\frac{R}{2\kappa^2} + \mathcal{L}\bigg],
\end{equation}
where $\kappa^2\equiv 8\pi G=1$. The ansatz employed here is magnetic monopole and spherical symmetry~\cite{Bronnikov:2000vy},
\begin{equation}
\label{A_mu_ms}
A_{t} = A_{r} = A_{\theta} =0, \,\,\,\,\,\,\,\,  \,\,\,\,\,\,\,\,  A_{\phi} = q (1-\cos\theta),
\end{equation}
and
\begin{equation}
\label{sss}
ds^2 = -f(r) dt^2 +f^{-1}(r) dr^2 + r^2 d\Omega^2.
\end{equation}
The energy-momentum tensor then becomes
\begin{eqnarray}
\label{T00}
T^t_t &=& T^r_r = - \mathcal{L} \\
\label{T22}
T^{\theta}_{\theta} &=& T^{\phi}_{\phi} = \frac{\mathcal{L}_F (F_{\theta\phi})^2}{r^4 \sin^2 \theta} - \mathcal{L}.
\end{eqnarray}
The Einstein tensor components according to our metric \eqref{sss} is given as
\begin{eqnarray}
\label{Rtt}
R_{tt} &=& \frac{f(r) f''(r)}{2} +\frac{f(r) f'(r)}{r} = -\kappa^2 f(r) T^{\theta}_{\theta}  \\
\label{Rrr}
R_{rr} &=& -\frac{f''(r)}{2 f(r)}-\frac{f'(r)}{r f(r)} = \kappa^2 f^{-1}(r) T^{\theta}_{\theta} \\
\label{Rthth}
R_{\theta\theta} &=& 1-r f'(r)-f(r) = -\kappa ^2 r^2 T^t_t
\end{eqnarray}
We then insert Eq. \eqref{T00} into Eq. \eqref{Rthth}, which let us obtain the general solution as
\begin{equation}
\label{fr}
f(r)= 1 +\frac{\kappa^2}{r} \int \rho(r) dr -\frac{C}{r}  \,\, \,\, , \,\,\,\, \rho(r) = \mathcal{L} r^2
\end{equation}
where $C$ is the integration constant. Since we are dealing with charged black hole, comparing our result with RN solution consequently assigns the mass of the black hole to the corresponding constant, giving $C=-2m$. On the other hand, the ansatz \eqref{A_mu_ms} directly gives magnetic field as $F_{\theta\phi} = q \sin \theta$. Putting the ENE and LNE lagrangian into the integral, we obtain the general solutions as
\begin{eqnarray}
\label{solusi}
f(r)=
\left\{\begin{array}{cl}
1-\frac{2 m}{r} +\frac{\kappa^2}{r} \int \big( e^{-\frac{Q^2}{2 \beta ^2 r^4}}-1 \big) \beta^2 r^2 dr ,\,\, \textrm{for ENE}. \\
1-\frac{2 m}{r} -\frac{8 \kappa^2}{r} \int \log \big( \frac{Q^2}{16 \beta ^2 r^4}+1 \big) \beta ^2 r^2 dr ,\,\, \textrm{for LNE}
\end{array} \right\}.
\end{eqnarray}
The integrals can be done analytically\footnote{We use Wolfram Mathematica v.12 to integrate it.}. Defining $\Gamma(a,b)$ as the incomplete gamma function, the explicit form of \eqref{solusi} can be written as
\begin{eqnarray}
\label{solusif}
f(r)=
\left\{\begin{array}{cl}
1-\frac{2 m}{r}-\frac{1}{6} \beta^2 r^2 \bigg[2 \sinh\left(\frac{q^2}{4 \beta ^2 r^4}\right) e^{-\frac{q^2}{4 \beta ^2 r^4}}+\sqrt[4]{2} \left(\frac{q^2}{\beta ^2 r^4}\right)^{3/4} \Gamma \left(\frac{1}{4},\frac{q^2}{2 r^4 \beta ^2}\right)\bigg],\,\, \textrm{for ENE}. \\
1-\frac{2 m}{r}-\frac{\sqrt{\beta}}{3 r} 2^{-1/4} q^{3/2} \bigg[\tan ^{-1}\left(1+\frac{2^{3/4} \sqrt{\beta } r}{\sqrt{q}}\right)-\tan ^{-1}\left(1-\frac{2^{3/4} \sqrt{\beta } r}{\sqrt{q}}\right)\\ + 2^{1/4}\left(\frac{\beta}{q}\right)^{3/2} r^3 \ln\left(1+\frac{q^2}{2r^4 \beta^2}\right)+ \ln\left(\sqrt{\frac{q - 2^{3/4} \sqrt{\beta q } r +\sqrt{2} \beta  r^2}{q+2^{3/4} \sqrt{\beta q} r+\sqrt{2} \beta  r^2}}\right)\bigg], \,\, \textrm{for LNE}
\end{array} \right\}.
\end{eqnarray}

To examine $f(r)$ behavior at infinite range, we expand our solution by exploring $\rho(r)$ in Eq.\eqref{fr} at $r \rightarrow \infty$, which give the asymptotic solutions as
\begin{eqnarray}
\label{sol_rinf}
f(r)=
\left\{\begin{array}{cl}
1-\frac{2 m}{r} +\frac{\kappa ^2 Q^2}{2 r^2} -\frac{\kappa ^2 Q^4}{40 \beta ^2 r^6} + \mathcal{O}(r^{-7}) ,\,\, \textrm{for ENE}. \\
1-\frac{2 m}{r} +\frac{\kappa ^2 Q^2}{2 r^2}-\frac{\kappa ^2 Q^4}{320 \beta ^2 r^6} + \mathcal{O}(r^{-7}) ,\,\, \textrm{for LNE}
\end{array} \right\}.
\end{eqnarray}
Furthermore, we also evaluate Eq.\eqref{fr} at $\beta \rightarrow \infty$ and find that the limit gives same results for both models:
\begin{equation}
\label{rn}
f(r)= 1 -\frac{2 m}{r} +\frac{\kappa ^2 Q^2}{2 r^2}.
\end{equation}
Setting $\kappa=1$ and rescaling $Q^2/2 \rightarrow Q^2$, our solution reverts back to RN solution.

Since our solutions are asymptotically flat, we can furthermore identify our mass and charge as
\begin{equation}
M_{ADM}=m, \textrm{and} \,\,\,\,  Q=q.
\end{equation} 

\begin{figure}[!h]
	\centering
	\begin{tabular}{cc}
		\includegraphics[height=6.3cm,keepaspectratio]{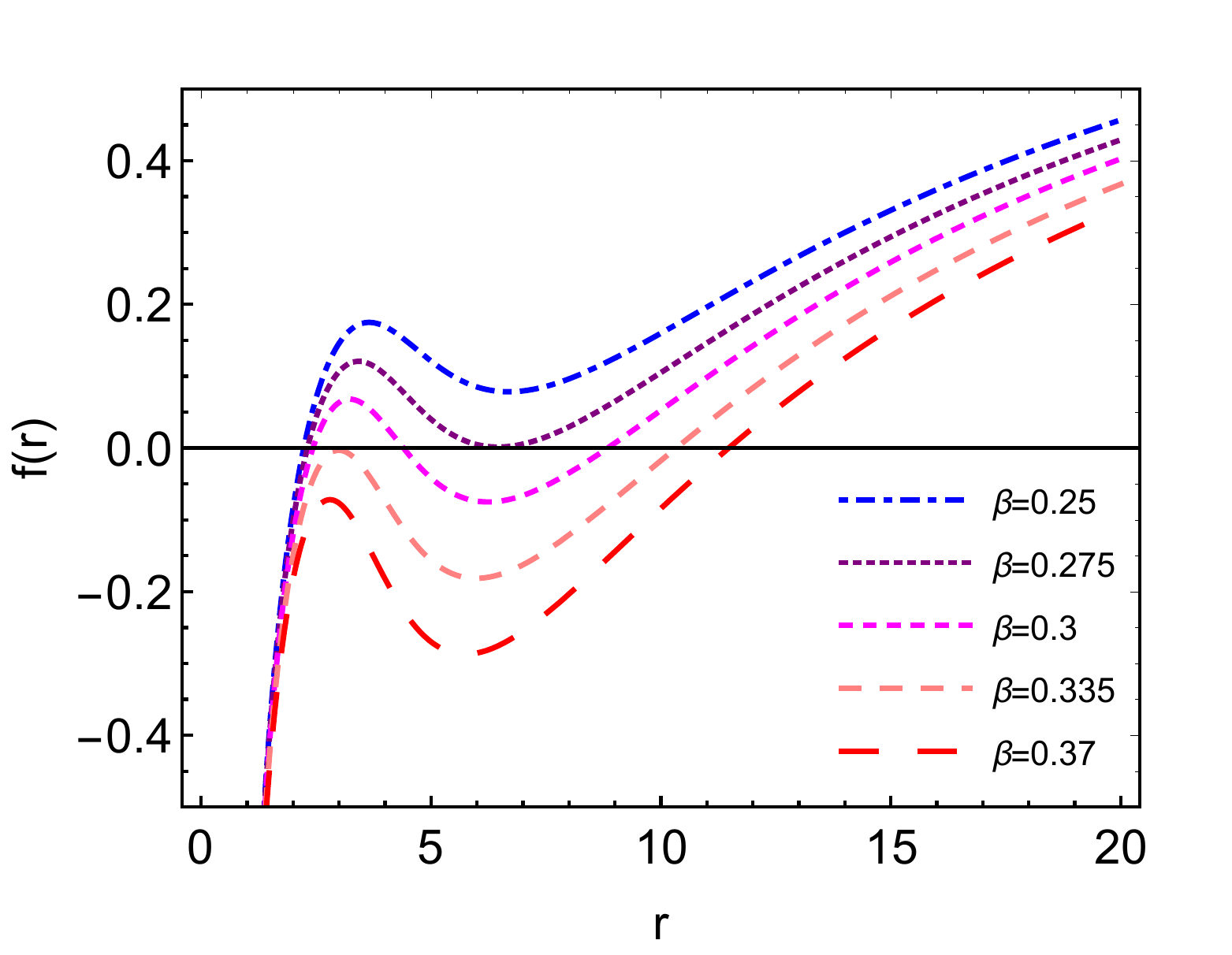} &		\includegraphics[height=6.3cm,keepaspectratio]{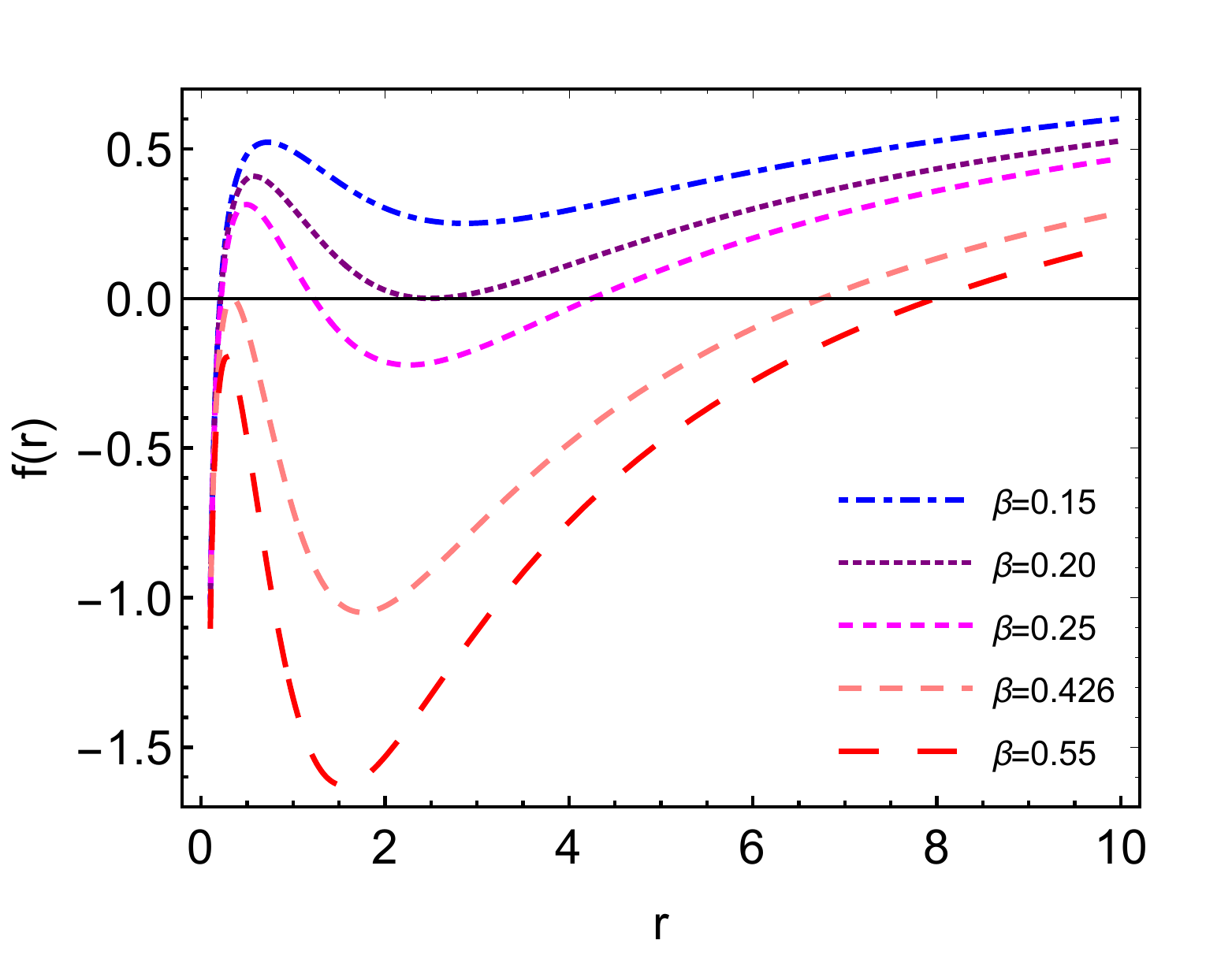}  
	\end{tabular}
	\caption{[Left] Plots of $f(r)$ of ENE with $M=1~\textrm{and}~Q=10$. [Right] Plots of $f(r)$ of LNE with $M=0.1~\textrm{and}~Q=4$.}
	\label{fQ}
\end{figure}

\begin{figure}[h!]
	\centering
	\includegraphics[scale=0.55]{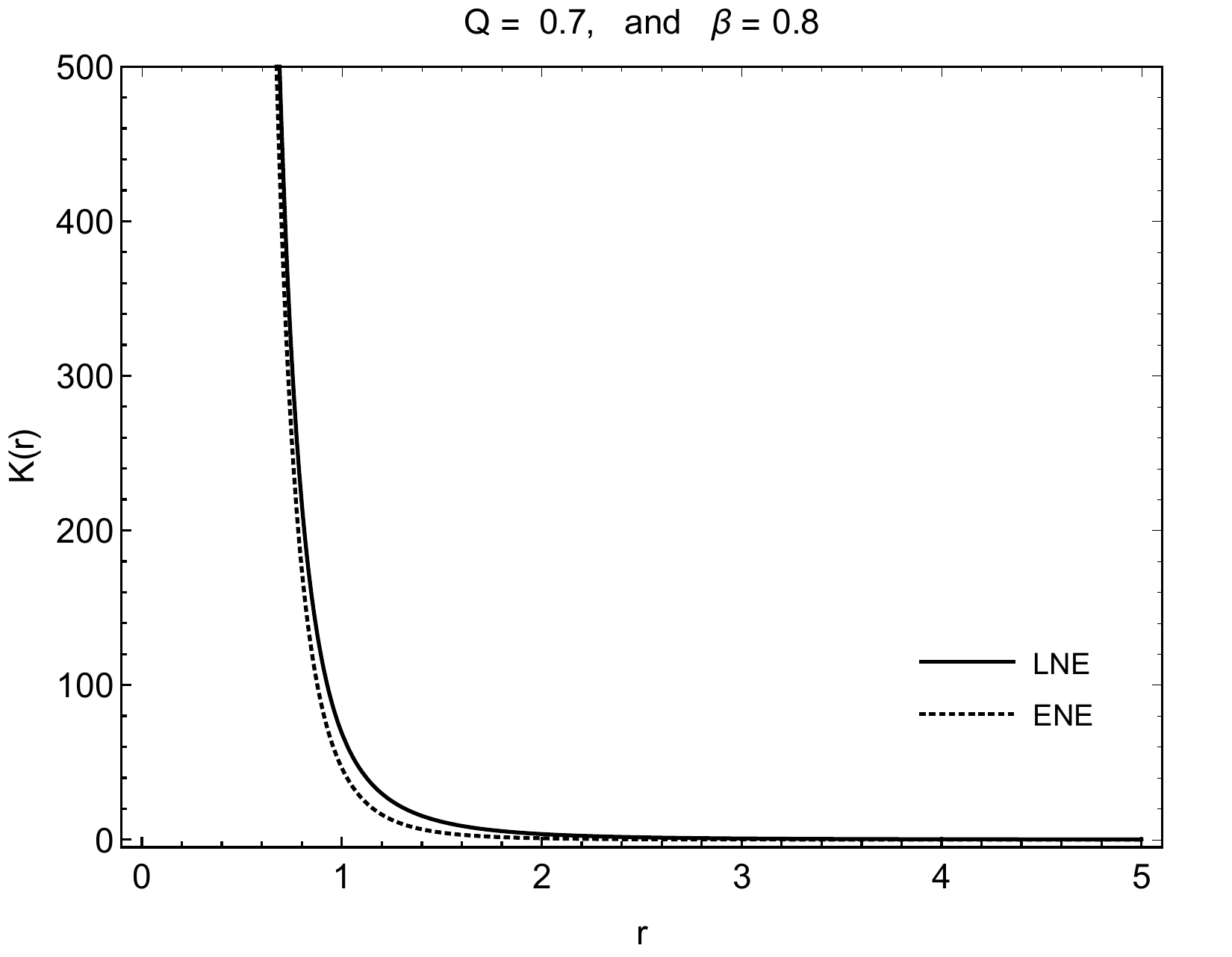}
	\caption{Kretschmann scalar of ENE and LNE with $ M=1 $. As we can see that these quantitiy diverge at the origin.}
	\label{fig:kre}
\end{figure}

The transcendental nature of the solutions makes it impossible to obtain the horizon(s) $f(r_h)=0$ analytically. 
\begin{figure}[!h]
	\centering
	\begin{tabular}{cc}
		\includegraphics[height=6.3cm,keepaspectratio]{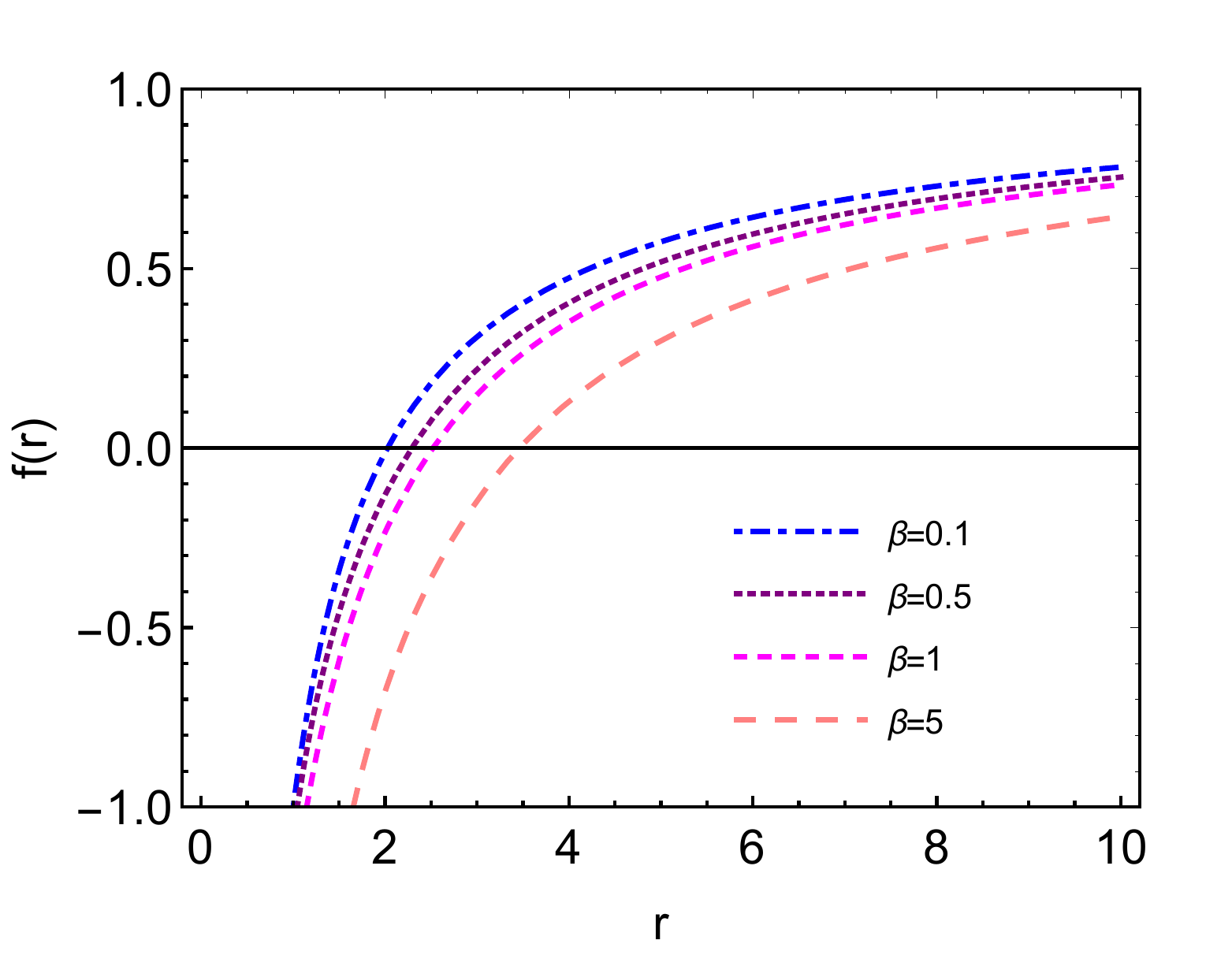} &		\includegraphics[height=6.3cm,keepaspectratio]{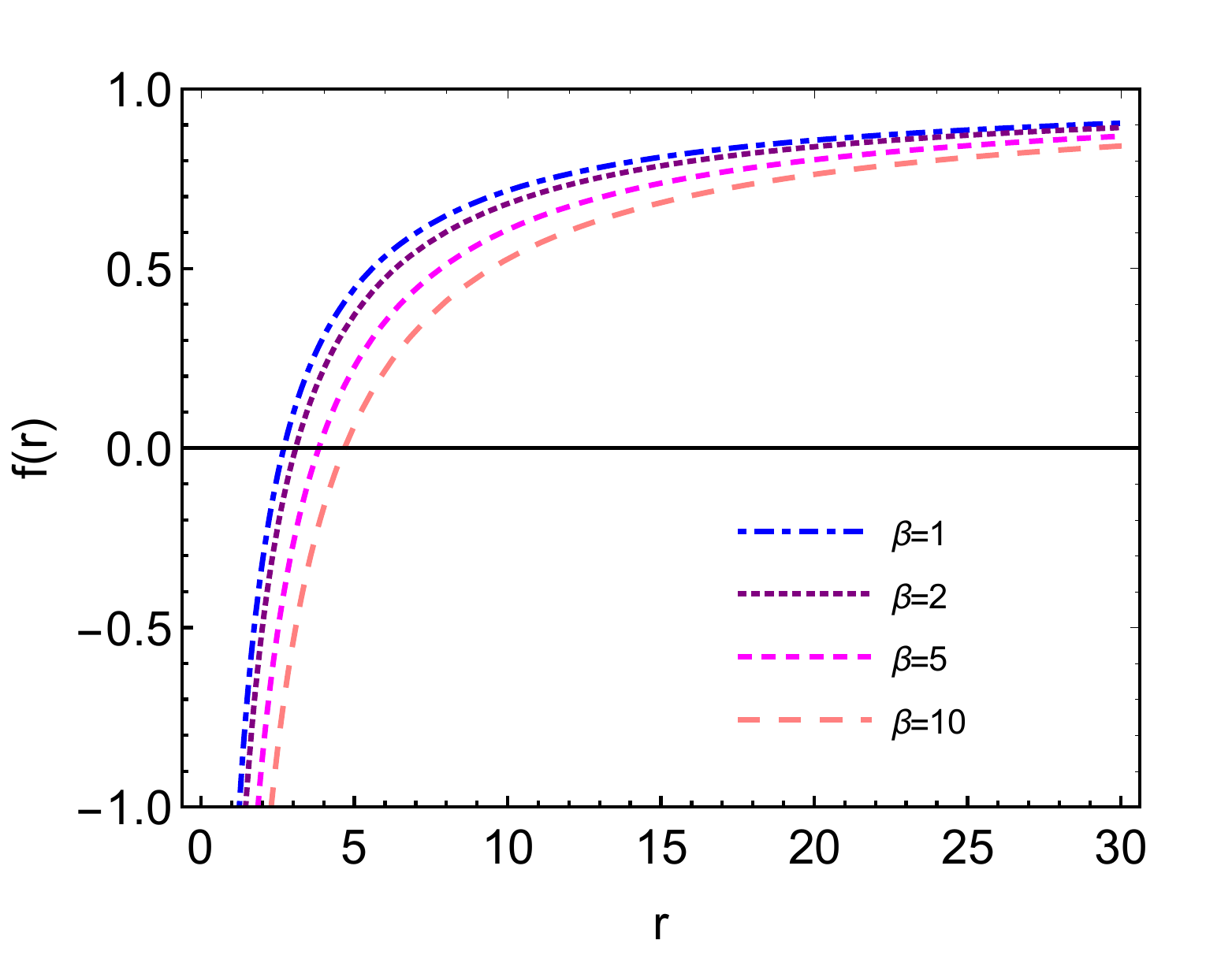}  
	\end{tabular}
	\caption{[Left] Plots of $f(r)$ of ENE with $M=Q=1$. [Right] Plots of $f(r)$ of LNE with similar parameter values.}
	\label{f}
\end{figure}
In Figs.~\ref{f}-\ref{fQ} we show typical plots of $ f(r) $ and the corresponding horizons for both cases. The horizon tends to shift to larger radius as $\beta$ increases. By varying the parameters solutions with two or three horizons can also be shown to exist. These horizons conceal singularity at the core, as can be seen from the diverging Kretschmann scalar shown in Fig.~\ref{fig:kre}. We also like to point out that unlike RN solution, the case of $Q>M$ does not producing naked singularity. While higher value of charge elevates the metric function, we find the limit $r\rightarrow 0$ always gives $f(r)=-\infty$. It ensure that at least one event horizon will exist regardless the amount of mass and charge of the black hole.


\section{Geodesics in One-horizon Case}
\subsection{Timelike Geodesics}
\label{tlg}

\begin{figure}[!h]
	\centering
	\begin{tabular}{cc}
		\includegraphics[height=6.5cm,keepaspectratio]{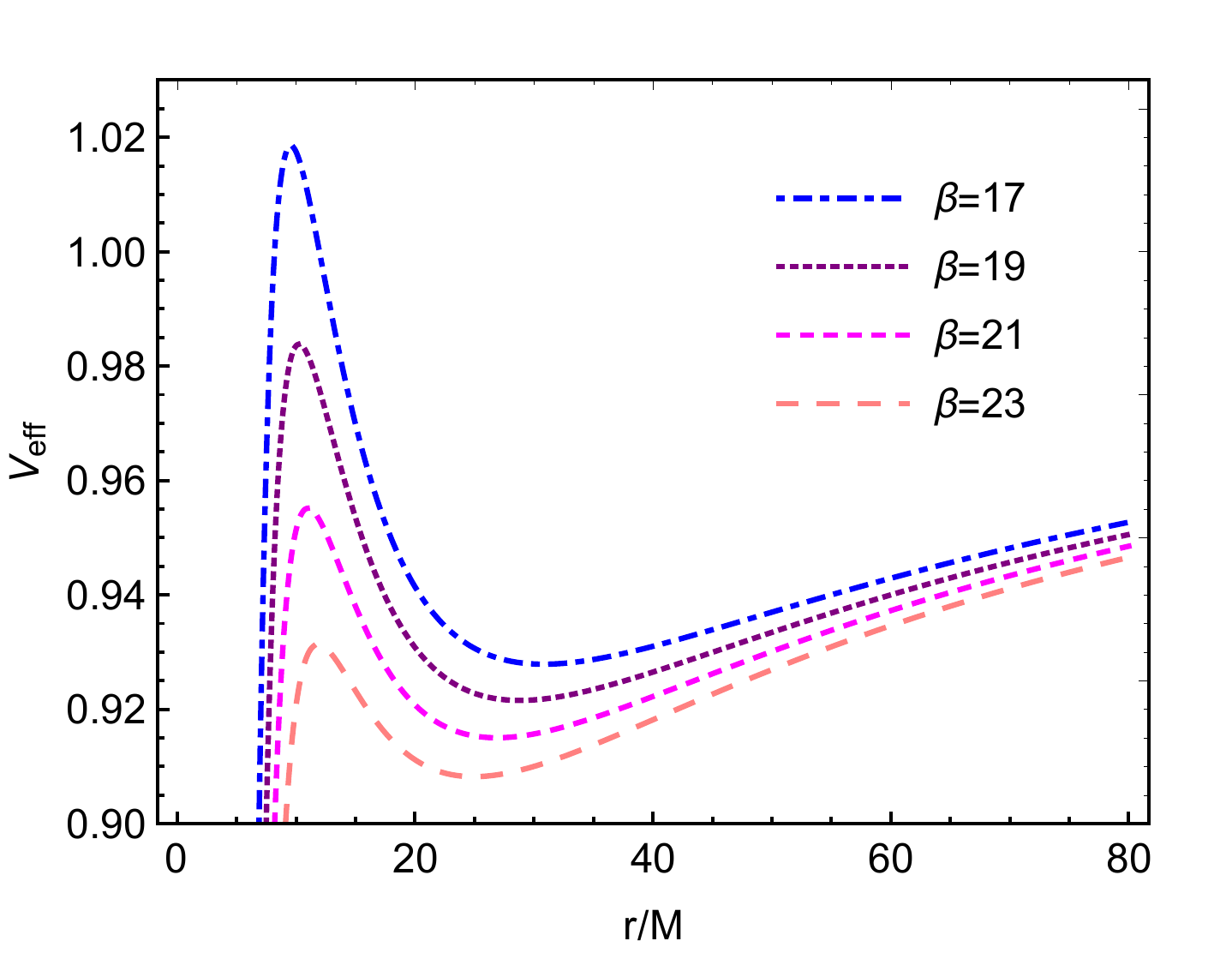} &		\includegraphics[height=6.5cm,keepaspectratio]{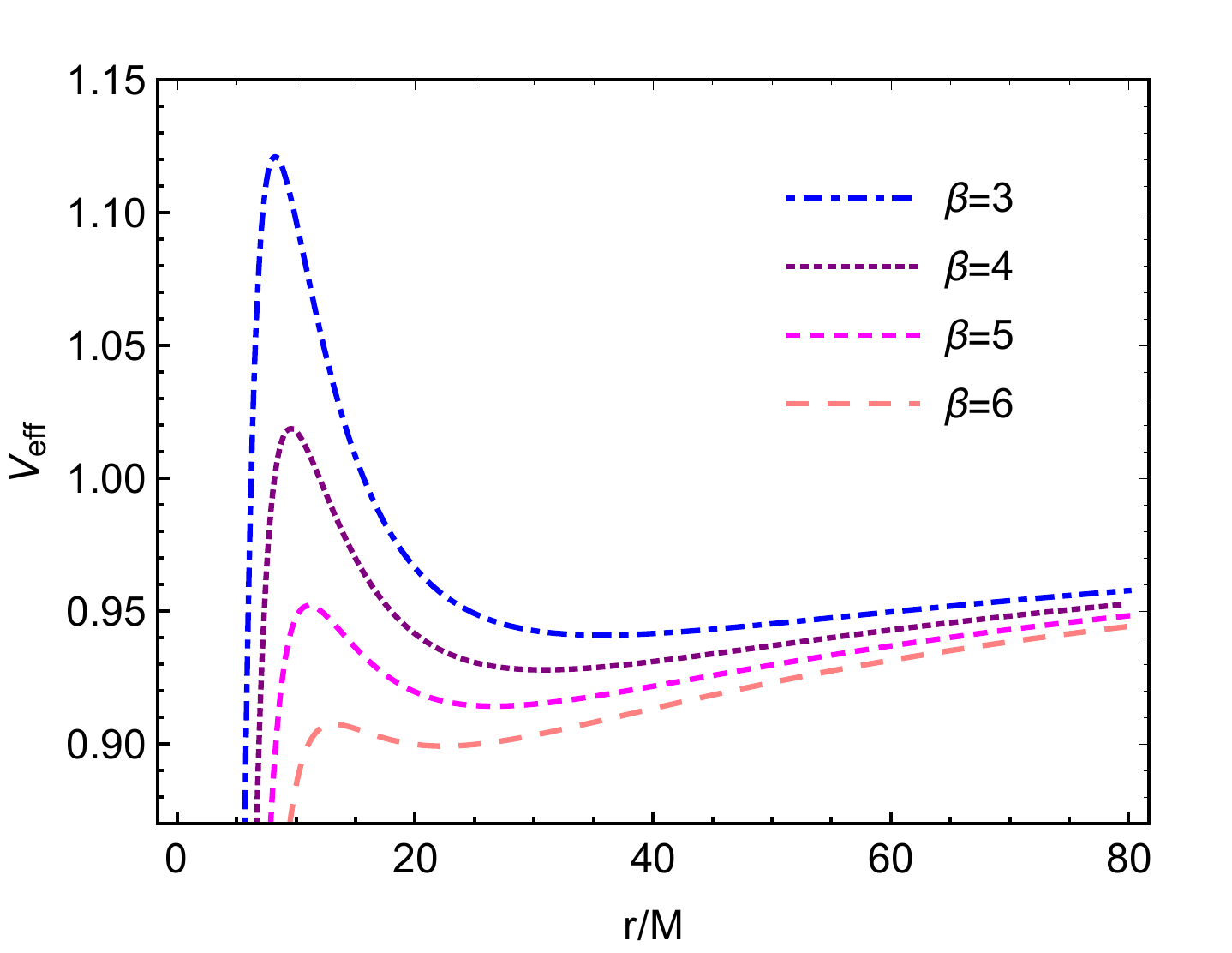}  
	\end{tabular}
	\caption{[Left] Plots of effective potential for massive particles of ENE with $Q=1$ and $L=0.01$. [Right] Plots of effective potential for massive particles of LNE with similar parameter values.}
	\label{tl}
\end{figure}

A test particle with mass $\mu$ and (electric/magnetic) charge $\epsilon$ around compact object can be described by the geodesics equation \cite{Breton:2001yk}
\begin{equation}
\label{geodesic}
\frac{d^2 x^{\nu}}{d\tau^2} +\Gamma_{\alpha\beta}^{\nu} \frac{dx^{\alpha}}{d\tau} \frac{dx^{\beta}}{d\tau} = -\frac{\epsilon}{\mu} F_{\sigma}^{\,\,\nu} \frac{dx^{\sigma}}{d\tau}.
\end{equation}
For our metric \eqref{sss}, the timelike geodesics on equatorial plane (due to spherical symmetry, $\theta = \pi/2$) can be written as
\begin{equation}
\label{tl_geo}
1= f \dot{t}^2 -f^{-1} \dot{r}^2 -r^2 \dot{\phi}^2.
\end{equation}
Further, the spherical symmetry and staticity imply two integrals of motion:
\begin{equation}
\label{dot_E}
\dot{t} = \frac{\mathbb{E}}{f} \,\,\,\, , \,\,\,\,
\dot{\phi} = \frac{\mathbb{L}}{r^2}.
\end{equation}
where $\mathbb{E}$ and $\mathbb{L}$ are the energy-and angular momentum-per unit mass of the test charged particles, respectively. Eq.~\eqref{tl_geo} can be rewritten as
\begin{equation}
\label{tg}
\dot{r}^2 = \mathbb{E}^2 - f(r) \bigg(\frac{\mathbb{L}^2}{r^2} + 1\bigg) .
\end{equation}
We can then easily define the ``one-dimensional" effective potential
\begin{equation}
\label{veff}
V_{eff}(r) = f(r) \bigg(\frac{\mathbb{L}^2}{r^2} + 1\bigg).
\end{equation}
Rescaling $r \rightarrow \frac{r}{M}$, $L \rightarrow \frac{M^2}{\mathbb{L}^2}$, $Q \rightarrow \frac{Q}{M}$, and $\beta \rightarrow \beta M$,

the effective potential and the orbital equation can be cast as
\begin{eqnarray}
\label{Veff_TL}
\bigg(\frac{dr}{d\phi}\bigg)^2 = \mathbb{E}^2 L r^4 - f(r) \left(r^2 +  L r^4\right) \equiv \mathcal{R}(r) \,\,\,\, , \,\,\,\, V_{eff}(r)= f(r)\left(\frac{1}{L r^2}+1\right).
\end{eqnarray}
Typical plots of $V_{eff}$ can be seen in Fig.~\ref{tl}.

\begin{figure}[!h]
	\centering
	\begin{tabular}{c}
		\includegraphics[height=6.5cm,keepaspectratio]{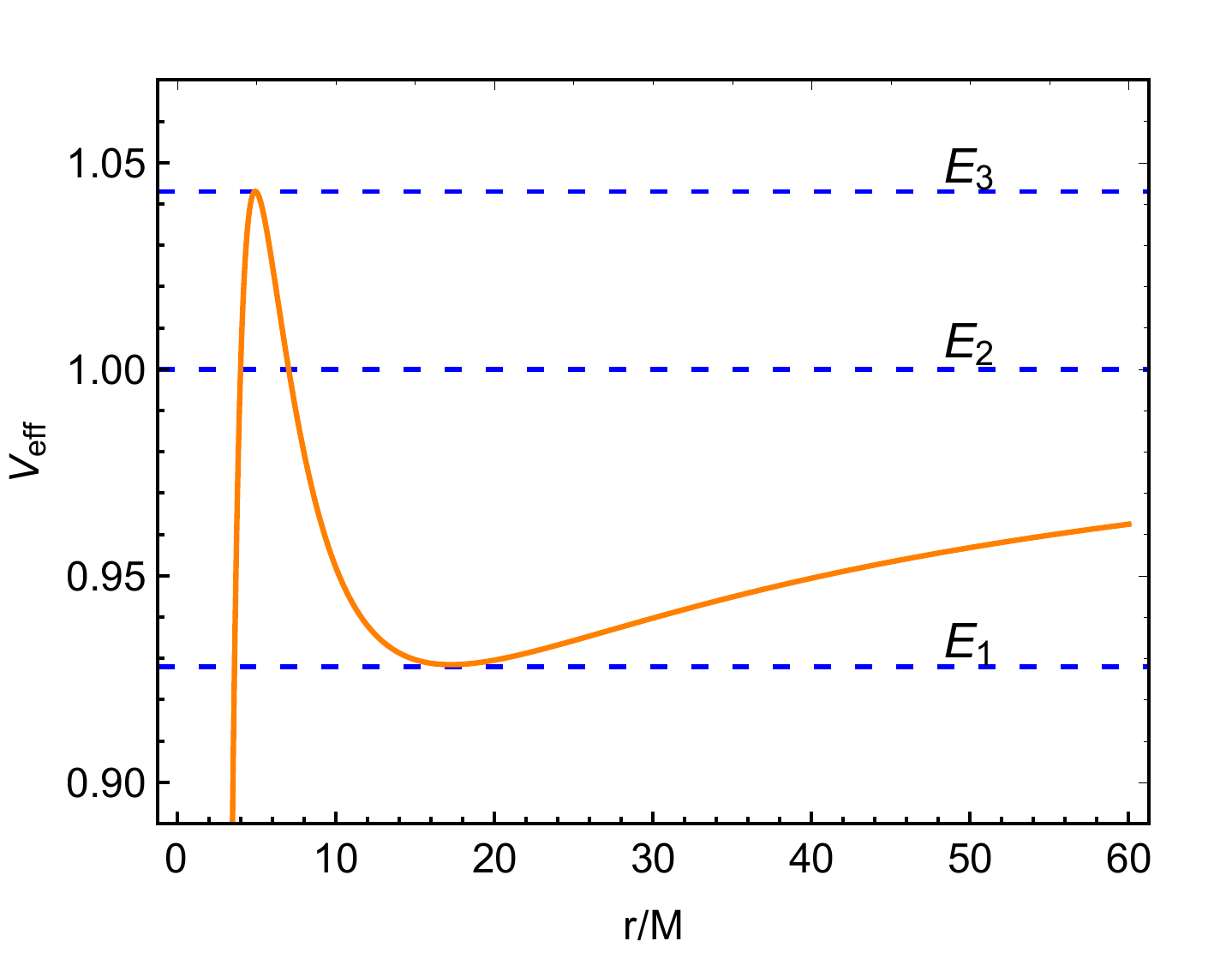}
	\end{tabular}
	\caption{A generic shape of $V_{eff}$ for LNE BH, divided by three energy levels into four regions. Each region determines the possible orbit types of the test particles.}
	\label{tl_}
\end{figure}

The orbital equation \eqref{Veff_TL} can be solved numerically. But even without doing so, the nature of orbits can be extracted by utilizing the shapes of the effective potential. An example of various energy level in one-horizon LNE black hole is shown in Fig. \ref{tl_}. As in the case of Schwarzchild and RN which have extensively been discussed in \cite{Hackmann:2010tqa}, here we can also identify types of orbits by their characteristics:

\begin{enumerate}[label=\roman*., itemsep=0pt, topsep=0pt]
	\item {\it Flyby orbit}: particle comes from $\infty$, approaching a periapsis $r_p$ and goes back to $\infty$.
	\item {\it Bound orbit}: particle oscillates between its periapsis and apoapsis ($r_p \leq r \leq r_a$) with $r_{EH} < r_p < r_a < \infty$.
	\item {\it Terminating bound orbit}: particle starts in the range of $r_{EH} < r_a < \infty$ and falls into singularity.
	\item {\it Terminating escape orbit}: particle comes from $\infty$ and falls into singularity.
\end{enumerate}
There are also the {\it Unstable Circular Orbit} (UCO) and the {\it Stable circular Orbit} (SCO), corresponding to the cases when the energy is exactly at the top of the potential barrier (the local maximum) and at its valley (local minimum), respectively.

The appearance of those orbits depends on the amount of real solutions of $\mathcal{R}(r)$ in \eqref{Veff_TL}. Since the energy of test particles directly affect the shape of orbit, it is therefore necessary to provide the contribution of $\mathbb{E}^2$ in the classification. By varying the parameters $Q, \beta, E$ and $L$, and comparing them to energy levels in Fig. \ref{tl_}, we can list the four regions as below.
\begin{enumerate}[label=\arabic*., itemsep=0pt, topsep=0pt]
	\item region I ($E_2 < \mathbb{E}^2 < E_3$): 2 positive real solutions, result in flyby and terminating bound orbits,
	\item region II ($\mathbb{E}^2 > E_3$): 0 positive real solutions, result in terminating escape orbits,
	\item region III ($\mathbb{E}^2 < E_1$): 1 positive real solution, results in terminating bound orbits,
	\item region IV ($E_1 < \mathbb{E}^2 < E_2$): 3 positive real solutions, result in bound and terminating bound orbits.
\end{enumerate}

\begin{figure}[!h]
	\centering
	\begin{tabular}{cccc}
		\includegraphics[height=3.6cm,keepaspectratio]{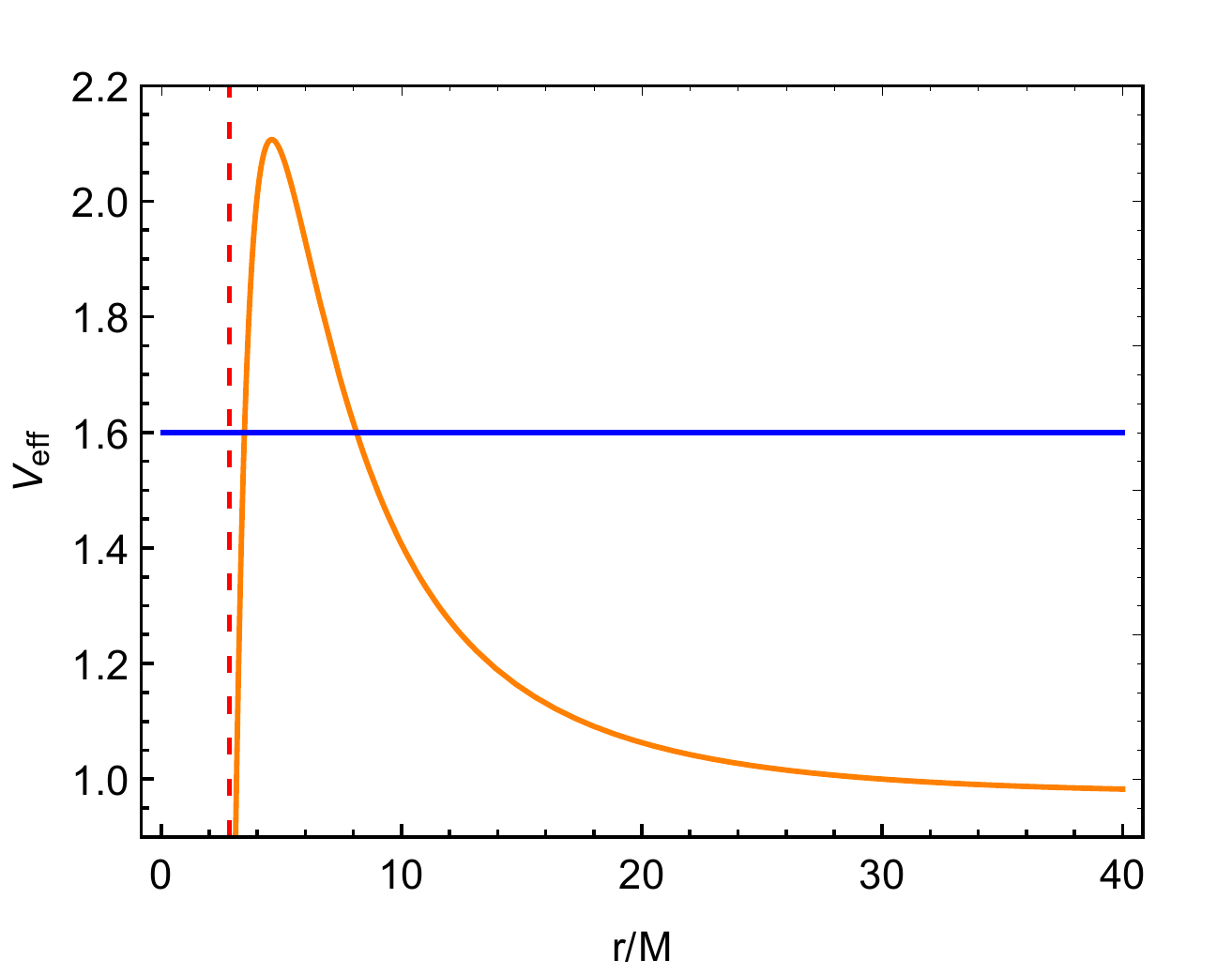}&	\includegraphics[height=3.6cm,keepaspectratio]{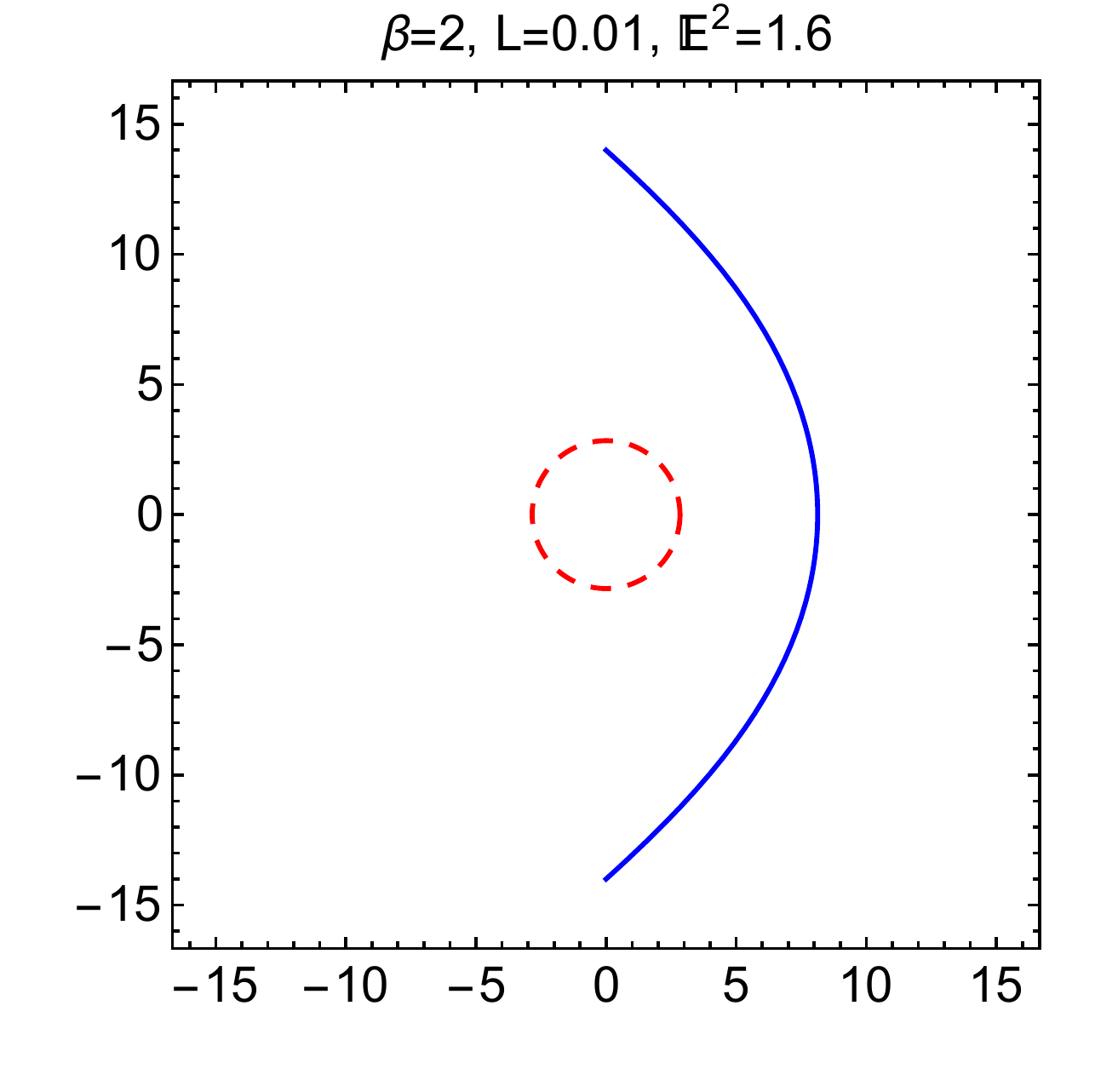}&
		\includegraphics[height=3.6cm,keepaspectratio]{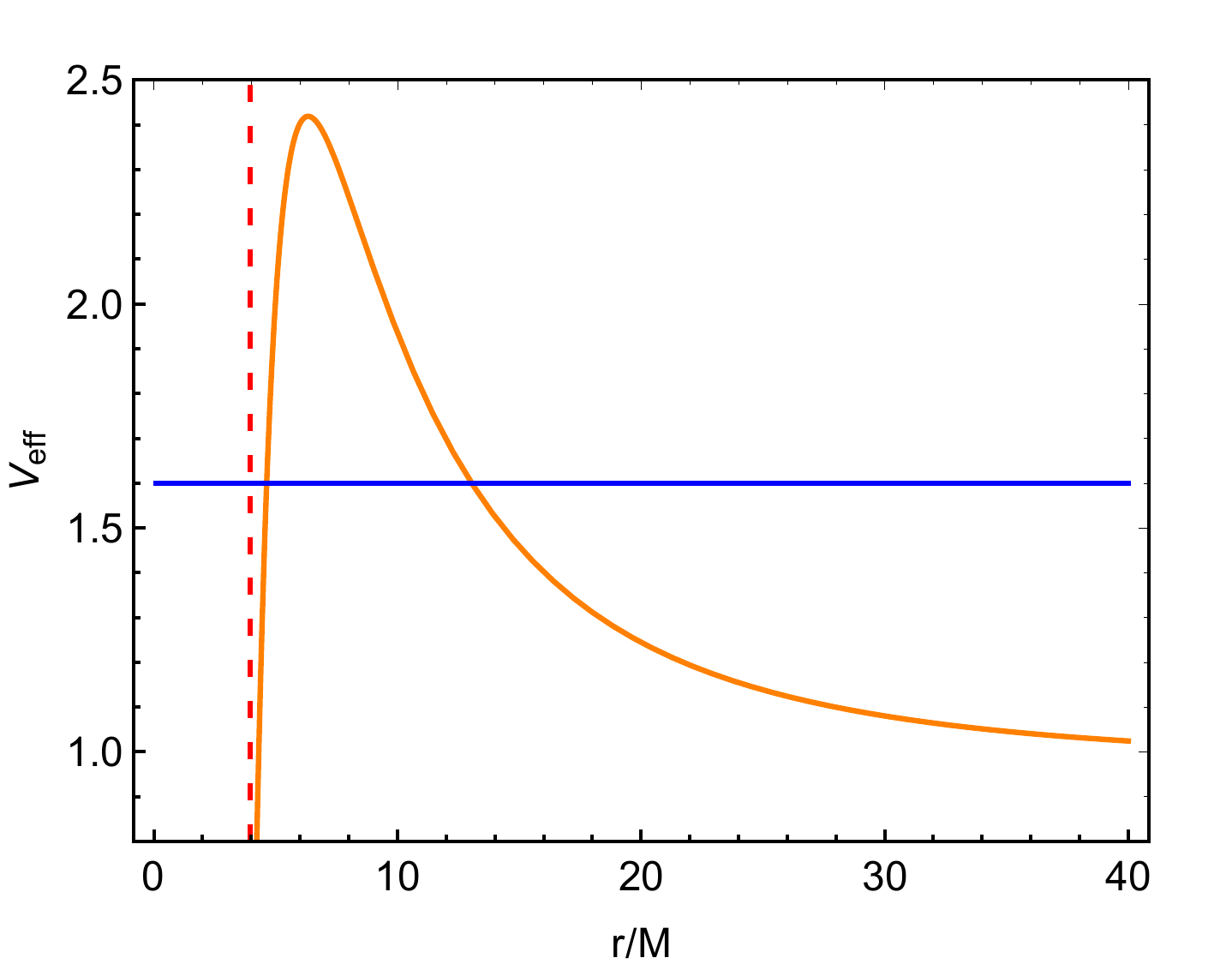}&	\includegraphics[height=3.6cm,keepaspectratio]{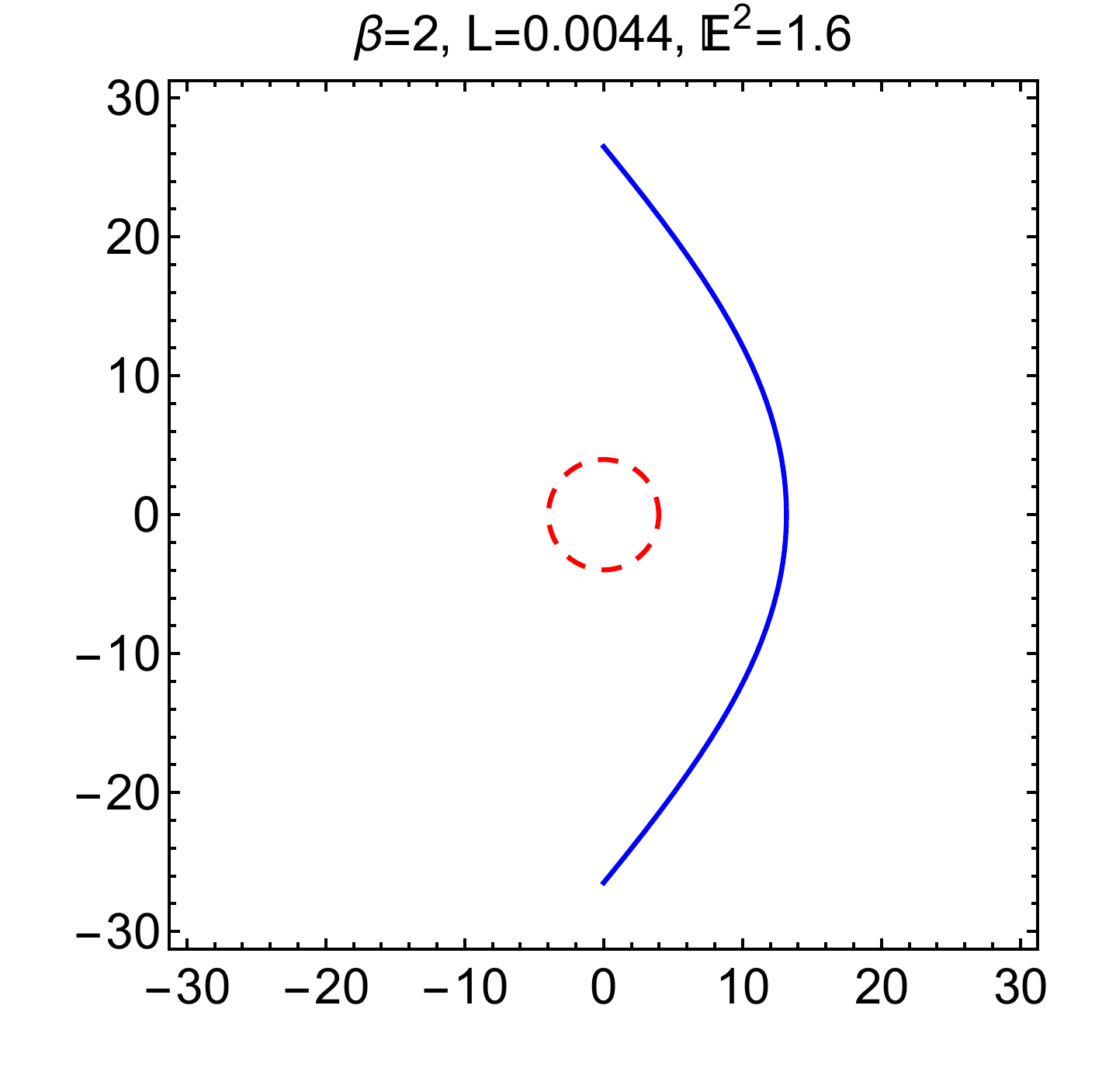}  
	\end{tabular}
	\begin{tabular}{cccc}
		\includegraphics[height=3.6cm,keepaspectratio]{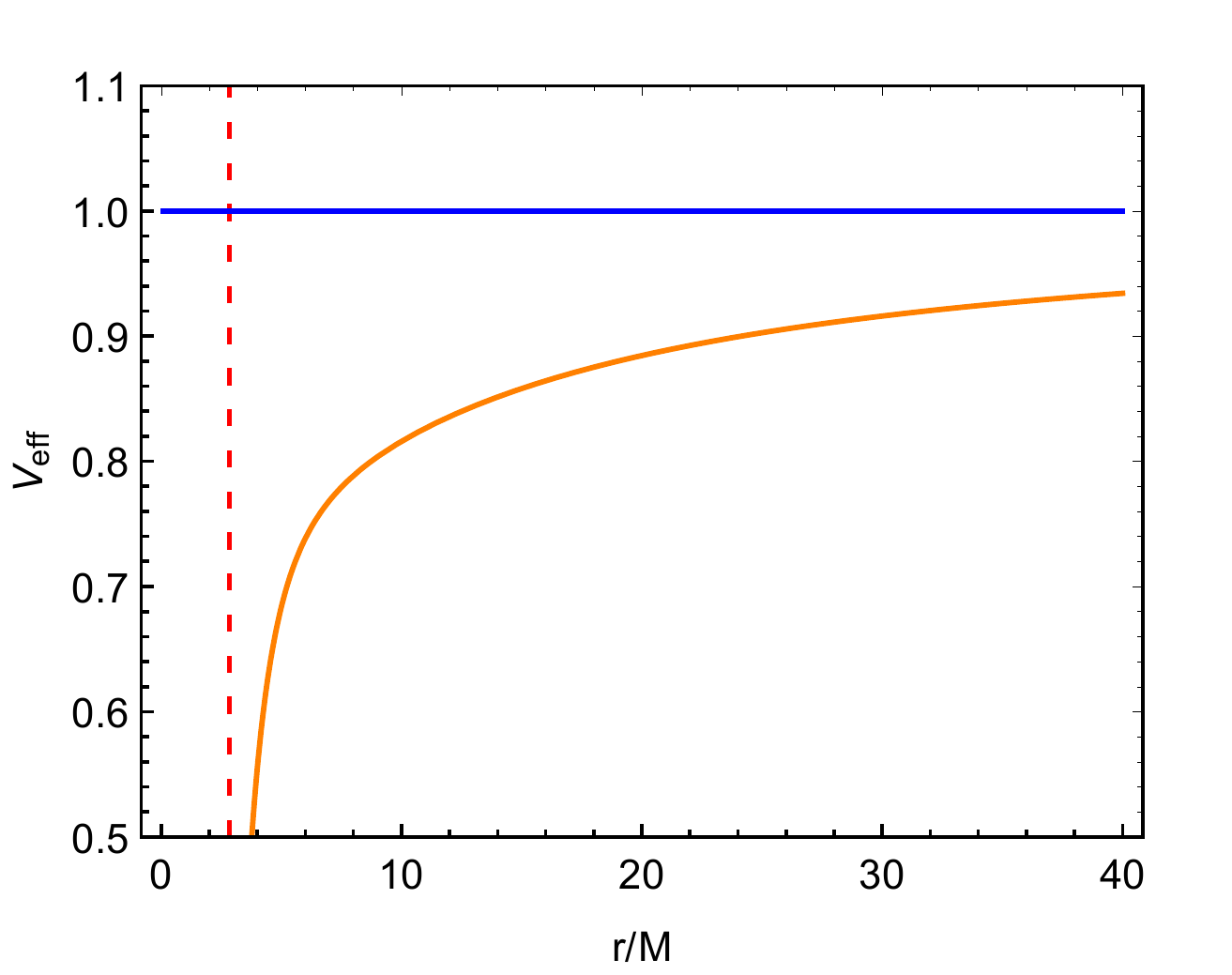}&	\includegraphics[height=3.6cm,keepaspectratio]{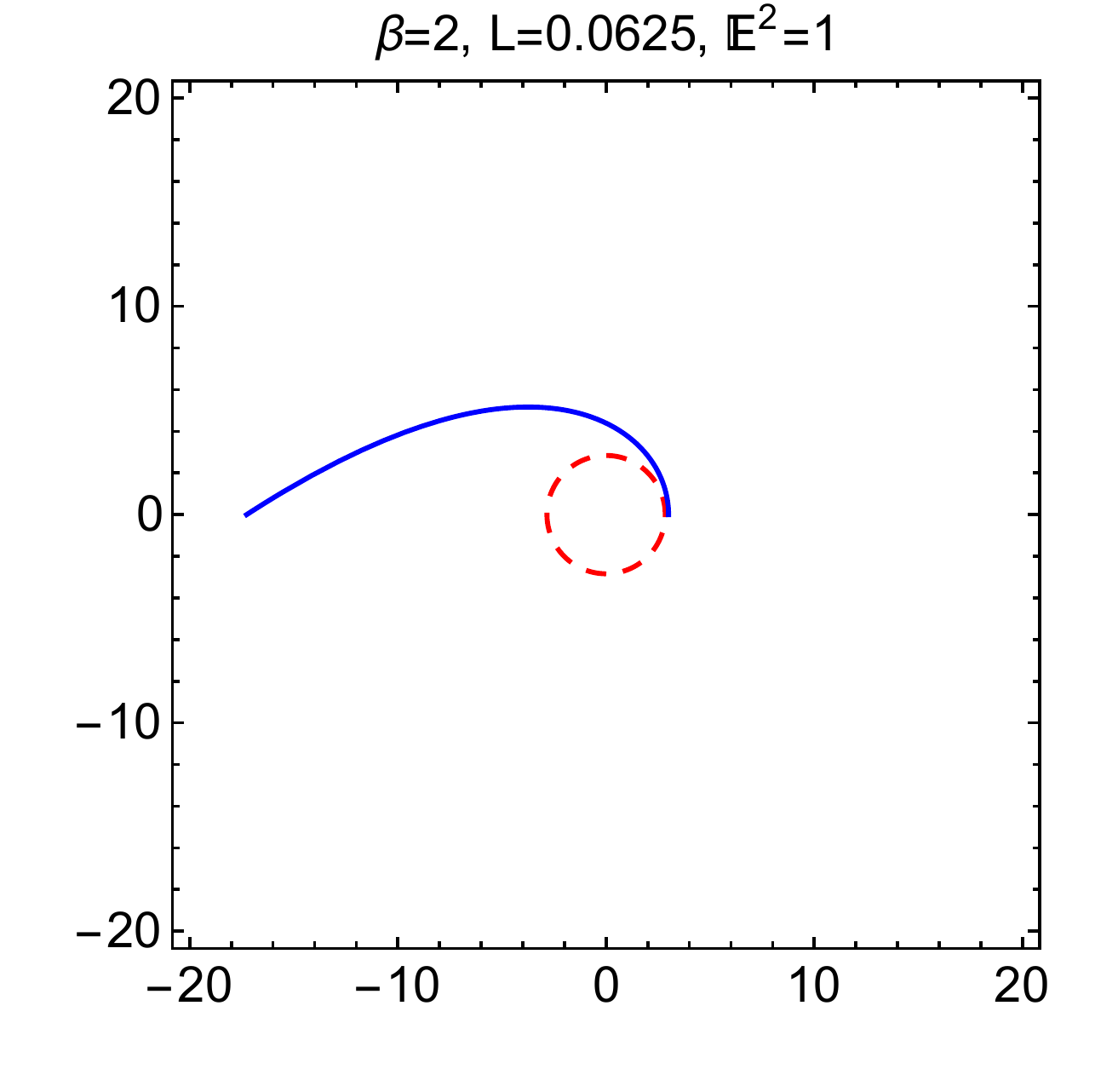}&
		\includegraphics[height=3.6cm,keepaspectratio]{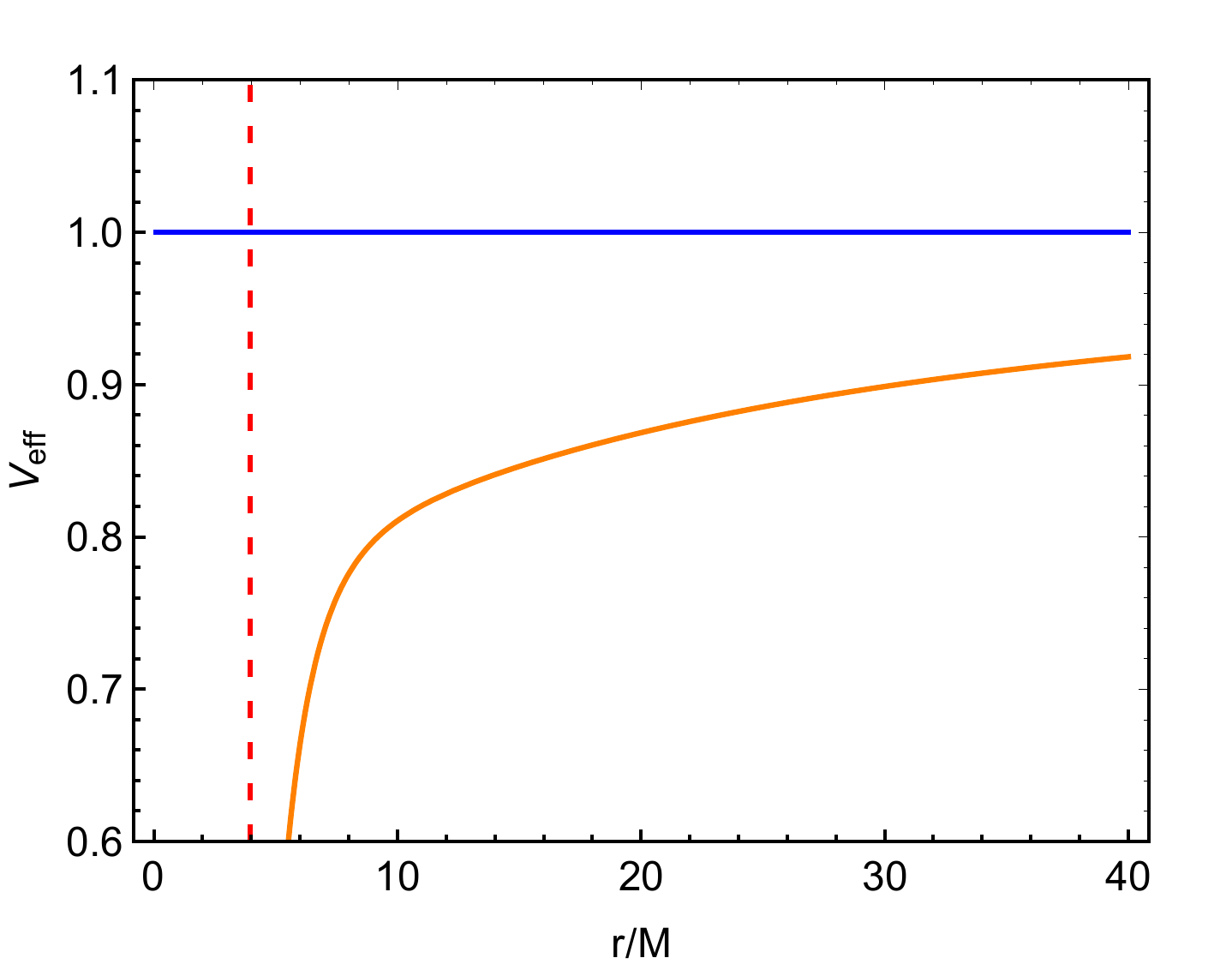}&	\includegraphics[height=3.6cm,keepaspectratio]{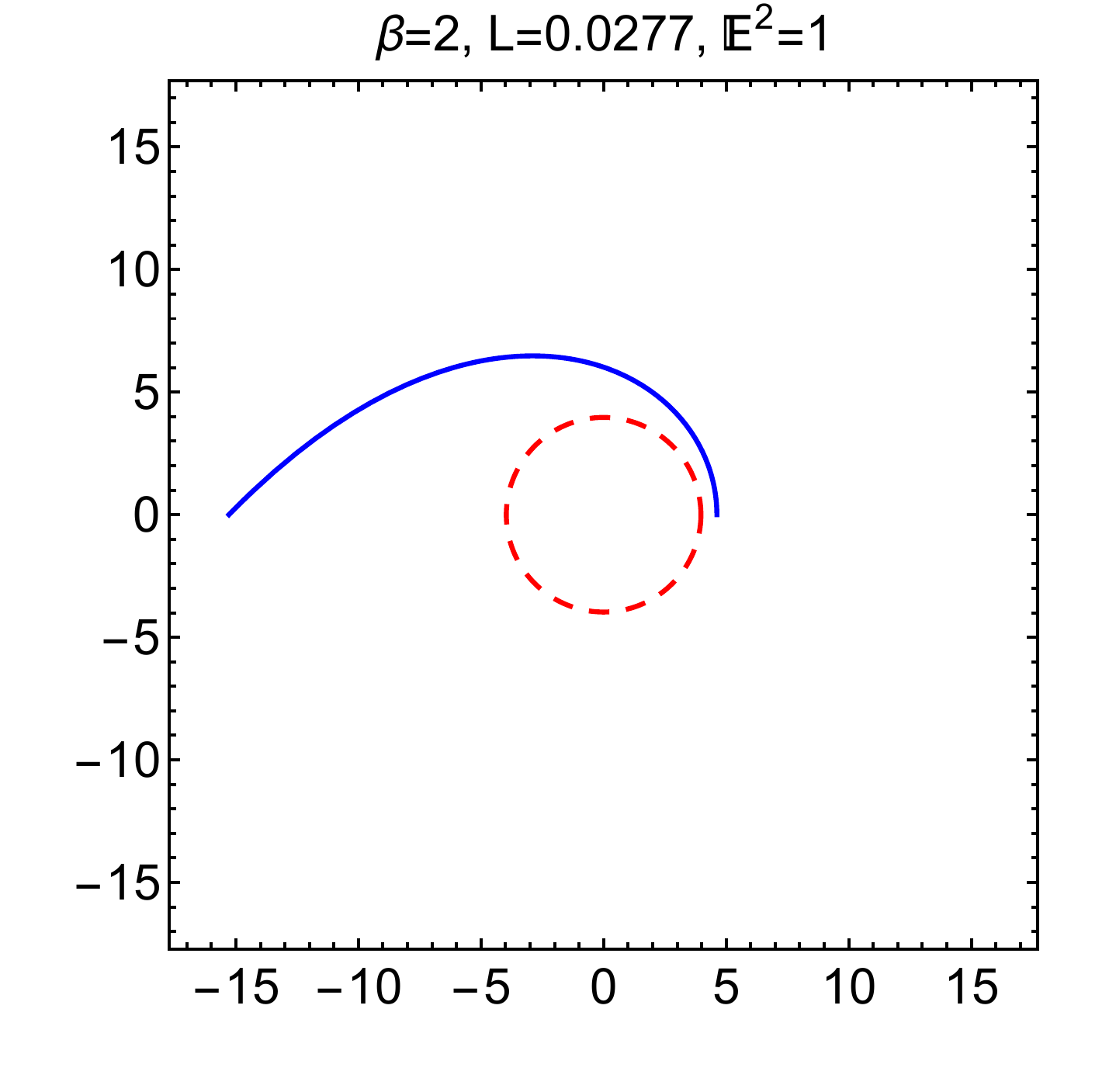} 
	\end{tabular}
	\begin{tabular}{cccc}
		\includegraphics[height=3.64cm,keepaspectratio]{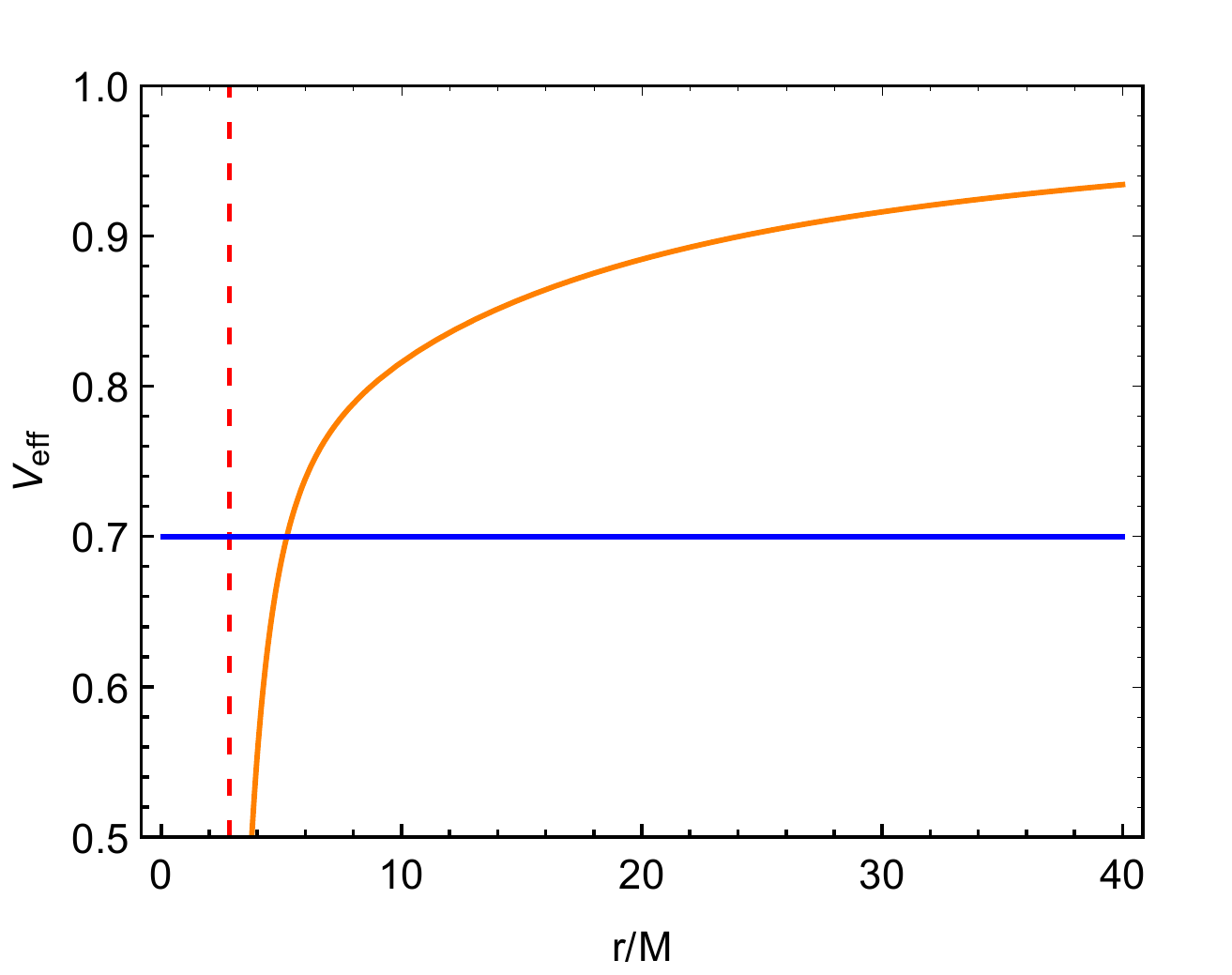}&	\includegraphics[height=3.64cm,keepaspectratio]{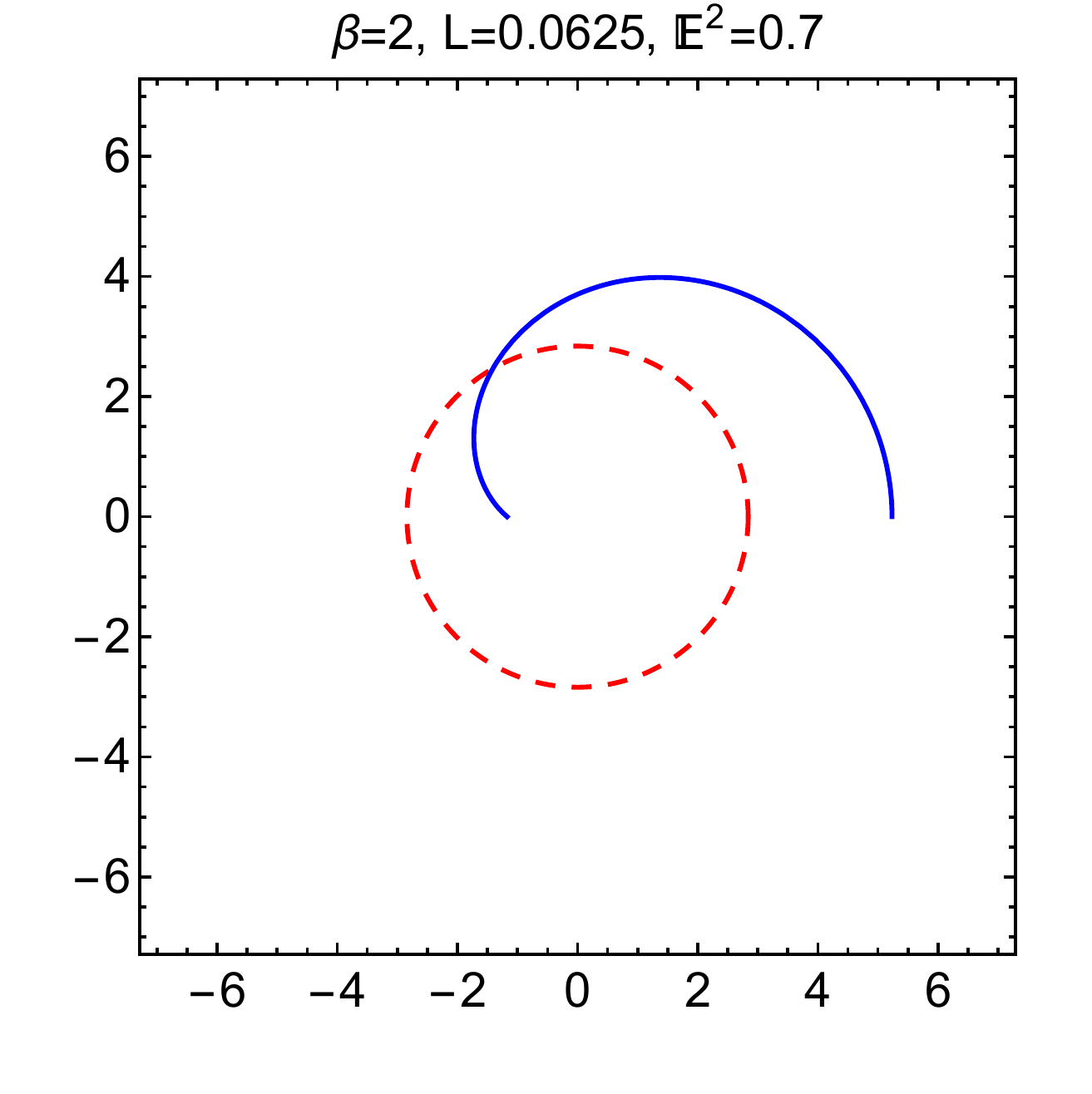}&
		\includegraphics[height=3.6cm,keepaspectratio]{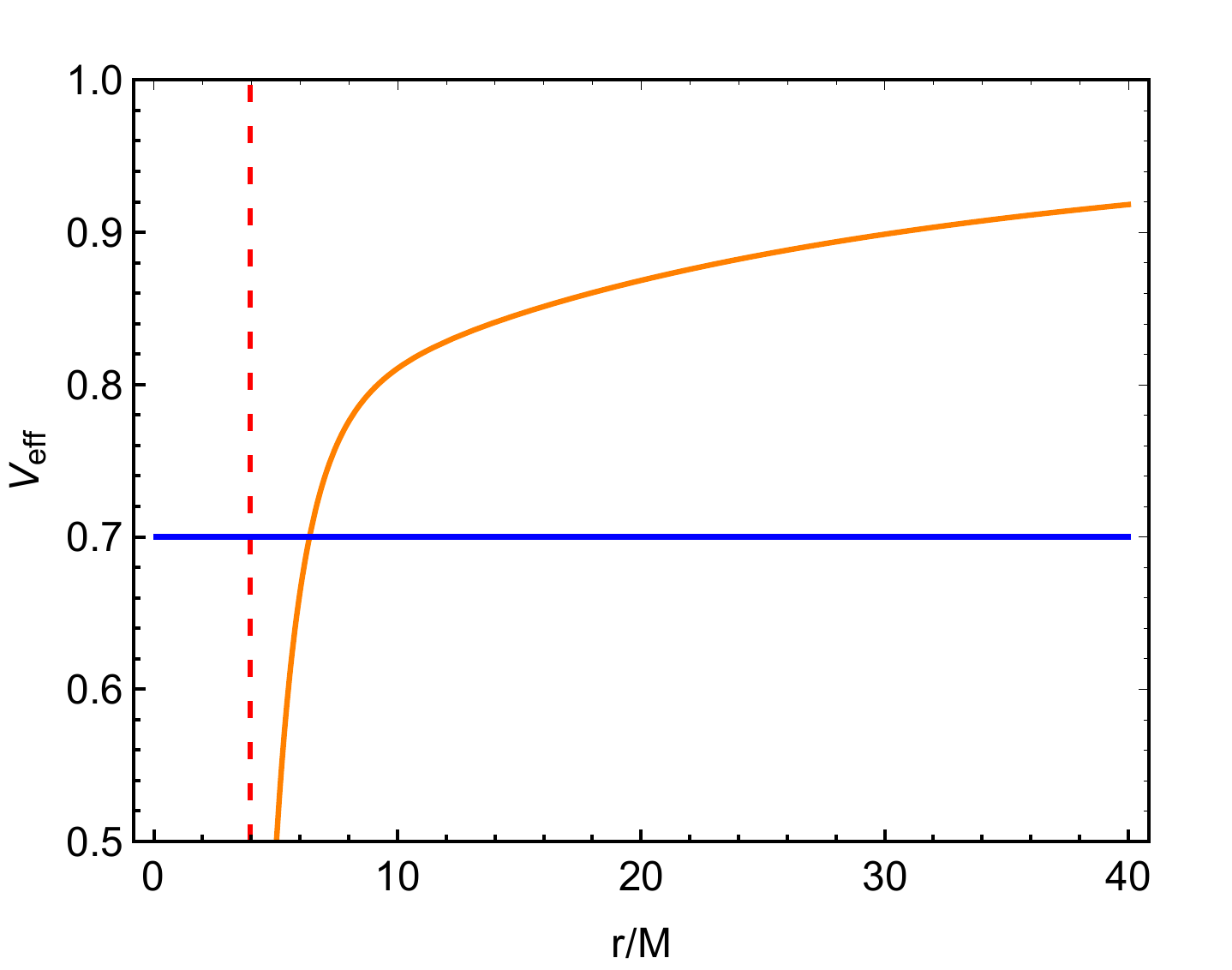}&	\includegraphics[height=3.6cm,keepaspectratio]{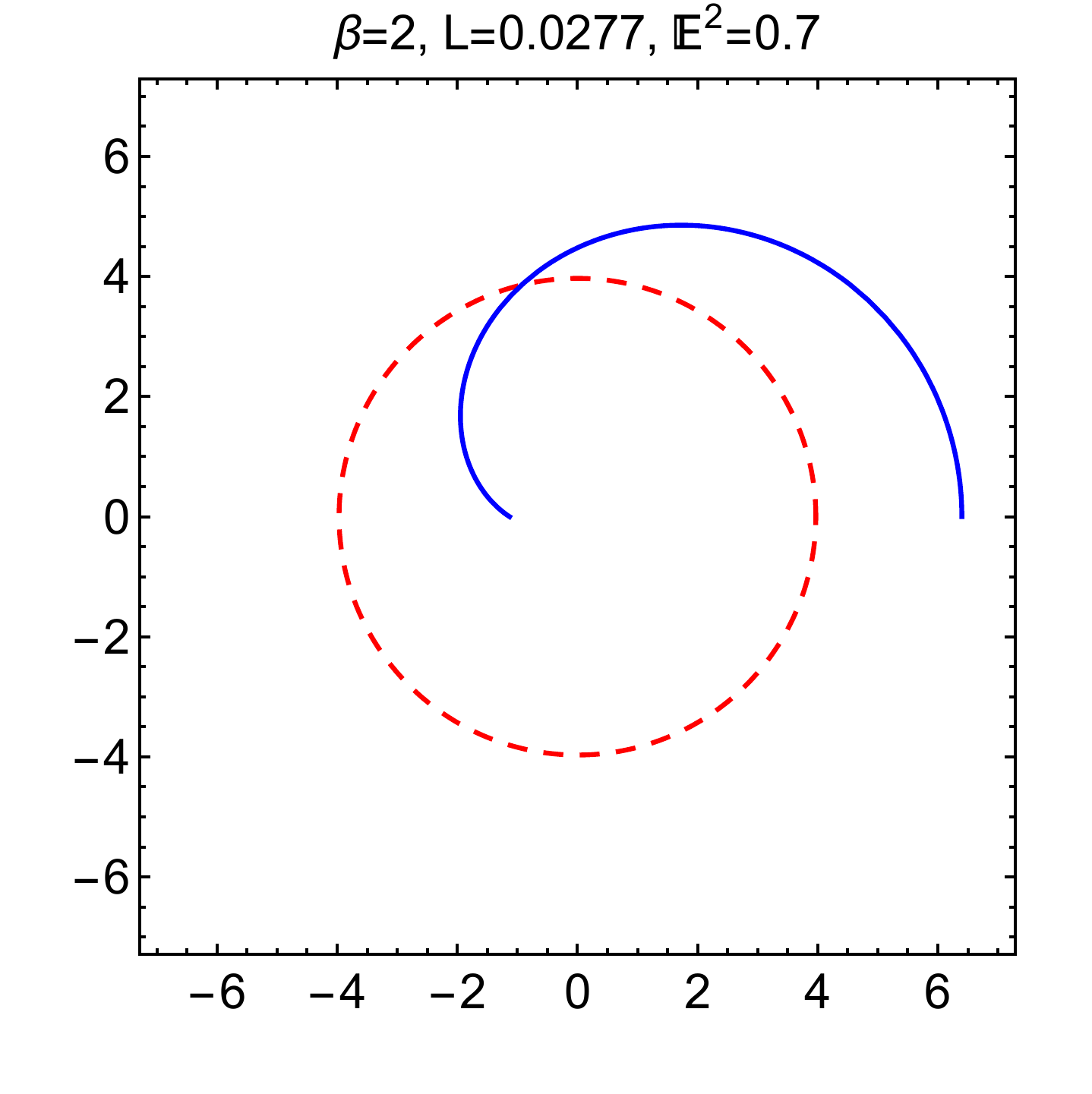} 
	\end{tabular}
	\begin{tabular}{cccc}
		\includegraphics[height=3.56cm,keepaspectratio]{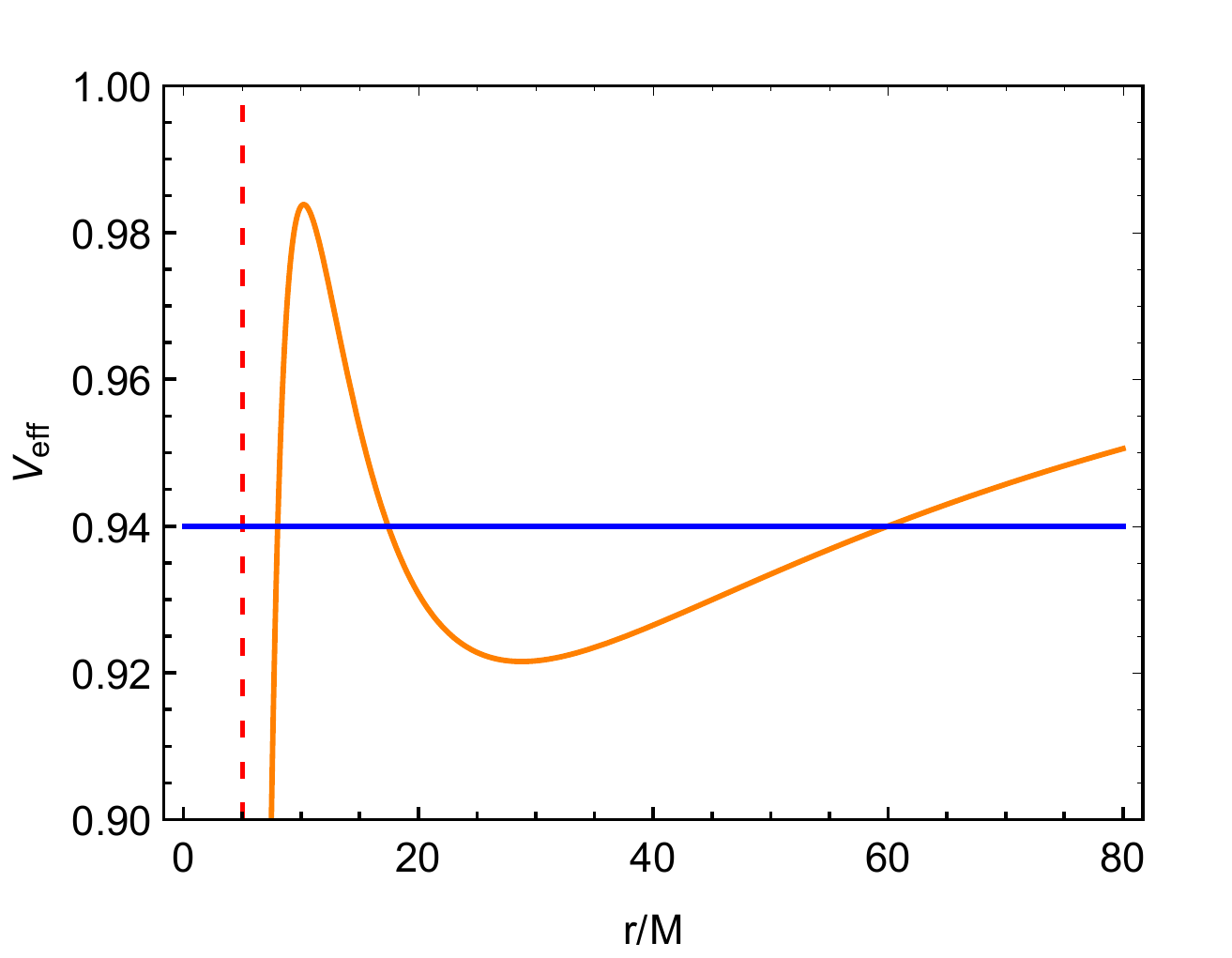}&	\includegraphics[height=3.56cm,keepaspectratio]{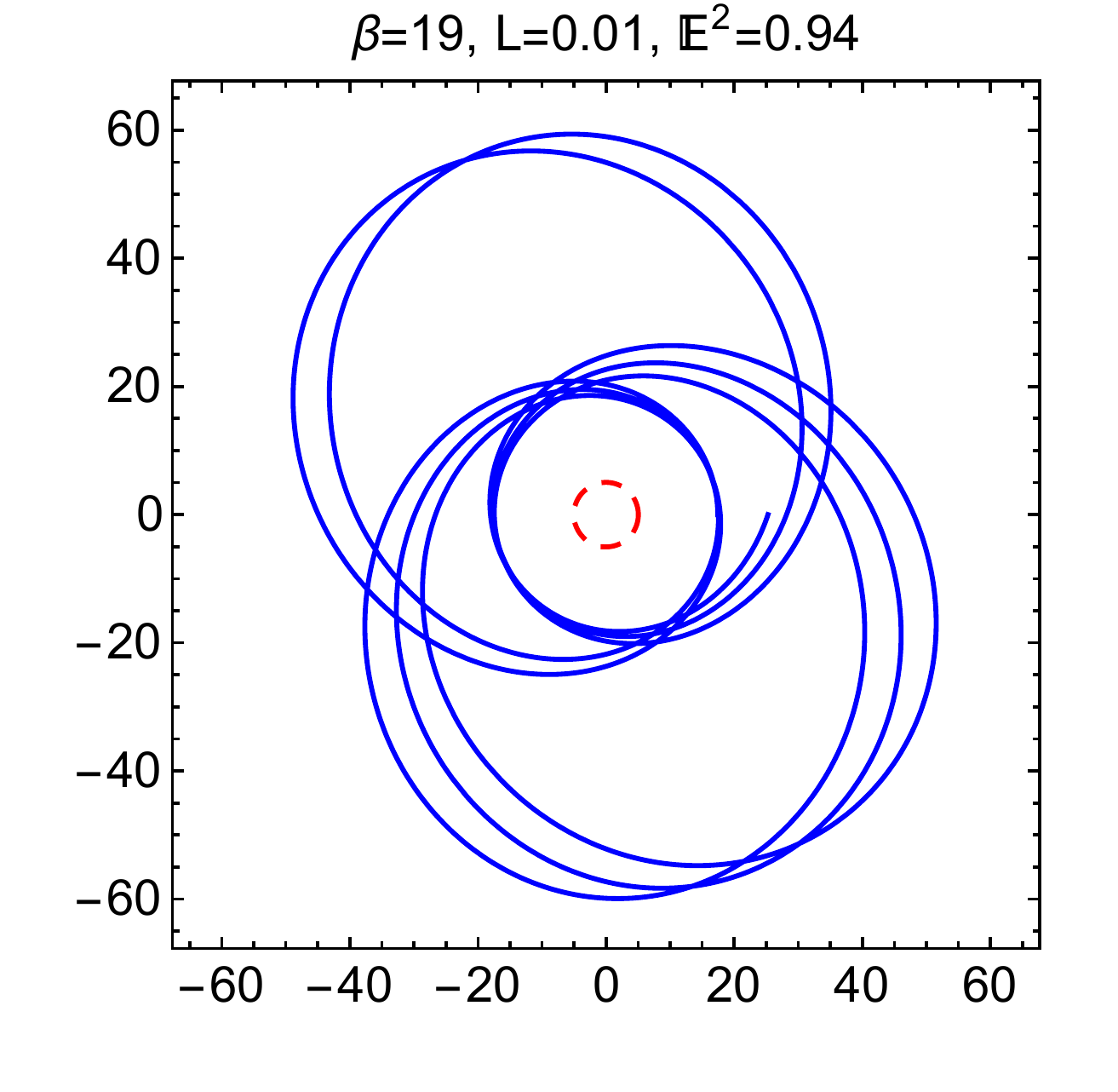}&
		\includegraphics[height=3.6cm,keepaspectratio]{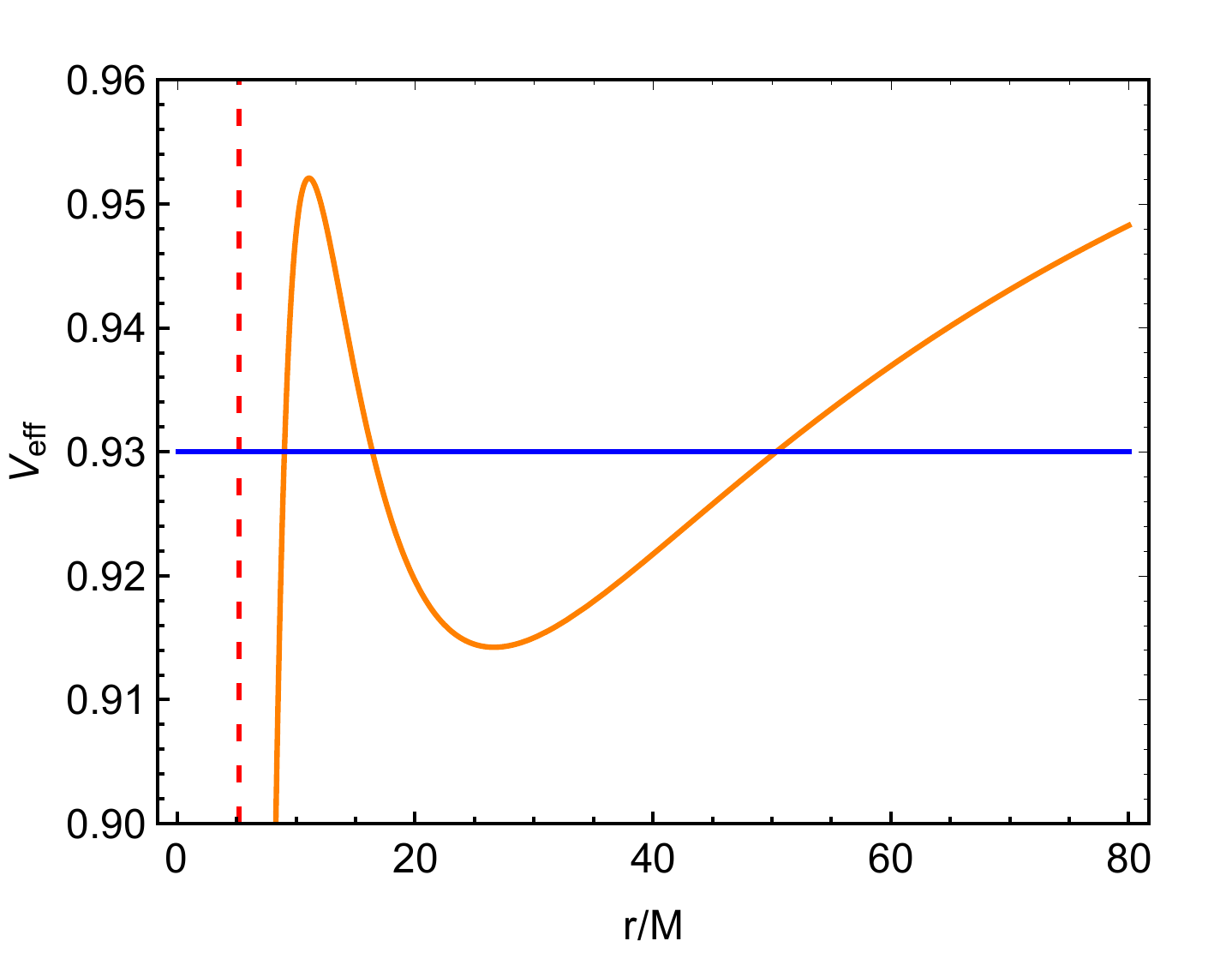}&	\includegraphics[height=3.6cm,keepaspectratio]{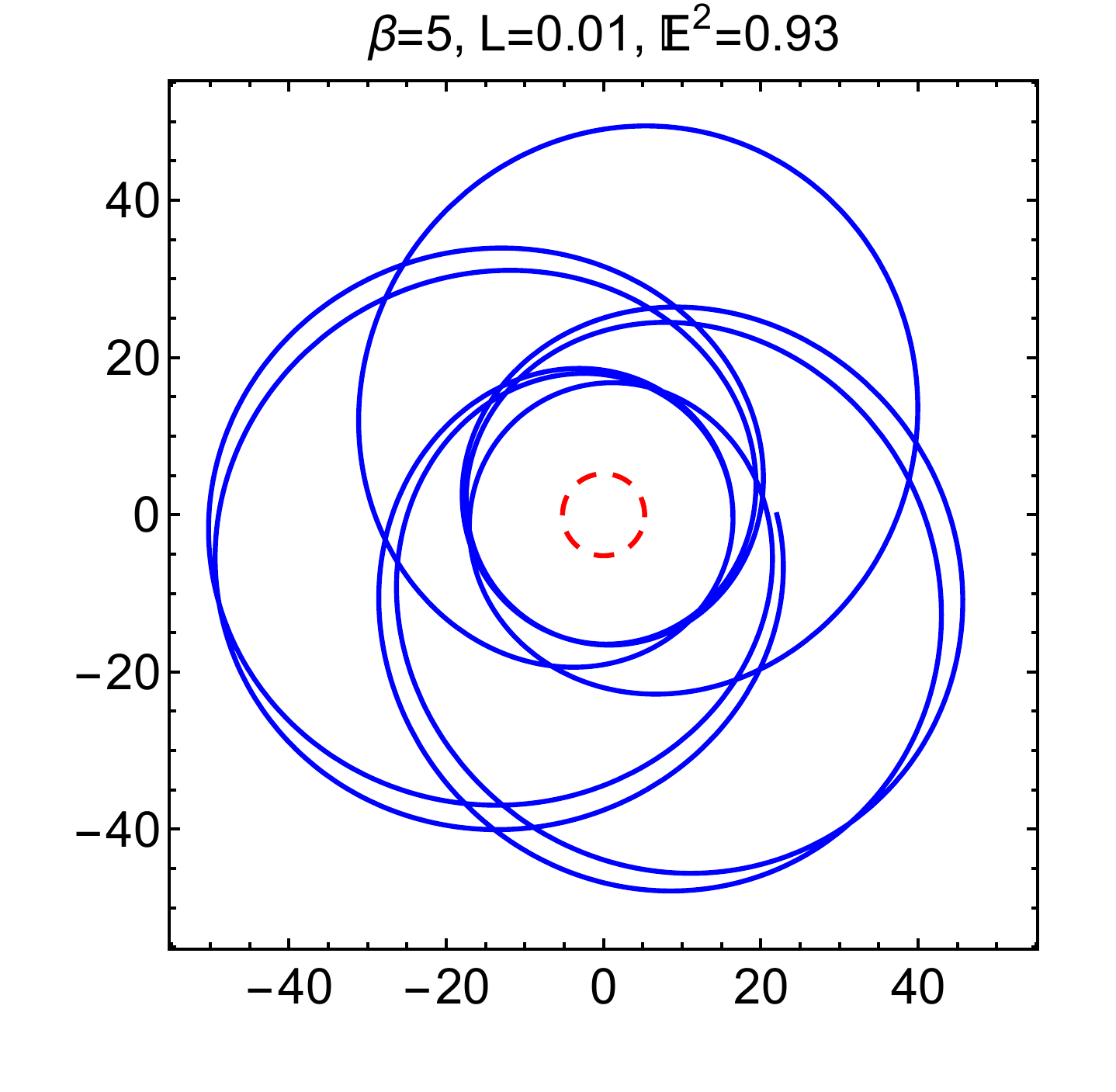} 
	\end{tabular}	
	\begin{tabular}{cccc}
	\includegraphics[height=3.56cm,keepaspectratio]{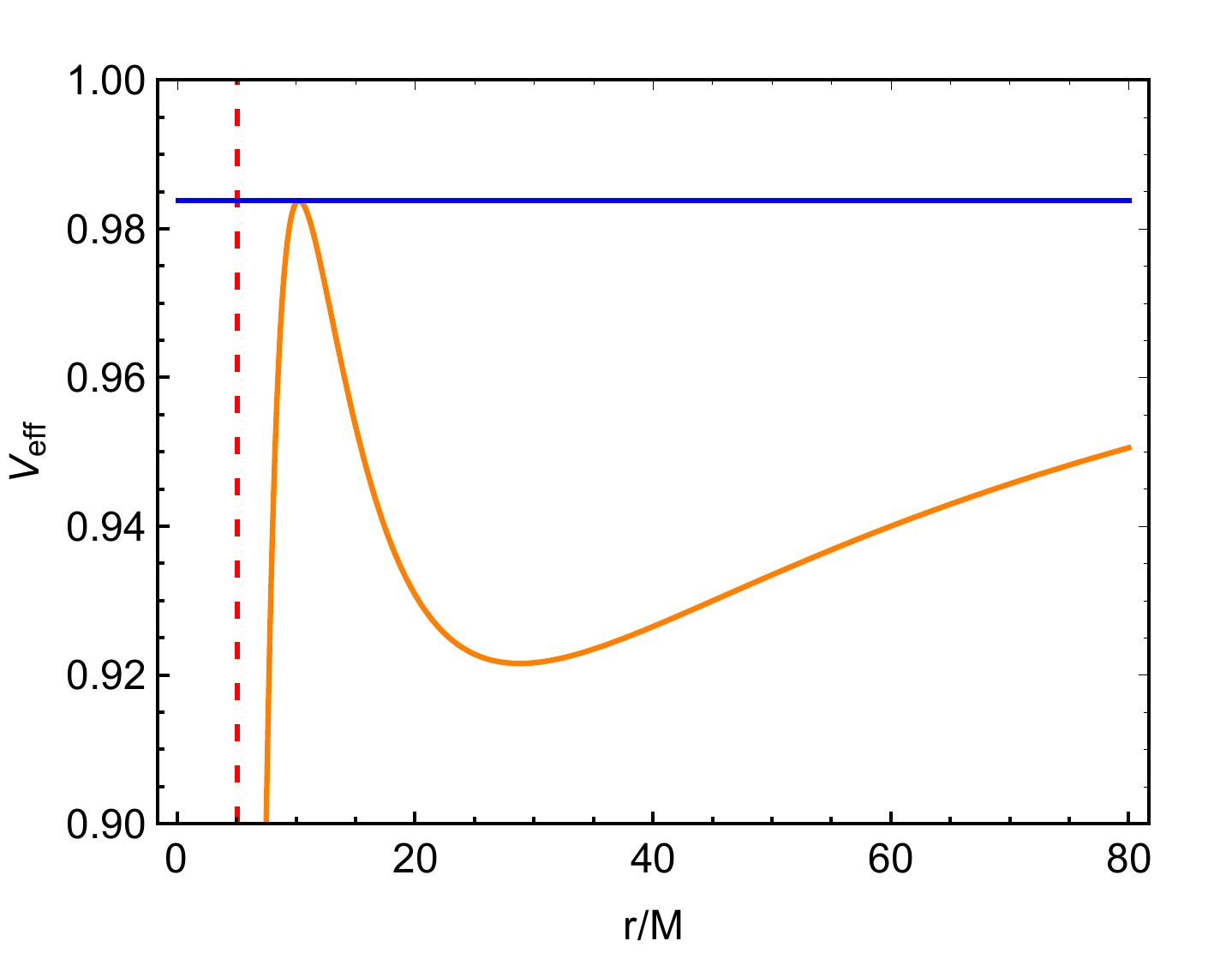}&	\includegraphics[height=3.56cm,keepaspectratio]{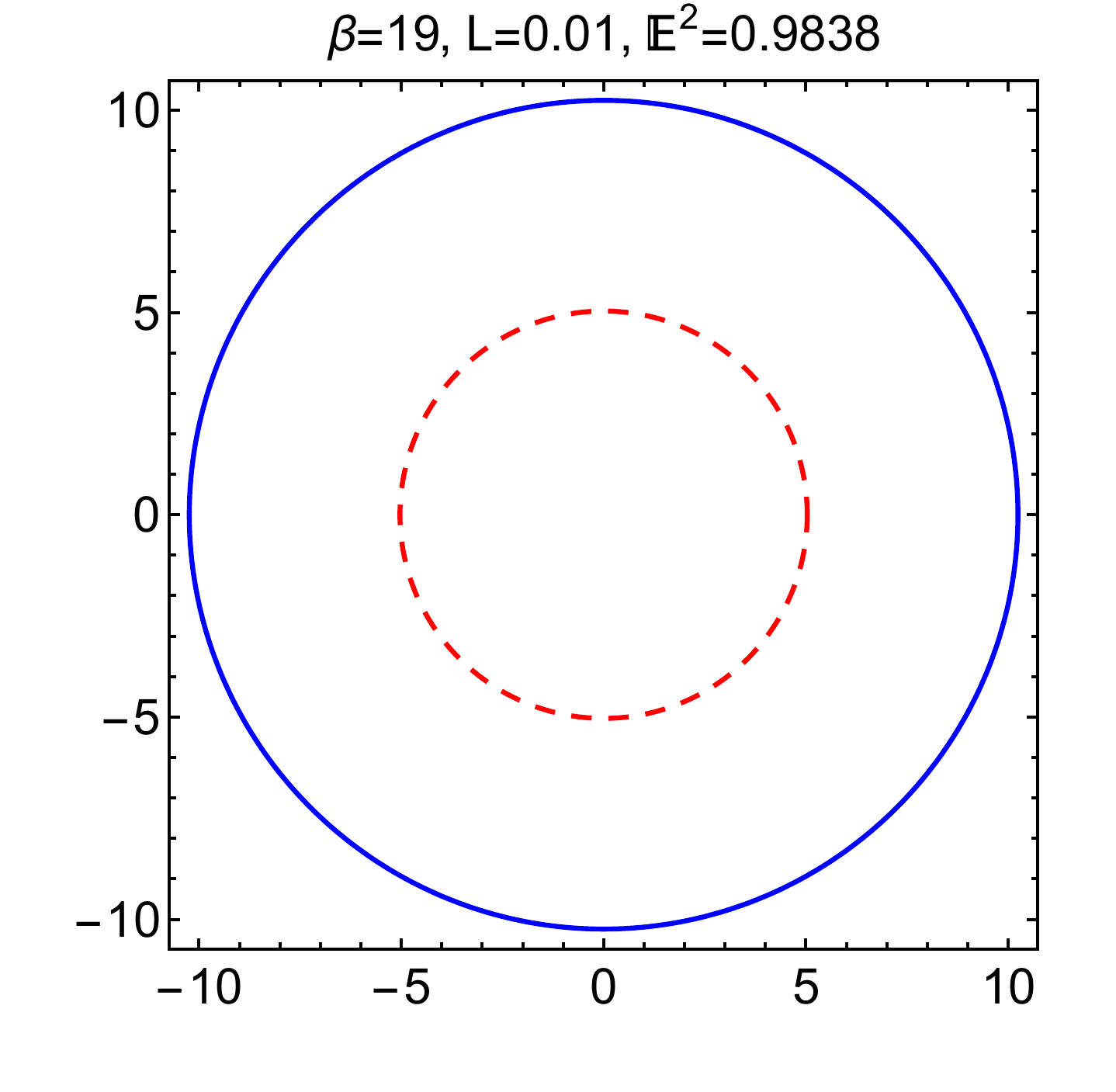}&
	\includegraphics[height=3.6cm,keepaspectratio]{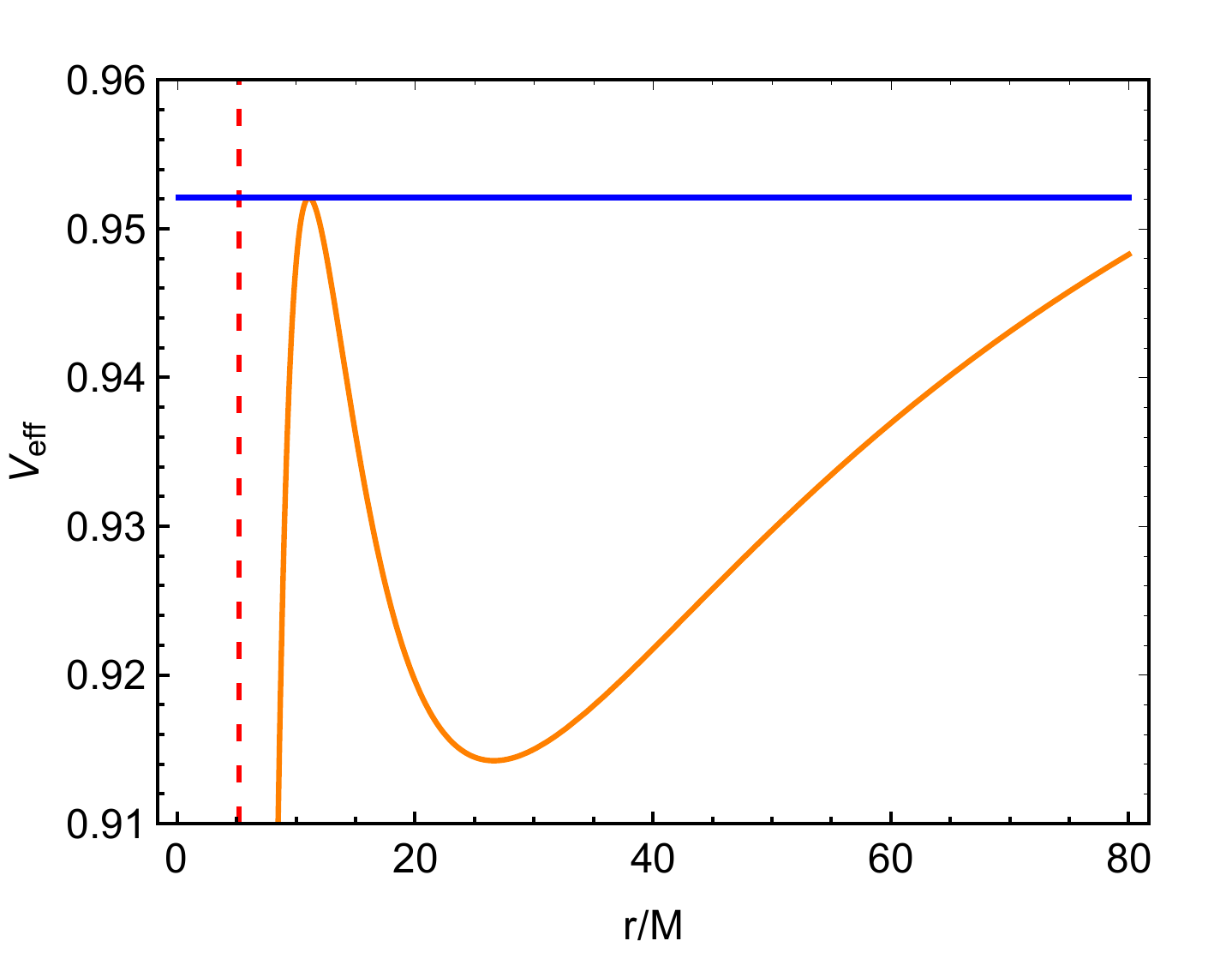}&	\includegraphics[height=3.6cm,keepaspectratio]{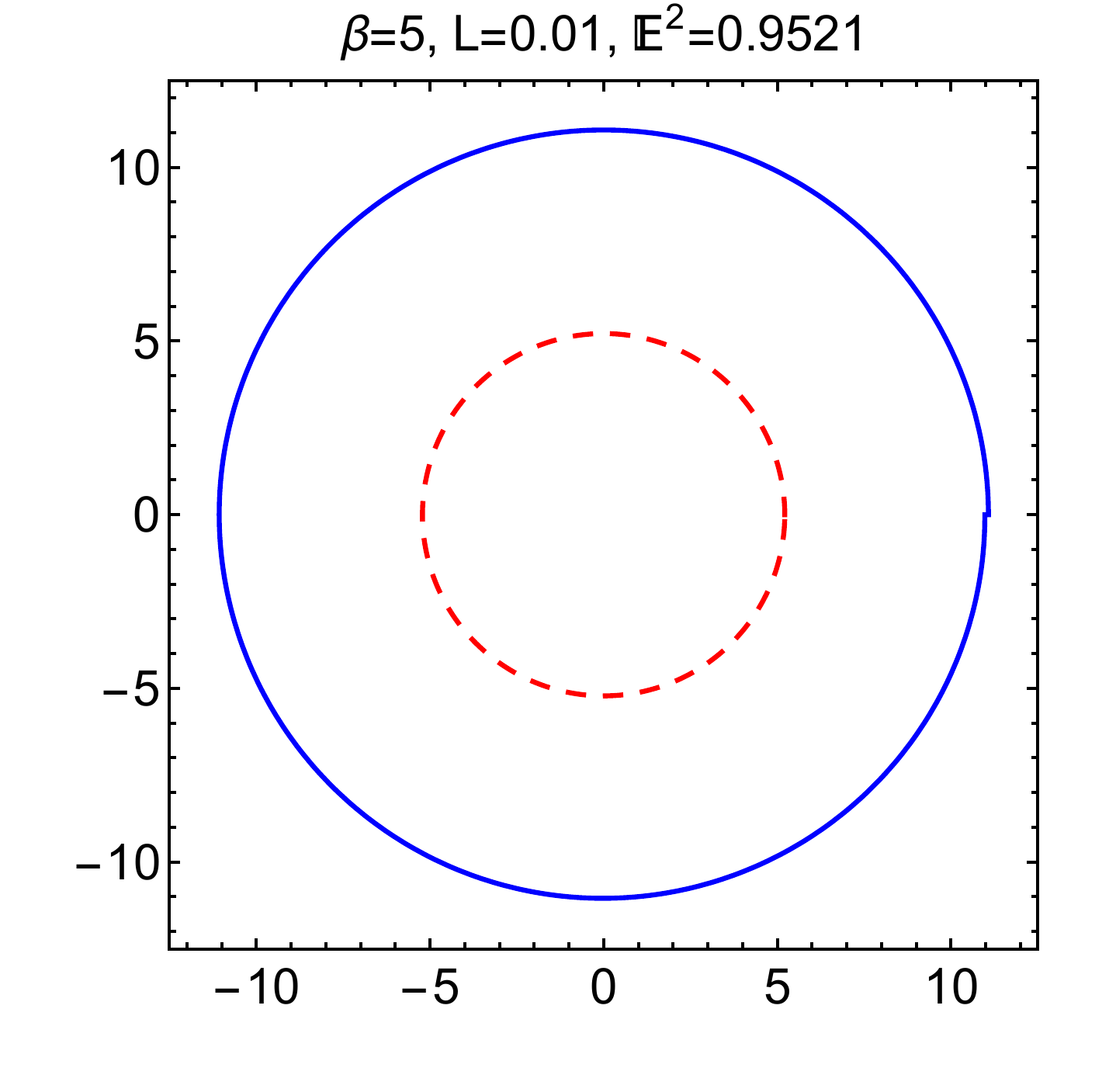} 
\end{tabular}		
	\caption{The possible regions I-IV and the special case of circular orbit in ENE (first and second columns) and LNE black hole (third and fourth columns) as pairs of an effective potential scenario and its corresponding possible orbit for the case of massive particles.}
	\label{EL_orbit_tl}
\end{figure}

The resulting orbits in all four regions and the particular case of unstable circular orbits for both ENE and LNE model are then displayed in Fig. \ref{EL_orbit_tl}. From the form of EoM \eqref{Veff_TL}, it is shown that the bound orbits are hardly closed. While there are two possible orbits in region I and IV, we only plot the distinctive case ones, which are flyby orbit in region I and bound orbit in region IV. It is observed that region I-III is able to exist in relatively small $\beta$, while region IV requires larger value of $\beta$. It is important to note that all the timelike orbits displayed here are exactly the same as in the case of RN geodesics.

\subsection{Null Geodesics}

\label{ng}

While massive particles propagate along their timelike geodesic described by the metric solutions, Novello {\it et al} showed that in NLED photon follows the null geodesic of its {\it effective} geometry given by~\cite{Novello:1999pg}
\begin{equation}
\label{effg}
g^{\mu\nu}_{eff}=\mathcal{L_F}g^{\mu\nu}-4\mathcal{L_{FF}}F^{\mu}_{\ \alpha}F^{\alpha\nu}.
\end{equation}
After some algebra, the general form of {\it conformally-rescaled} effective line element can be written as
\begin{equation}
\label{dseff}
ds^2_{eff} = - f(r) dt^2+ f(r)^{-1} dr^2+ h(r) r^2 d\Omega^2.
\end{equation}
where $h(r)$ is a factor given as
\begin{eqnarray}
\label{h}
h(r)=
\left\{\begin{array}{cl}
\beta ^2 r^4(\beta ^2 r^4-4 Q^2)^{-1},& \textrm{ENE}\\
1+16 Q^2(32 \beta ^2 r^4-15 Q^2)^{-1},& \textrm{LNE}.
\end{array} \right\}
\end{eqnarray}
The null rays in the mentioned line element \eqref{dseff} follow the trajectories given by
\begin{equation}
\label{null_deff}
0= f \dot{t}^2 -f^{-1} \dot{r}^2 -hr^2 \dot{\phi}^2.
\end{equation}
The orbital equation and effective potential $V_{eff}$ can be extracted out as
\begin{eqnarray}
\label{Veff_null}
\bigg(\frac{dr}{d\phi}\bigg)^2 = \mathbb{E}^2 L r^4 - \frac{f(r) r^2}{h} = \mathcal{R}(r) \,\,\,\, , \,\,\,\, V_{eff}=\frac{f(r)}{h(r) L r^2}.
\end{eqnarray}
\begin{figure}[!h]
	\centering
	\begin{tabular}{cc}
		\includegraphics[height=6.5cm,keepaspectratio]{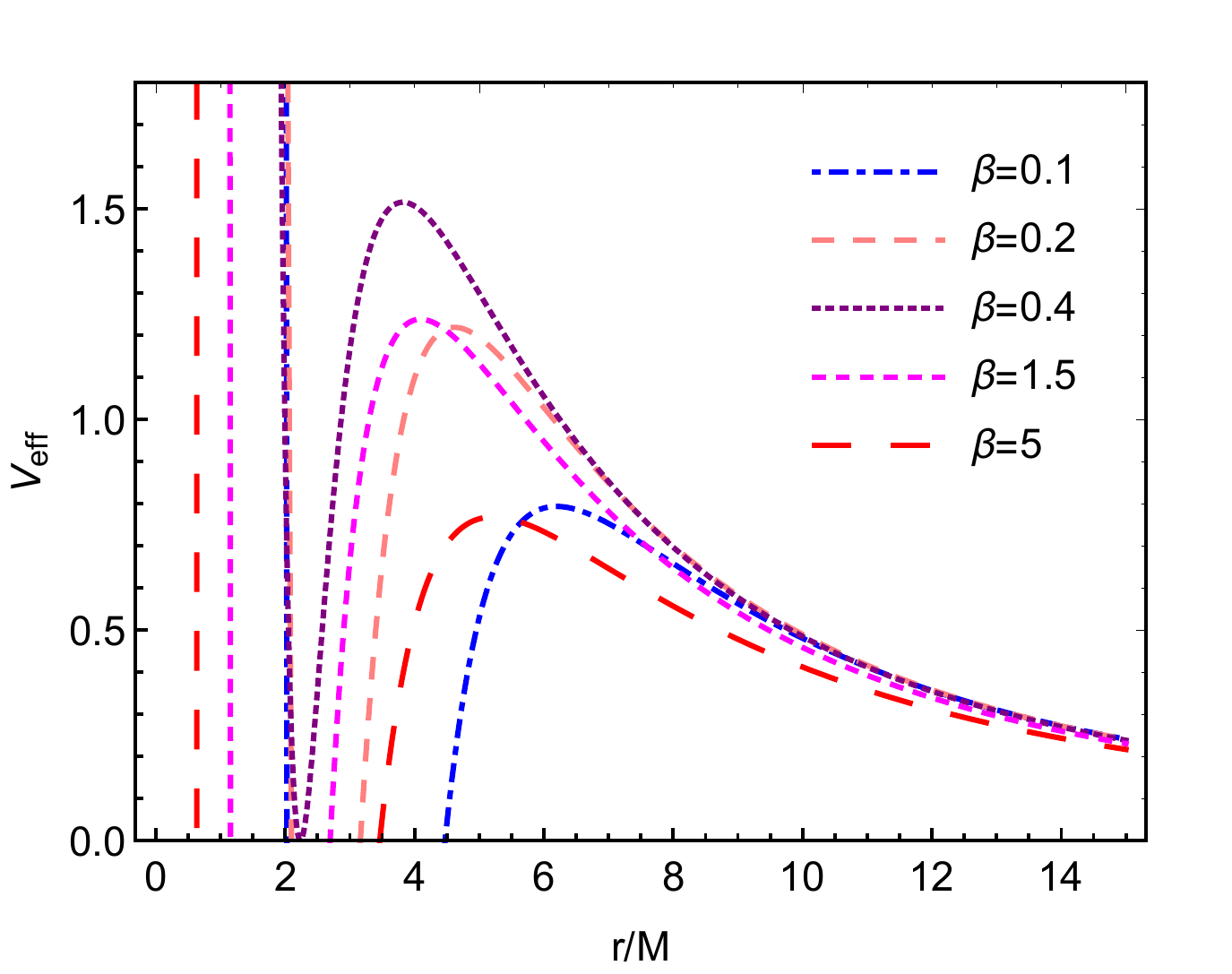} &	\includegraphics[height=6.5cm,keepaspectratio]{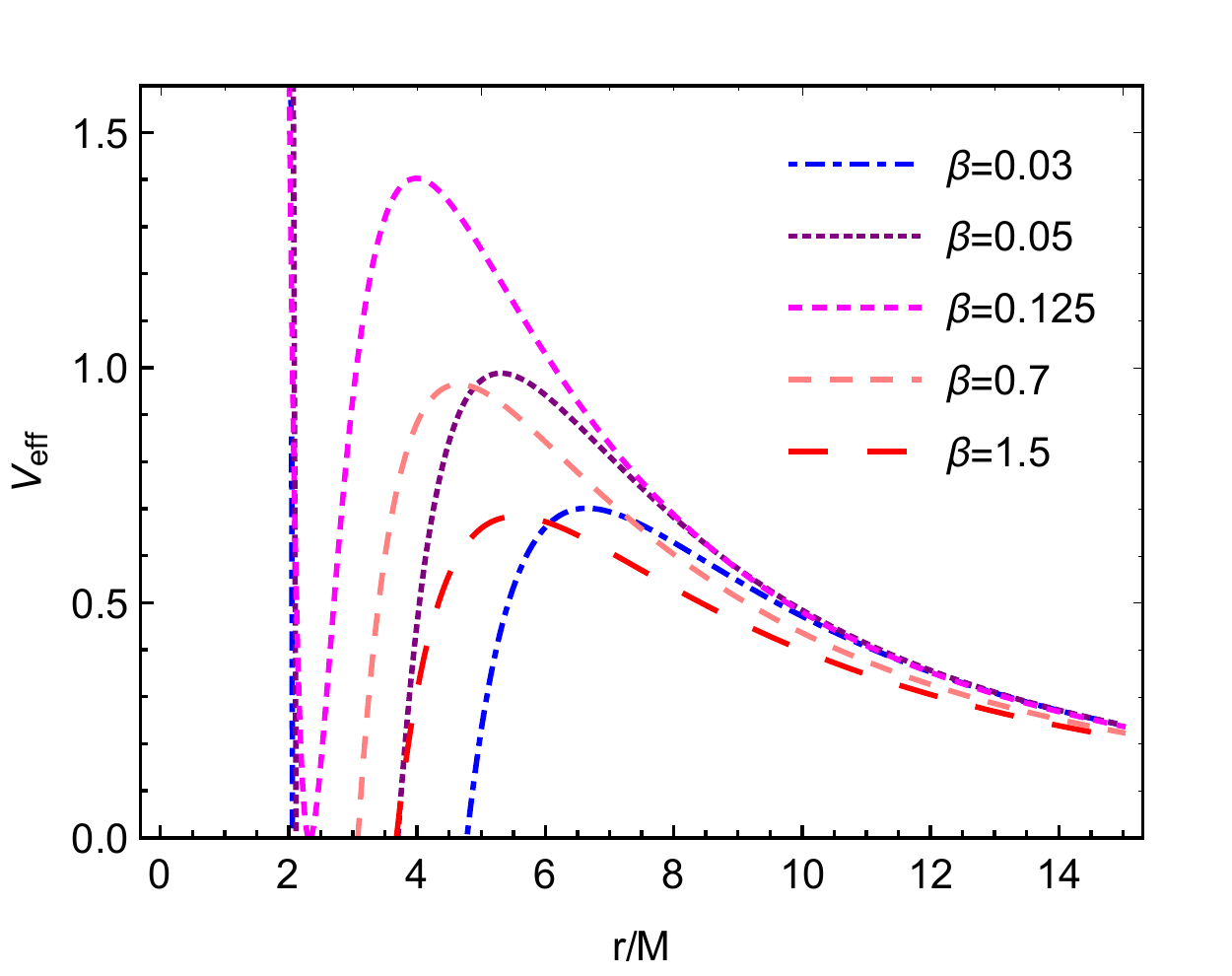}  
	\end{tabular}
	\caption{[Left] Plots of effective potential for light particles of ENE with $Q=1$ and $\mathbb{L}=0.0156$. [Right] Plots of effective potential for light particles of LNE with similar parameter values.}
	\label{null}
\end{figure}

\begin{figure}[!ht]
	\centering
	\begin{tabular}{cccc}
		\includegraphics[height=3.6cm,keepaspectratio]{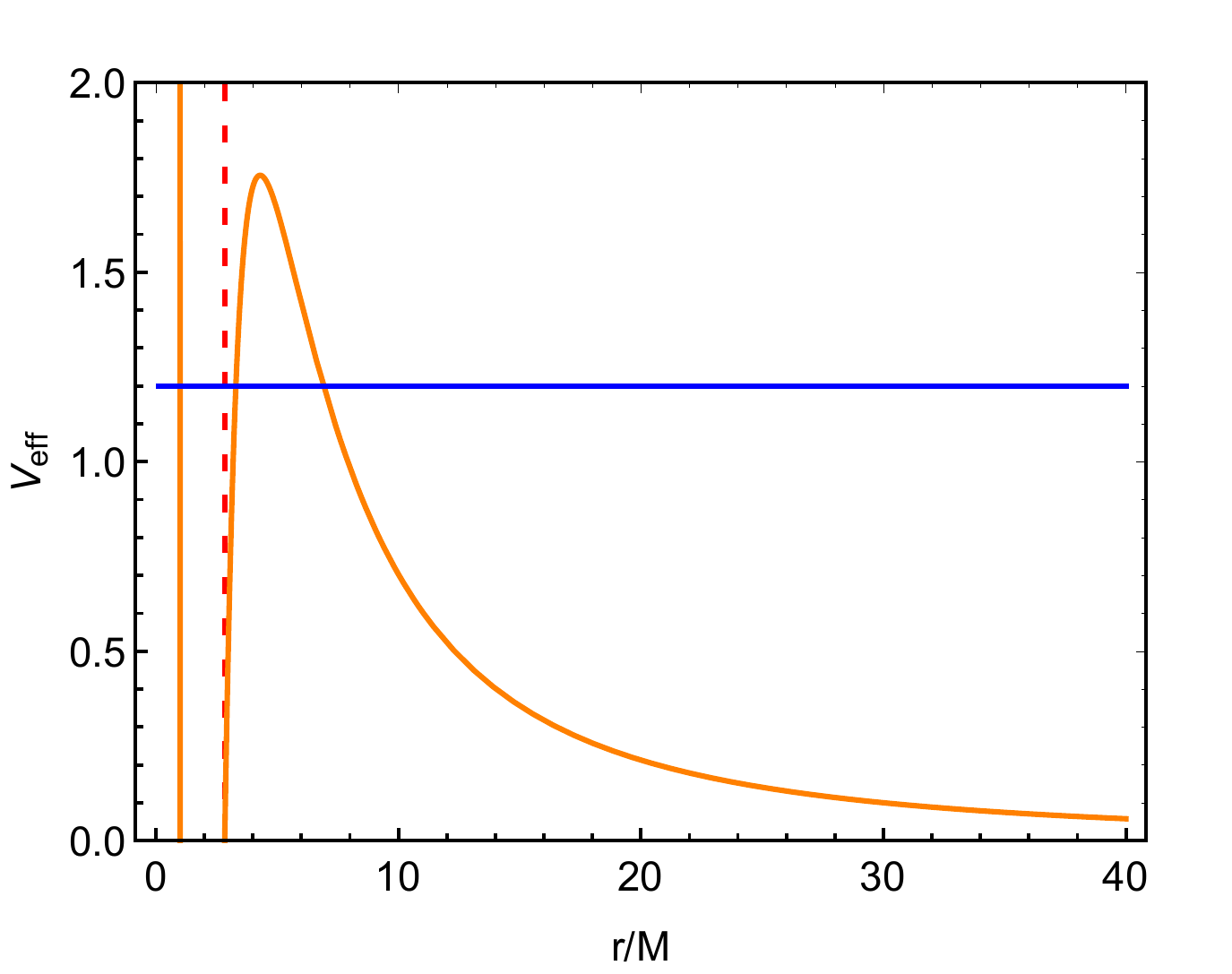}&	\includegraphics[height=3.6cm,keepaspectratio]{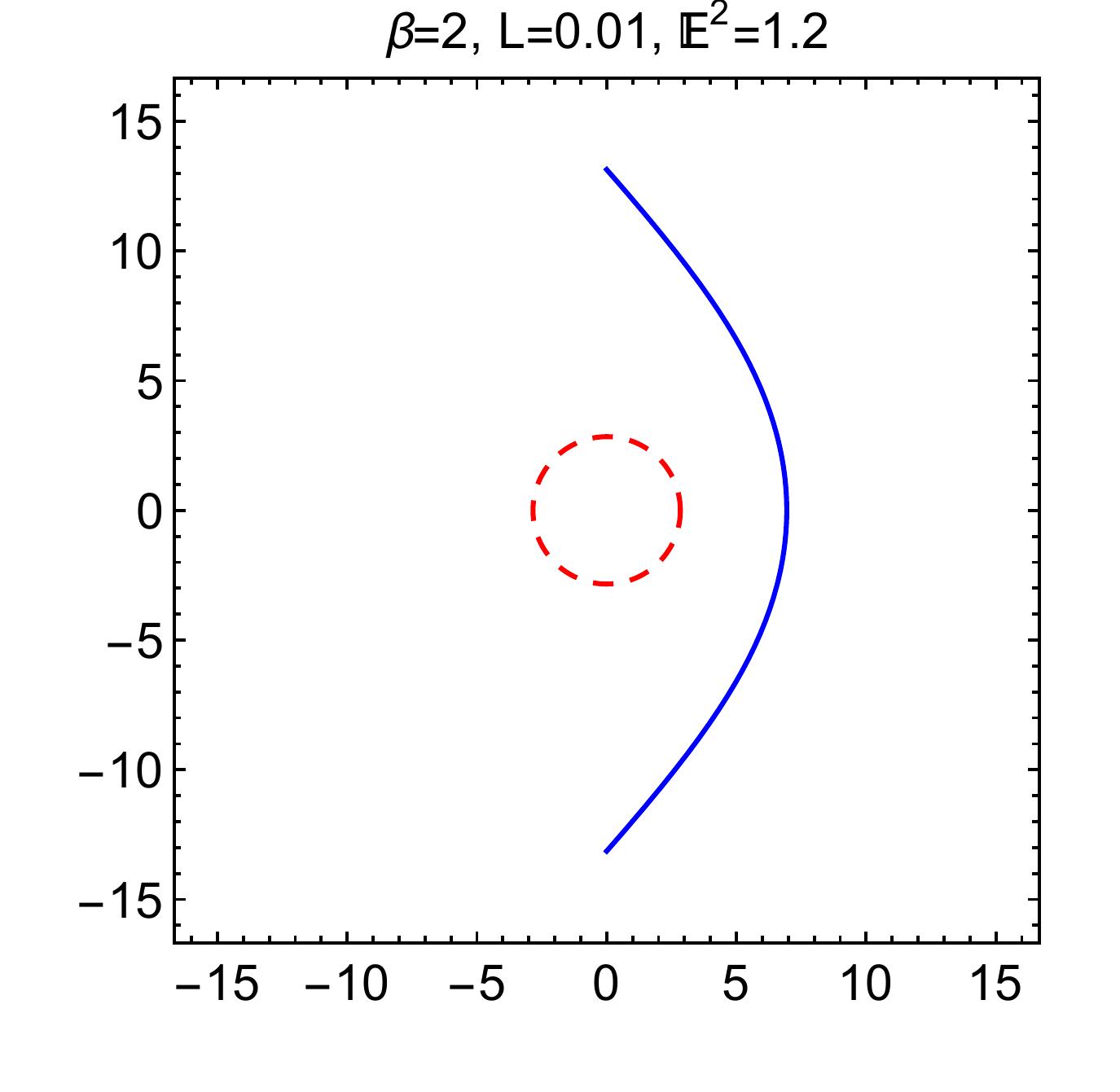}&
		\includegraphics[height=3.6cm,keepaspectratio]{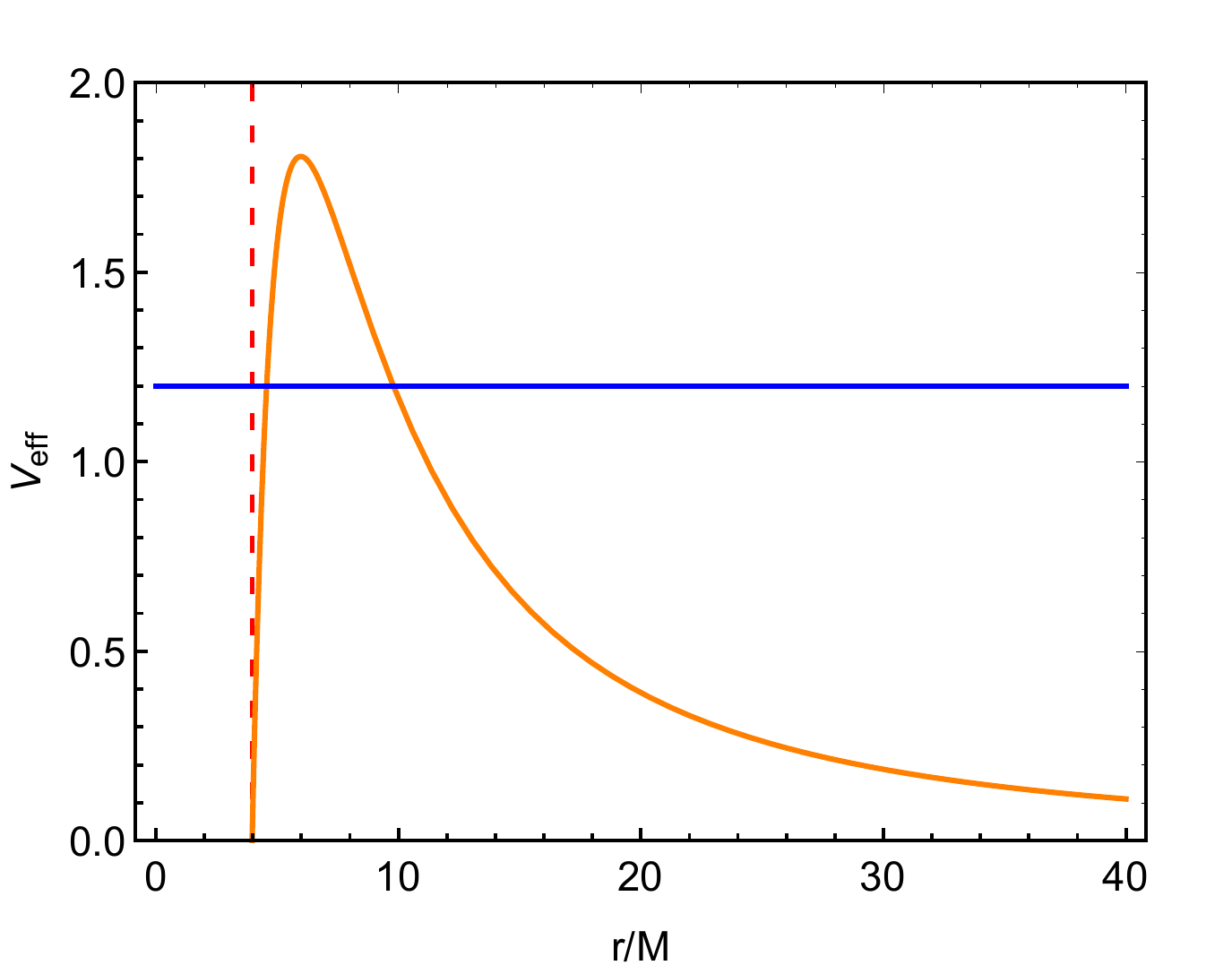}&	\includegraphics[height=3.6cm,keepaspectratio]{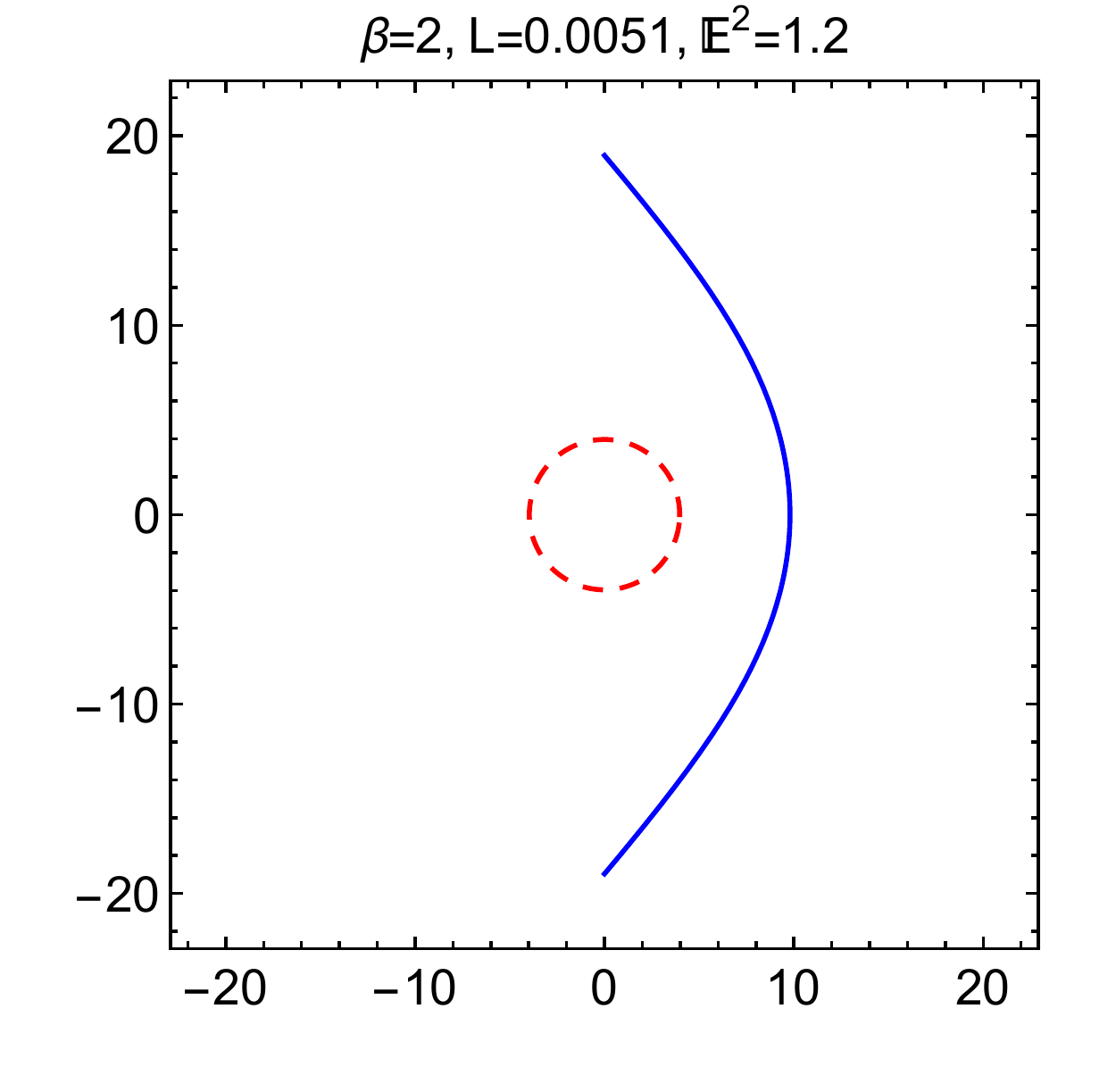}  
	\end{tabular}
	\begin{tabular}{cccc}
		\includegraphics[height=3.65cm,keepaspectratio]{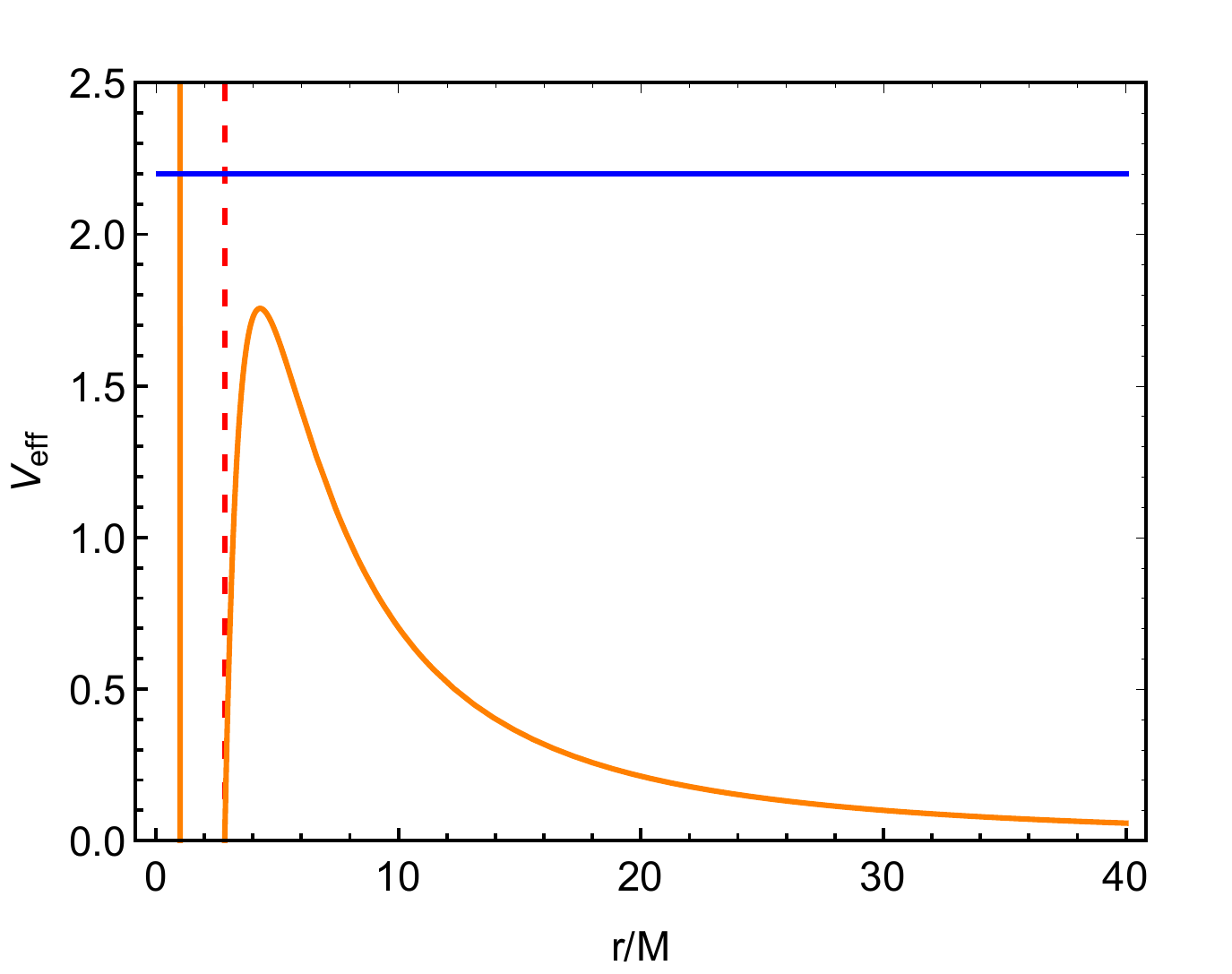}&	\includegraphics[height=3.65cm,keepaspectratio]{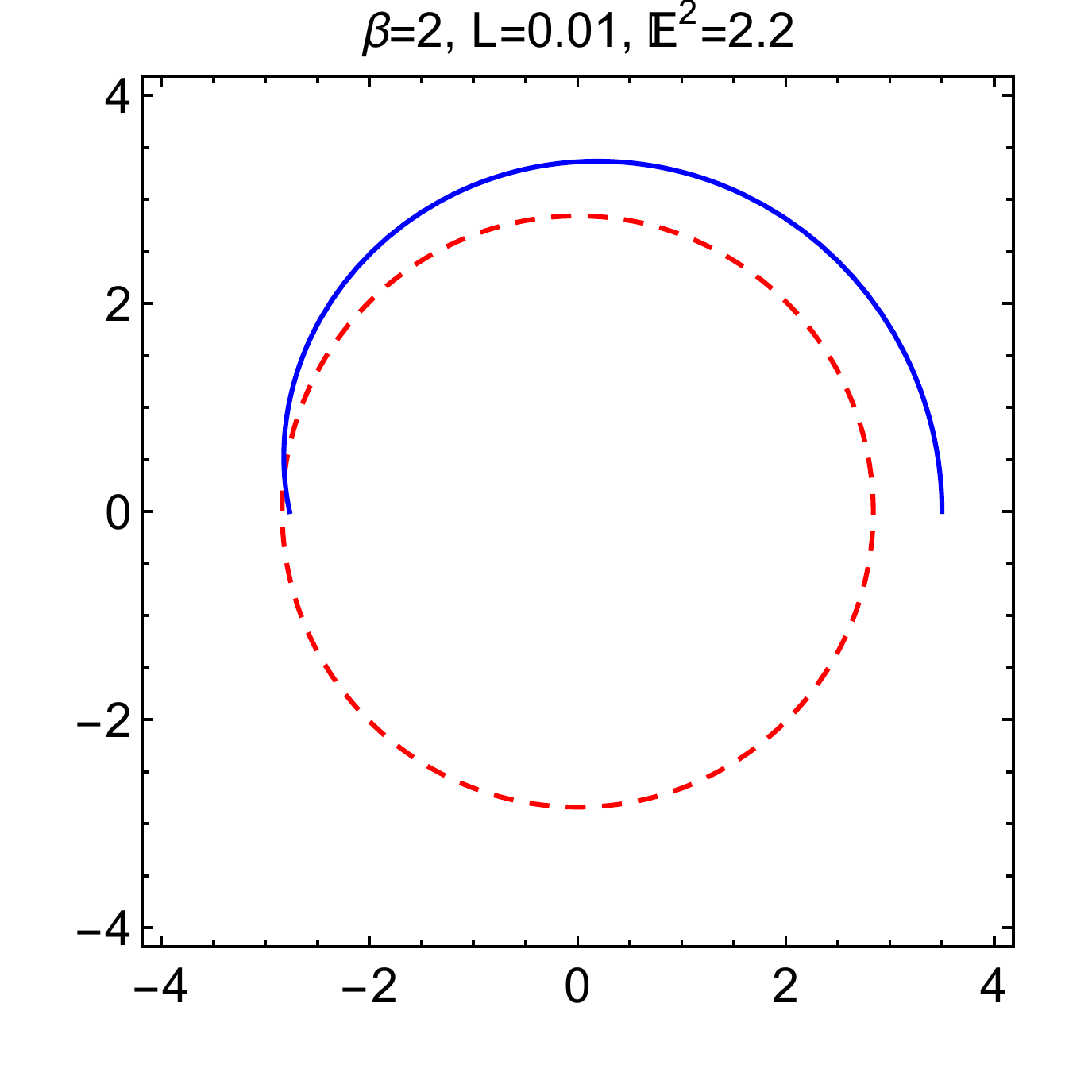}&
		\includegraphics[height=3.65cm,keepaspectratio]{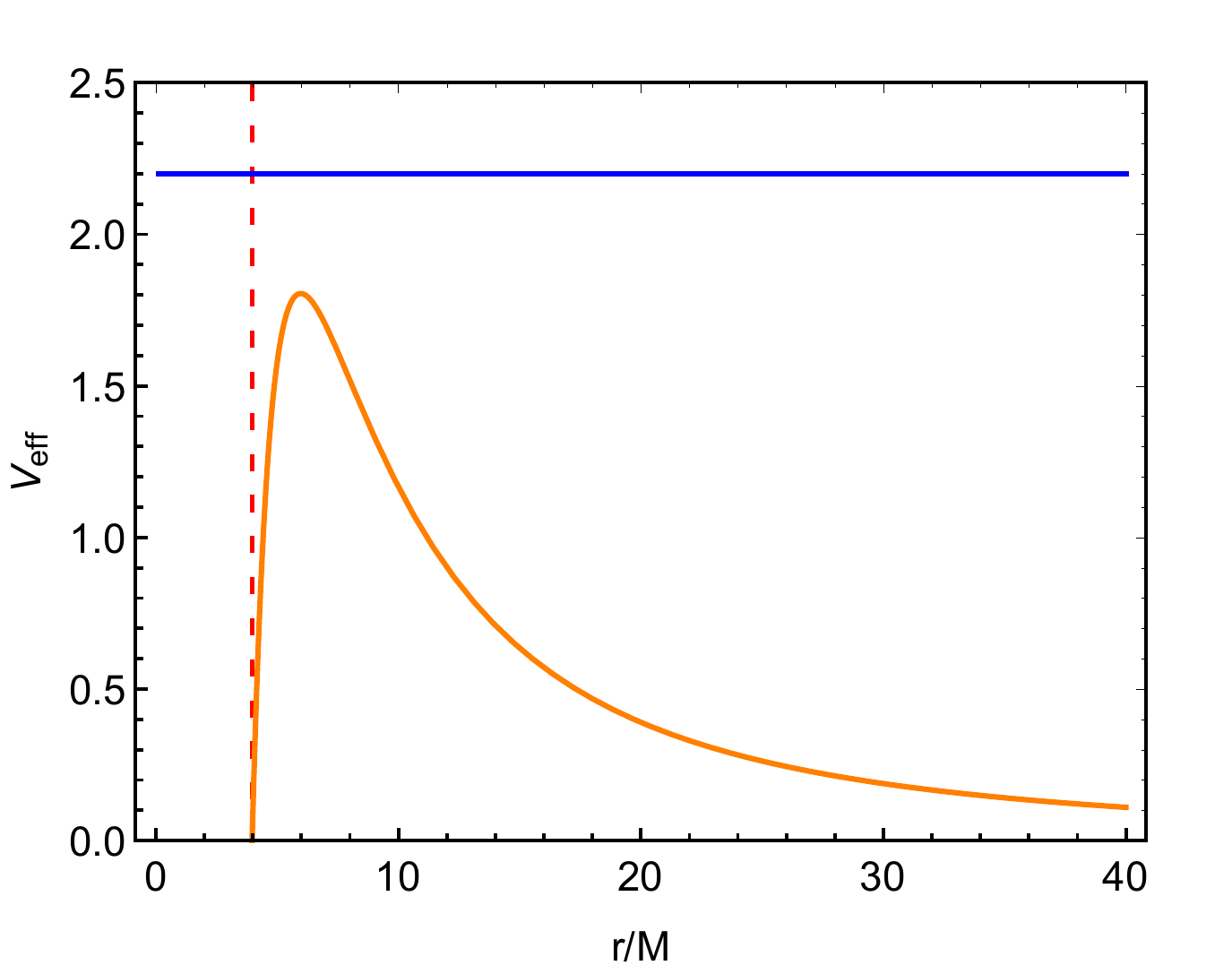}&	\includegraphics[height=3.65cm,keepaspectratio]{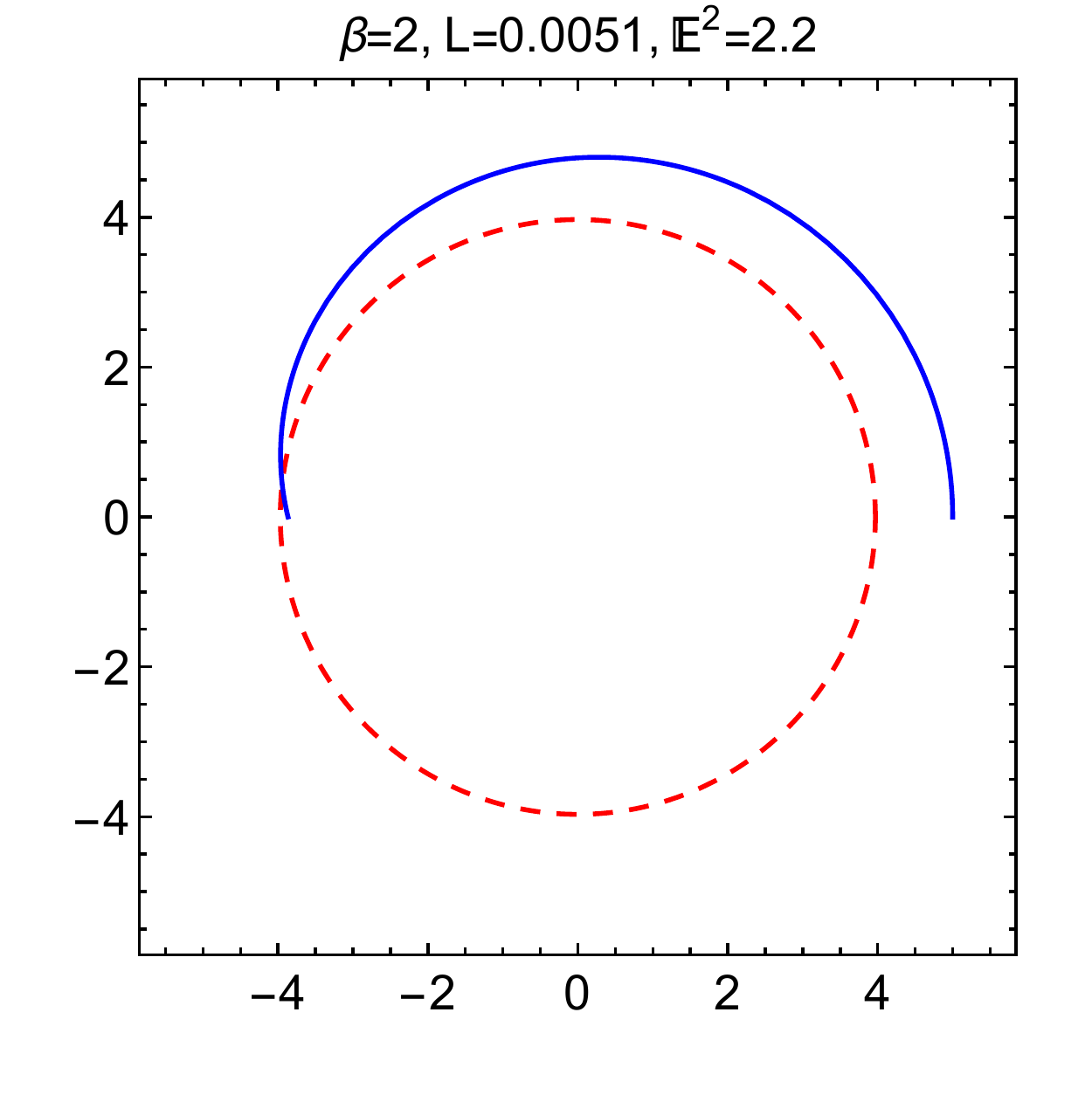}
	\end{tabular}
	\begin{tabular}{cccc}
	\includegraphics[height=3.6cm,keepaspectratio]{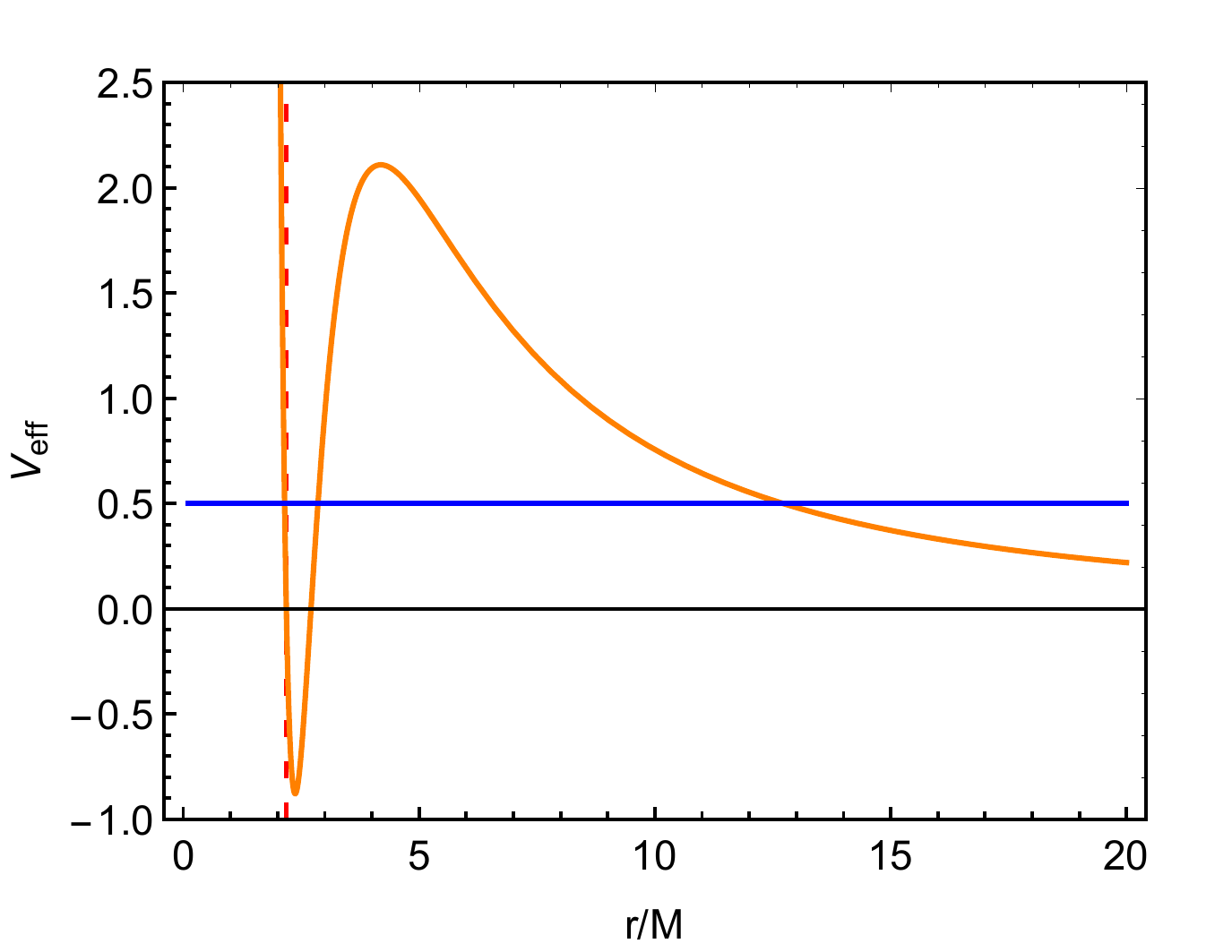}&	\includegraphics[height=3.6cm,keepaspectratio]{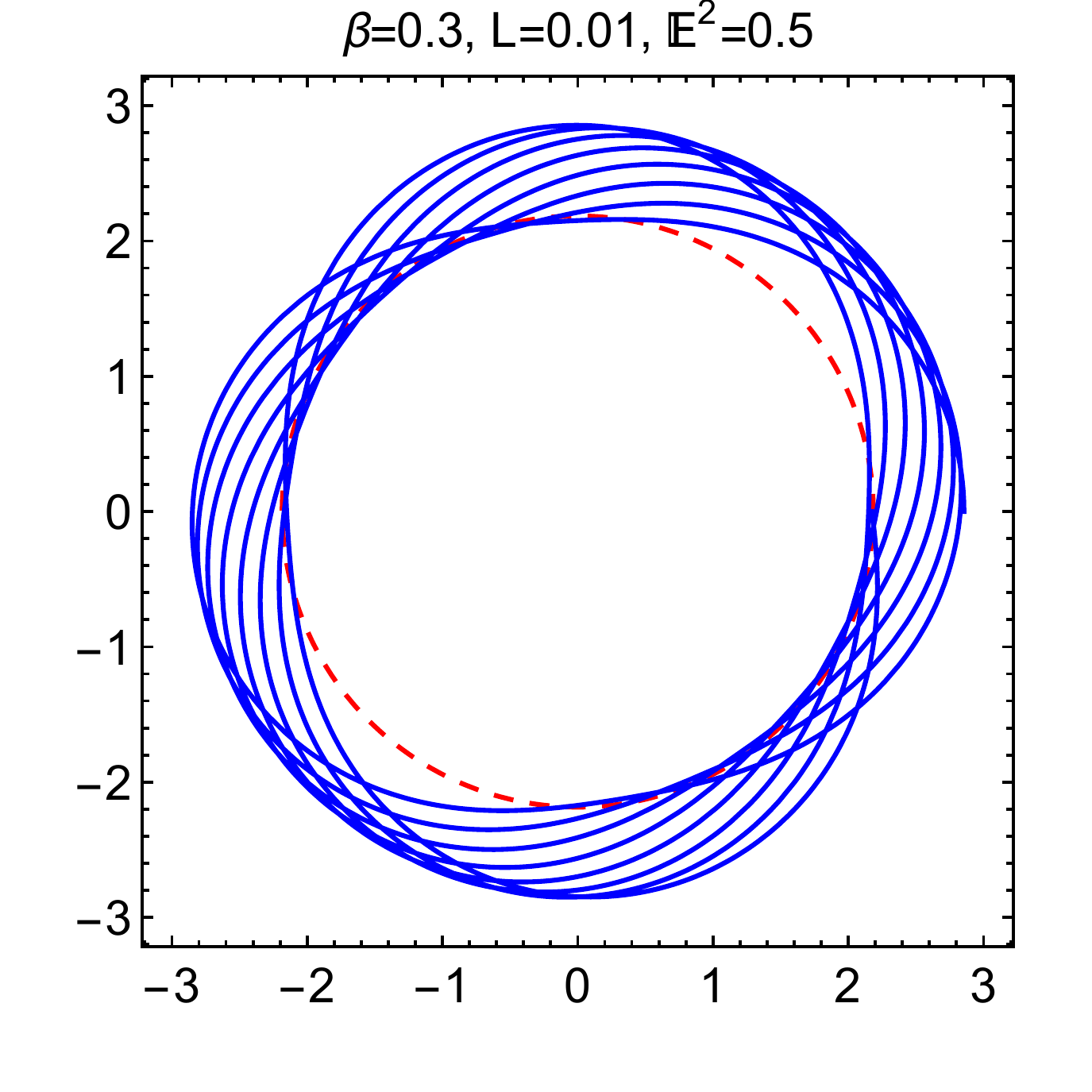}&
	\includegraphics[height=3.6cm,keepaspectratio]{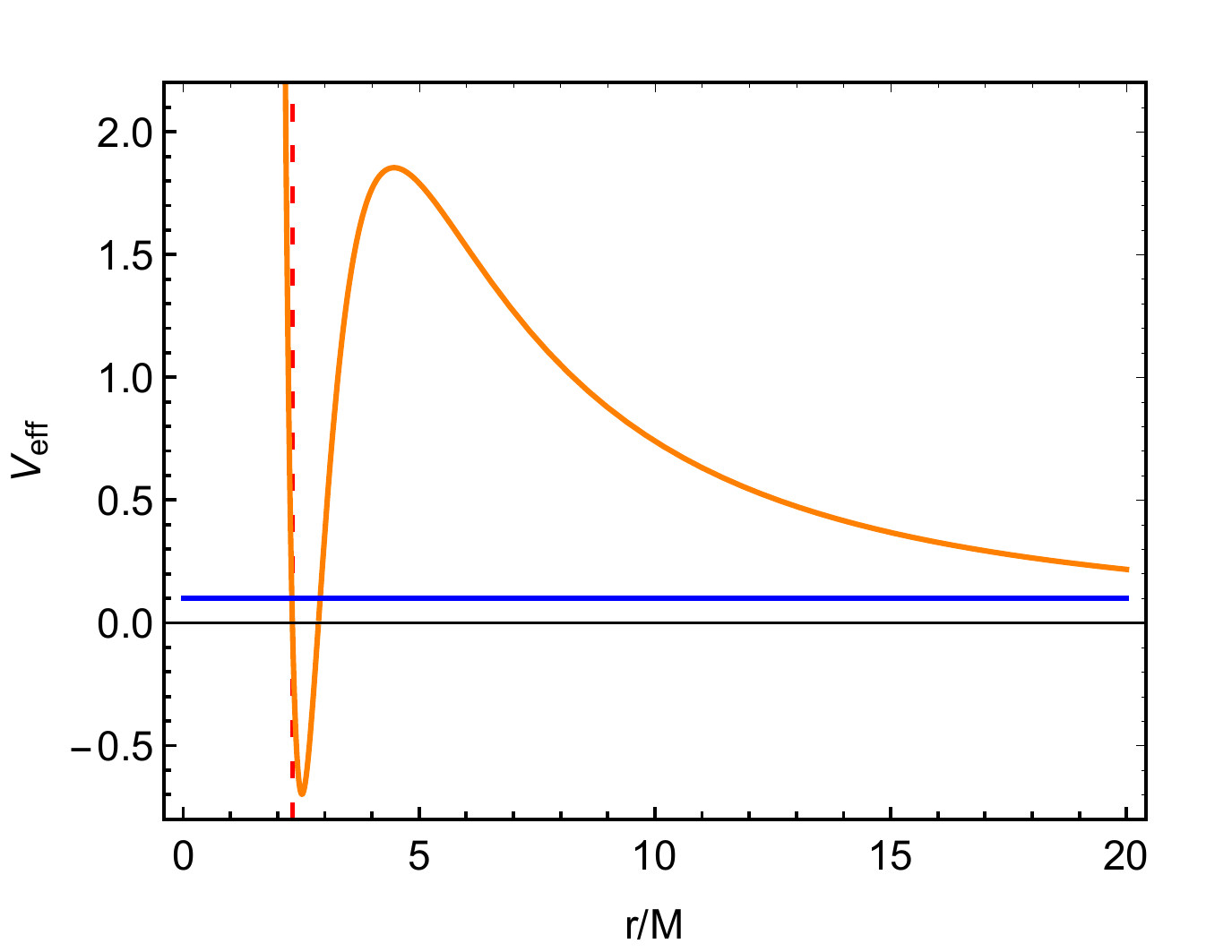}&	\includegraphics[height=3.6cm,keepaspectratio]{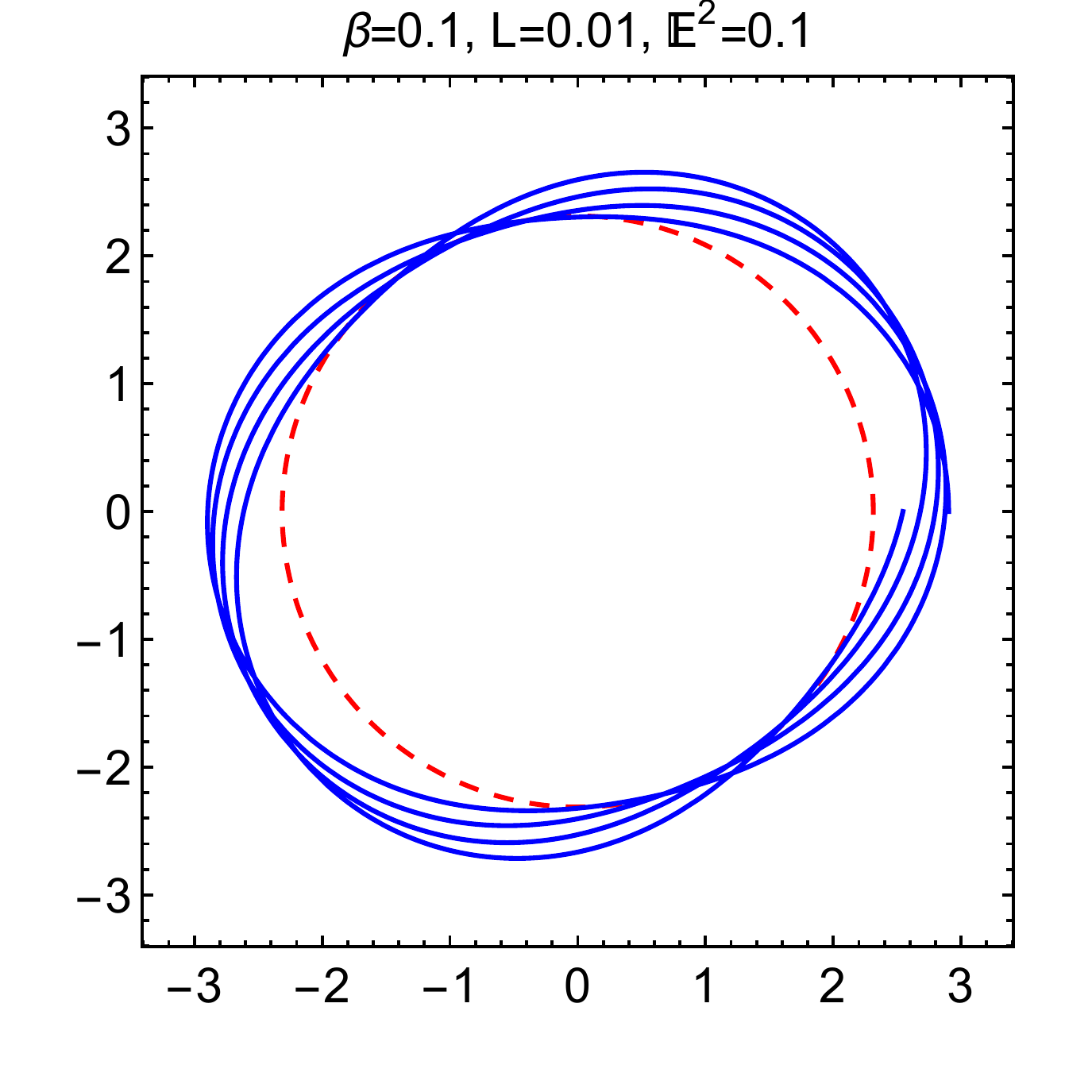}
	\end{tabular}			
	\caption{The possible regions I-III in ENE (first and second columns) and LNE black hole (third and fourth columns) as pairs of an effective potential scenario and its corresponding possible orbit for the case of light particles.}
	\label{EL_orbit_null}
\end{figure}

The $V_{eff}$ behavior for both models are shown in Fig.~\ref{null}. It has been pointed out in~\cite{Habibina:2020msd} that one possible genuine feature not present in linear electrodynamics (the RN black hole) is the appearance of local minimum located outside the horizon.
Discarding the solution behind the event horizon, there exist three regions based on the observed potentials, as shown in Fig.~\ref{EL_orbit_null}:
\begin{enumerate}[label=\arabic*., itemsep=0pt, topsep=0pt]
	\item region I: 2 positive real solutions, result in flyby and terminating bound orbits,
	\item region II: 0 positive real solutions, result in terminating escape and terminating bound orbits,
	\item region III: 2 positive real solutions, results in flyby and bound orbits.
\end{enumerate}
As in the timelike case, the nature of the orbital equation makes the bound orbits hardly closed. In the first two regions, the bound orbits are terminating since the one of the radius lies inside the horizon. The third region, however, is quite interesting. One of the radius coincides exactly with the event horizon. The resulting orbits are the precessed ellipse whose minor axes are the event horizon. This is a quite interesting situation that is not featured in the null scenario of RN geodesics. The potential barrier hold the light particles outside the event horizon and keep them on the precessed trajectories, preventing them from actually falling into singularity. 

\section{Geodesics in the Extremal Case}
\label{two}

The resulting metric solutions can also have multi horizons. It is seen in Fig.~\ref{fQ} that the flat charged ENE (and LNE) black holes can have to up to three horizons. We define the extremal solutions to be the cases where the second and the outer radii coincide, called the extremal solution.  The inner horizon is denoted as the true one ($r_1$) while the extremal one as $r_{2}$. To the best of our investigation the $V_{eff}$ enables local minima not exclusively outside the extremal $r_2$. 

\begin{figure}[!ht]
	\centering
	\begin{tabular}{cc}
		\includegraphics[height=4.7cm,keepaspectratio]{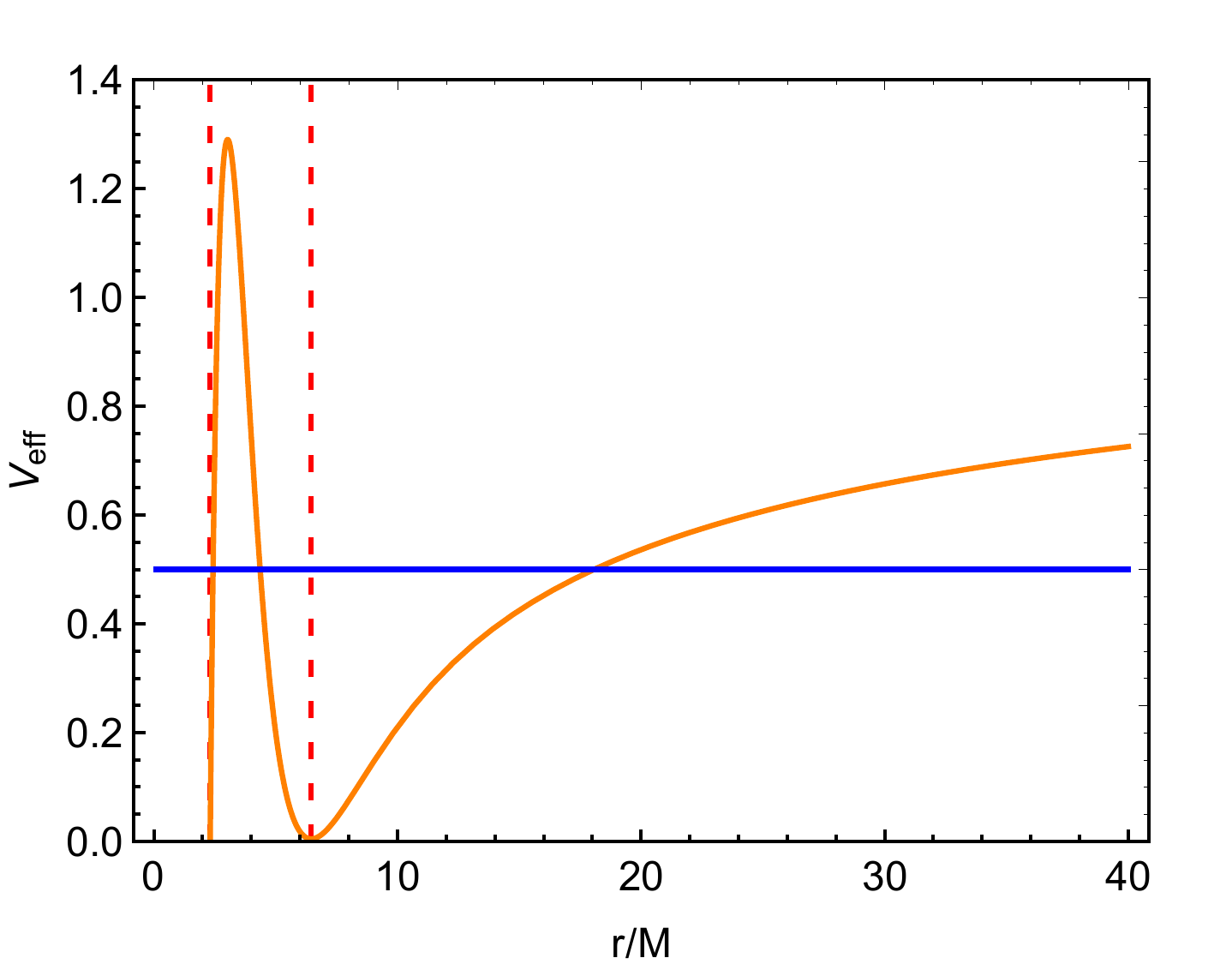}&	\includegraphics[height=4.7cm,keepaspectratio]{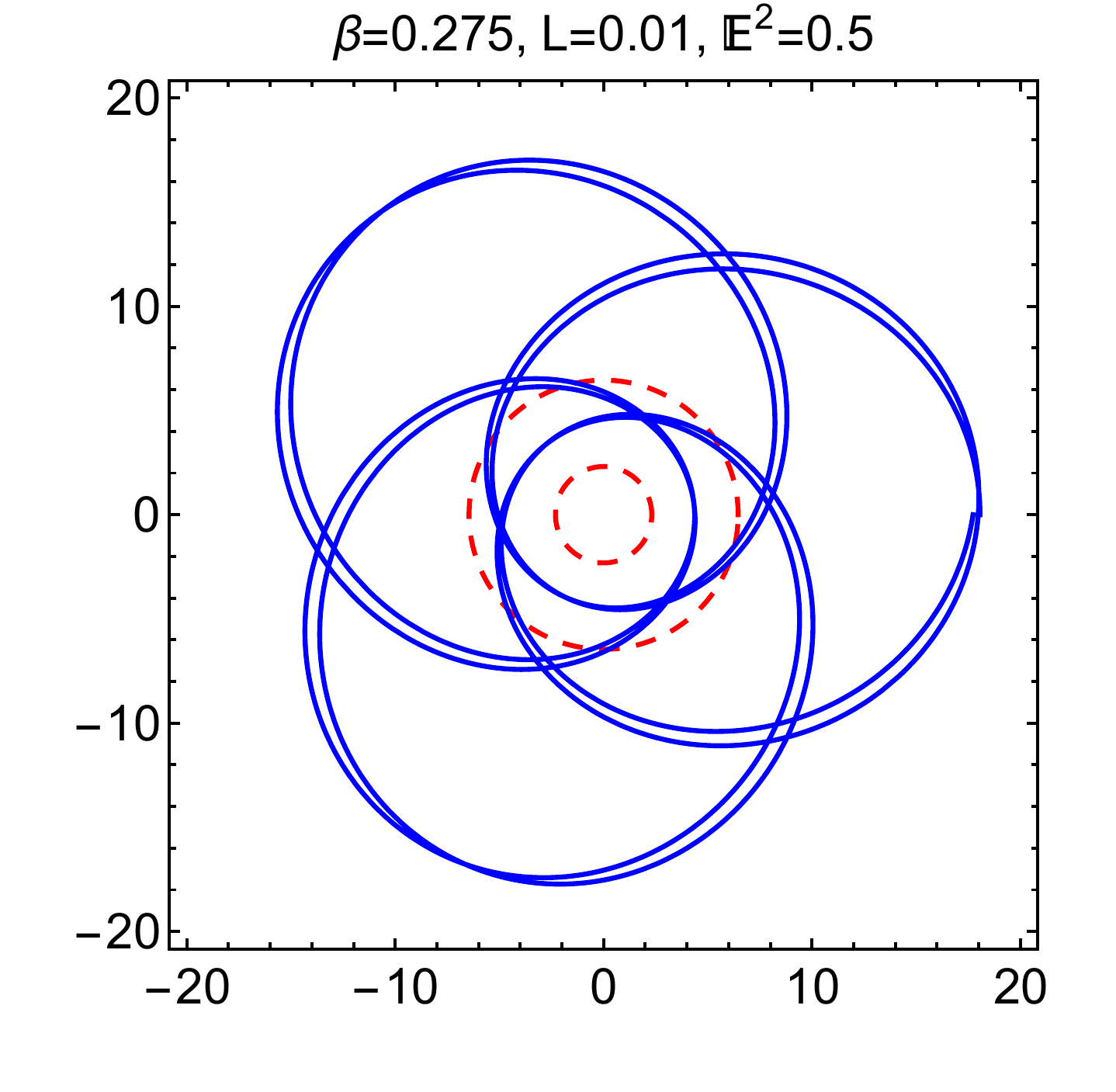}  
	\end{tabular}
	\begin{tabular}{cc}
		\includegraphics[height=4.7cm,keepaspectratio]{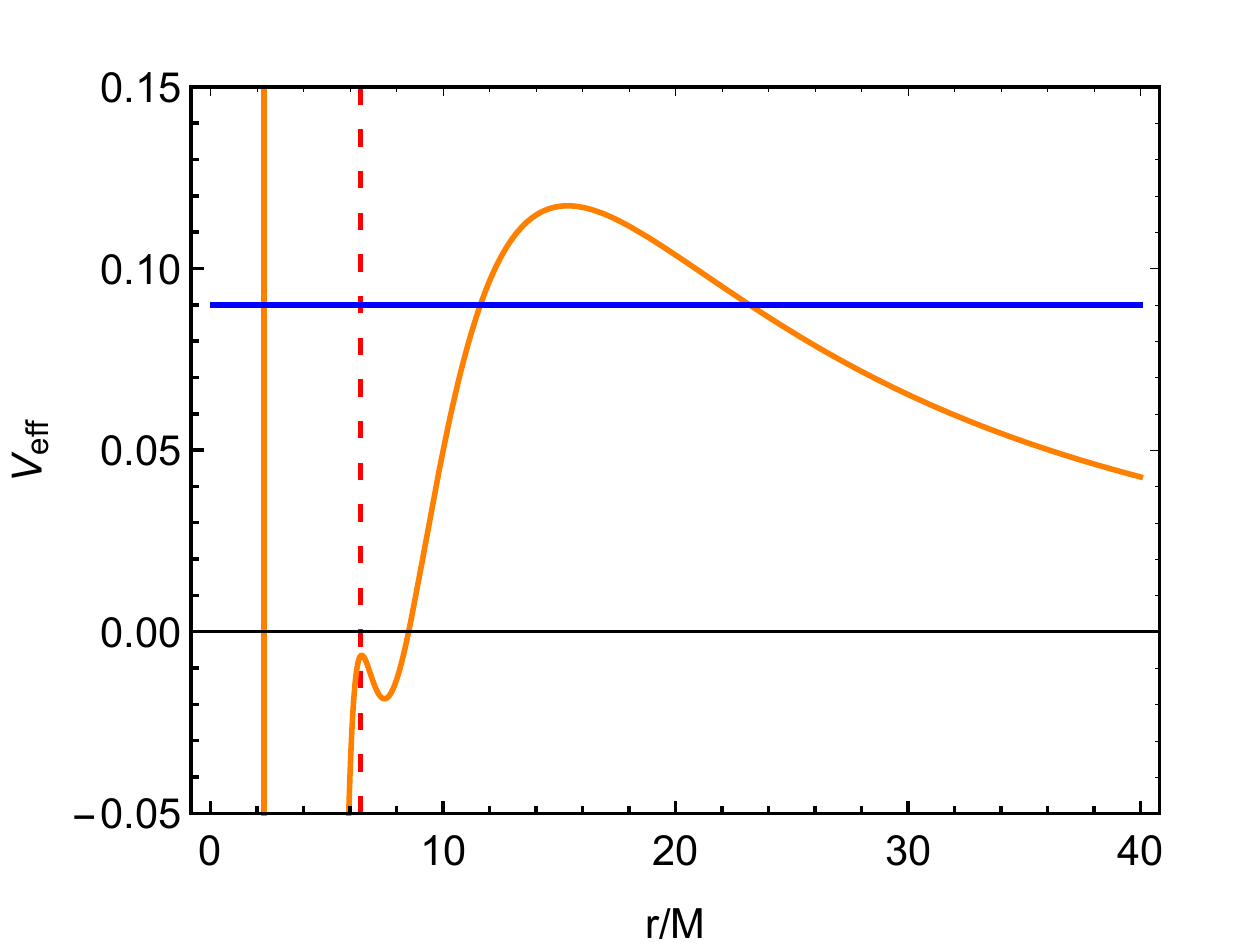}&	\includegraphics[height=4.7cm,keepaspectratio]{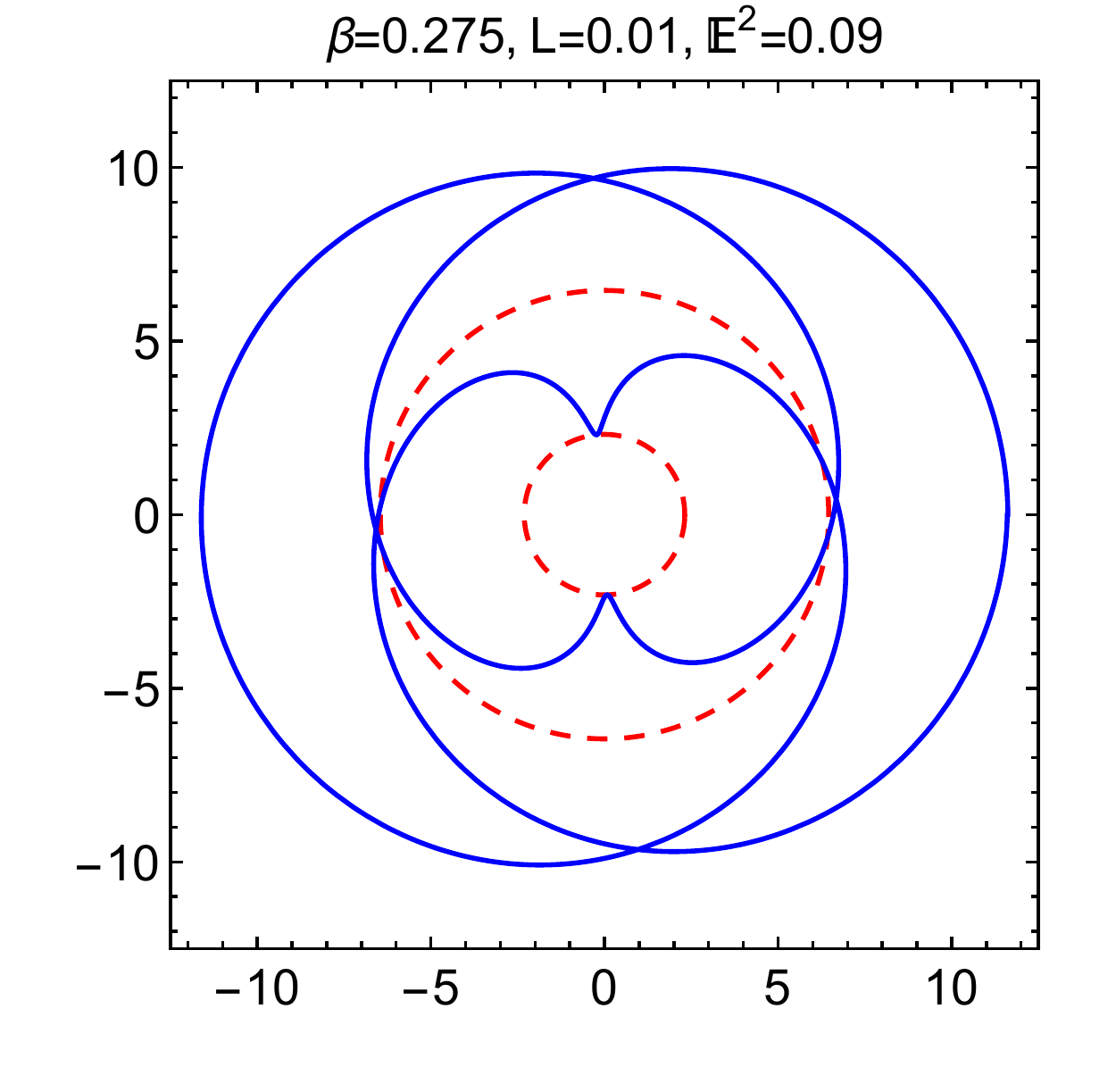} 
	\end{tabular}		
	\caption{The possible bound orbits in two-horizons ENE black hole as pairs of an effective potential scenario and its corresponding possible orbit for massive particles (top) and light particles (bottom).}
	\label{E_orbit_ext}
\end{figure}

In Fig.~\ref{E_orbit_ext} we show typical $V_{eff}$ for the extremal ENE. The timelike case (top left) is reminiscent to the null orbit of extremal RN, where the circular photon radius coincides with the extremal horizon~\cite{Pradhan:2010ws}. Particles manage to stay in a closed orbit, creating an epitrochoid orbit. It is important to note that the orbit does not coincide with the true horizon. The case for the null orbit (bottom left) is more subtle. The inner orbital radii lies on the true horizon $r_1$, and a litte curvature bounce is discovered precisely at $r_1$. The null trajectory interpolates between $r_1$ and the apsis outside $r_2$, forming an epicycloid bound orbit, which is a special case of epitrochoid. One can assume that the curvature bounce is the result of large value of charge that overcomes the gravitational pressure from the black hole mass. In both cases, the orbital path necessarily crosses the extremal horizon $r_2$.

Horizon-crossing orbit is a quite tricky condition. Unlike in vacuum spacetime, potential barrier due to the charge in charged spacetime does not allow the singularity to swallow the particles and instead drag the particles outward the horizon into another universe. This leads to the premise of many-world orbits, in which the orbit is seen as infinite continuity of the patches of the spacetime \cite{Grunau:2010gd}. Our resulting orbits in Fig. \ref{E_orbit_ext} are no stranger to this circumstance. The similar orbits where particles cross the outer horizon and bounce from the inner horizon are shown to exist in higher dimensional RN spacetime \cite{Hackmann:2008tu}.

\begin{figure}[!ht]
	\centering2
	\begin{tabular}{cc}
		\includegraphics[height=4.7cm,keepaspectratio]{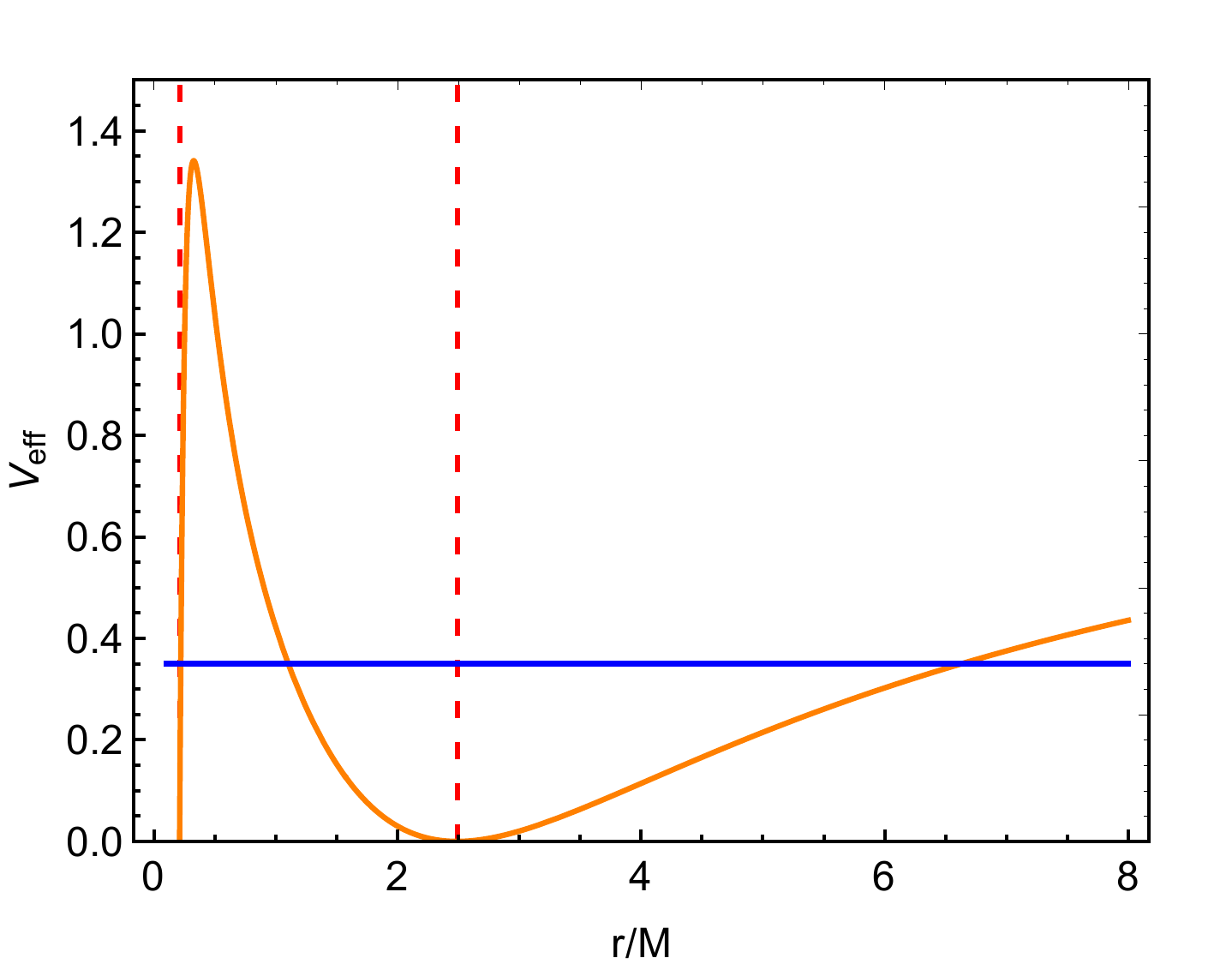}&	\includegraphics[height=4.7cm,keepaspectratio]{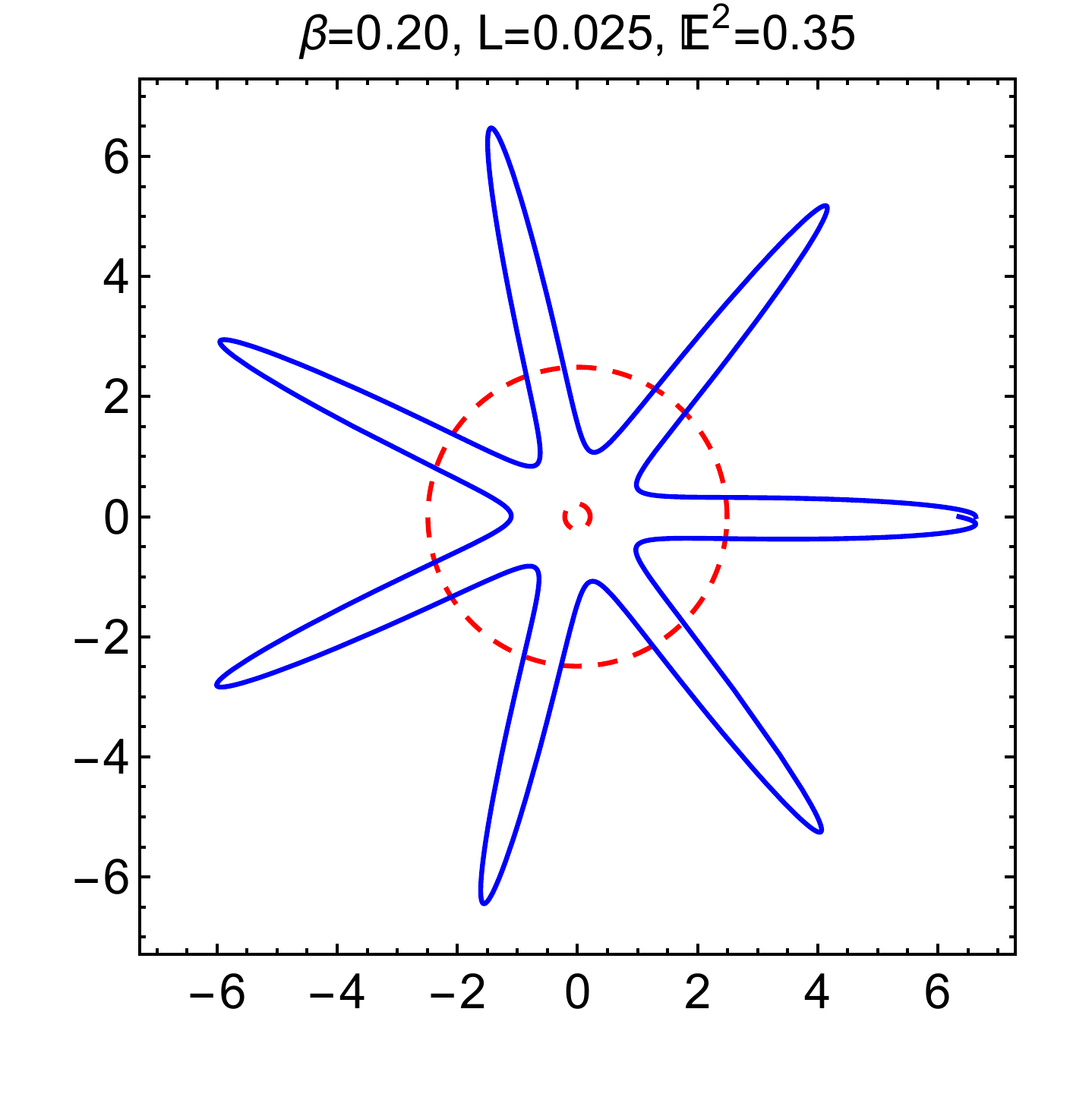}  
	\end{tabular}
	\begin{tabular}{cc}
		\includegraphics[height=4.7cm,keepaspectratio]{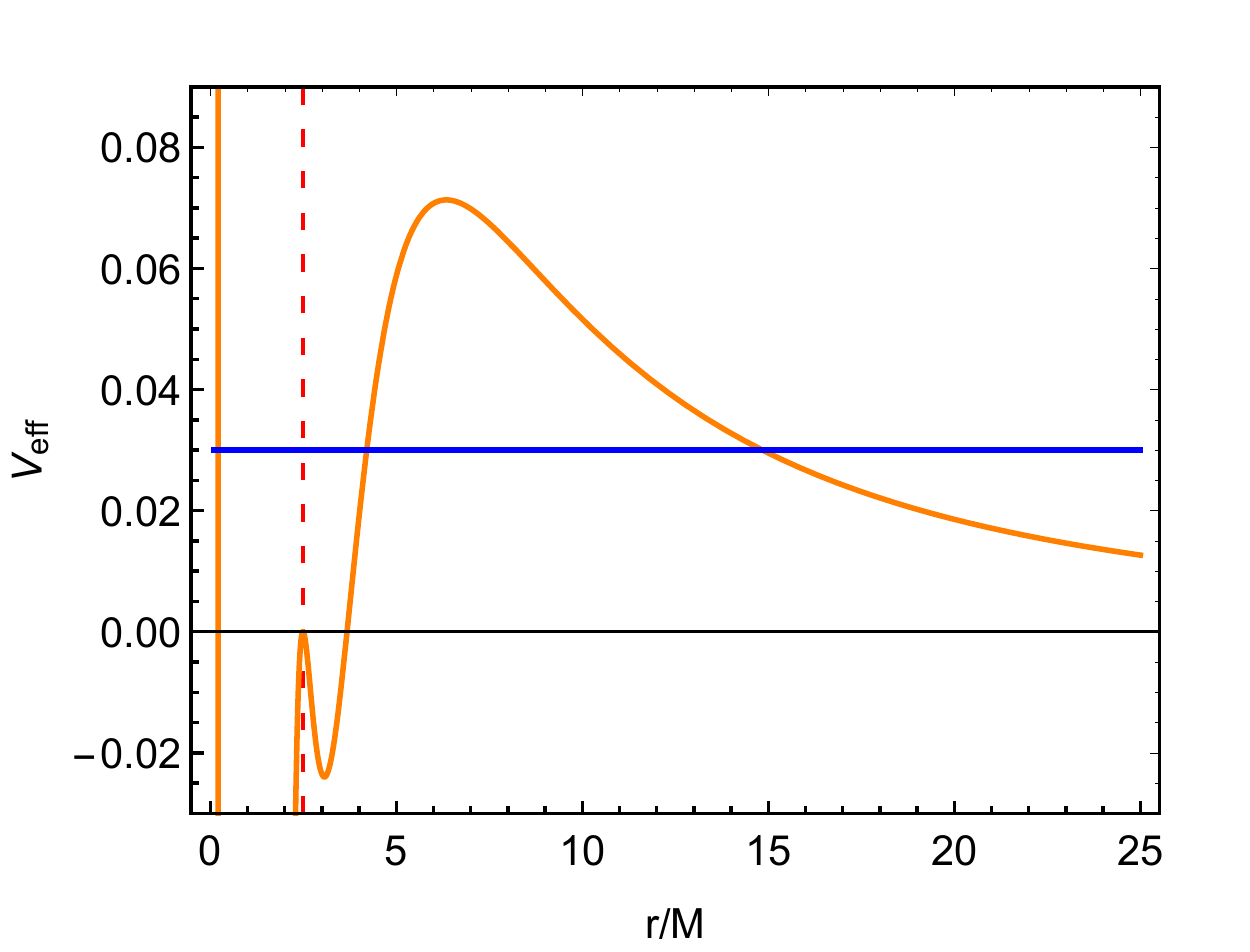}&	\includegraphics[height=4.7cm,keepaspectratio]{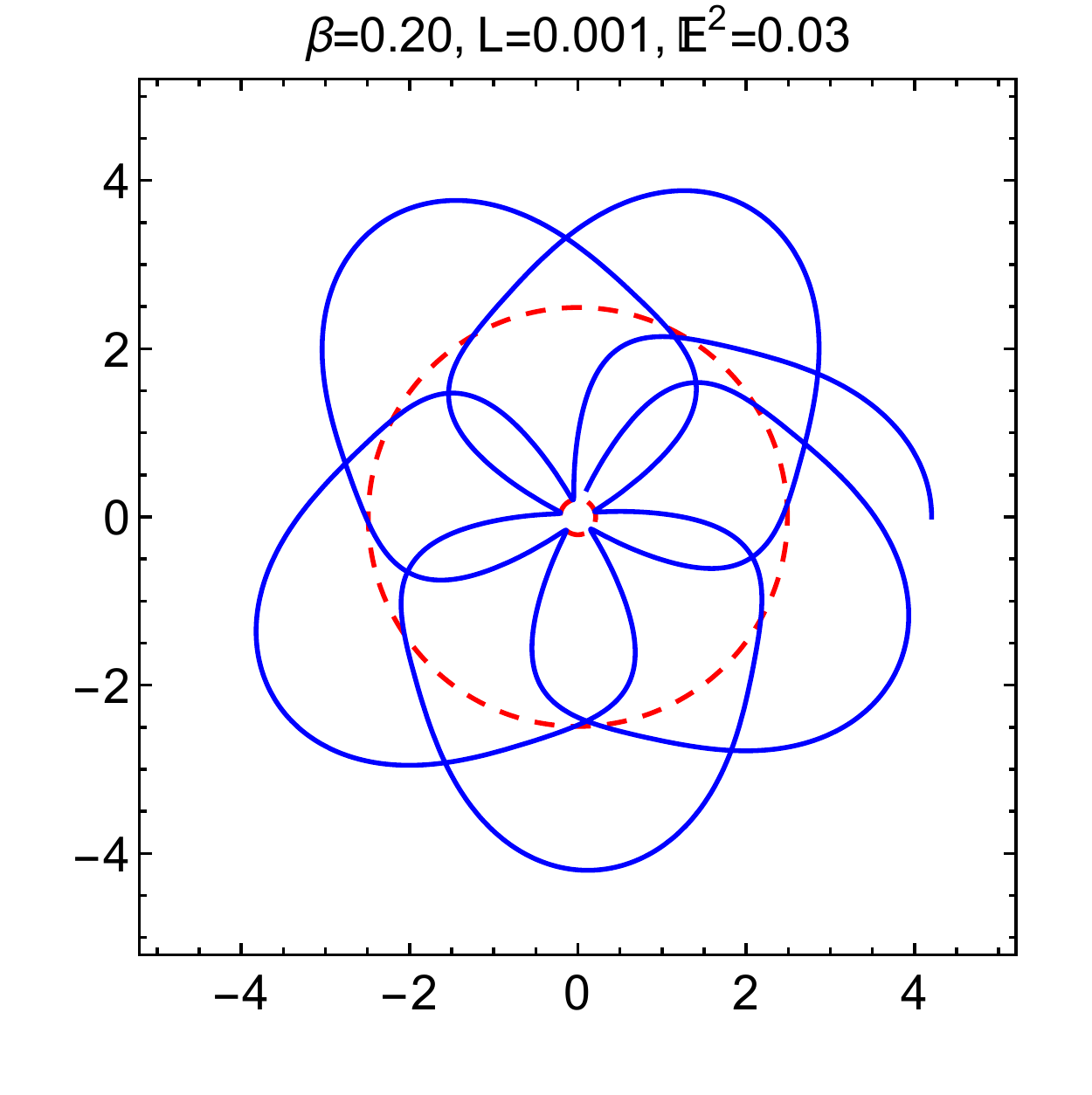} 
	\end{tabular}		
	\caption{The possible bound orbits in two-horizons LNE black hole as pairs of an effective potential scenario and its corresponding possible orbit for massive particles (top) and light particles (bottom).}
	\label{L_orbit_ext}
\end{figure}

Typical orbits of two-horizon LNE black hole are shown in Fig. \ref{L_orbit_ext}. For massive particles it is shown that the orbit is precessed star-polygon-shaped with large radii. On the other hand, the orbit of photon is similar to the ENE case where it travels from a radius outside the outer horizon and bounces from the inner horizon back to back, producing yet another epicycloid orbit. Unfortunately, we have not been able to study the structure of spacetime between the two horizon radii. At the moment, it is therefore unclear whether timelike or null bound orbits are physical in the extremal LNE black holes.

\section{Geodesics in Three-Horizon Case}
\label{three}

\begin{figure}[!ht]
	\centering
	\begin{tabular}{cc}
		\includegraphics[height=4.7cm,keepaspectratio]{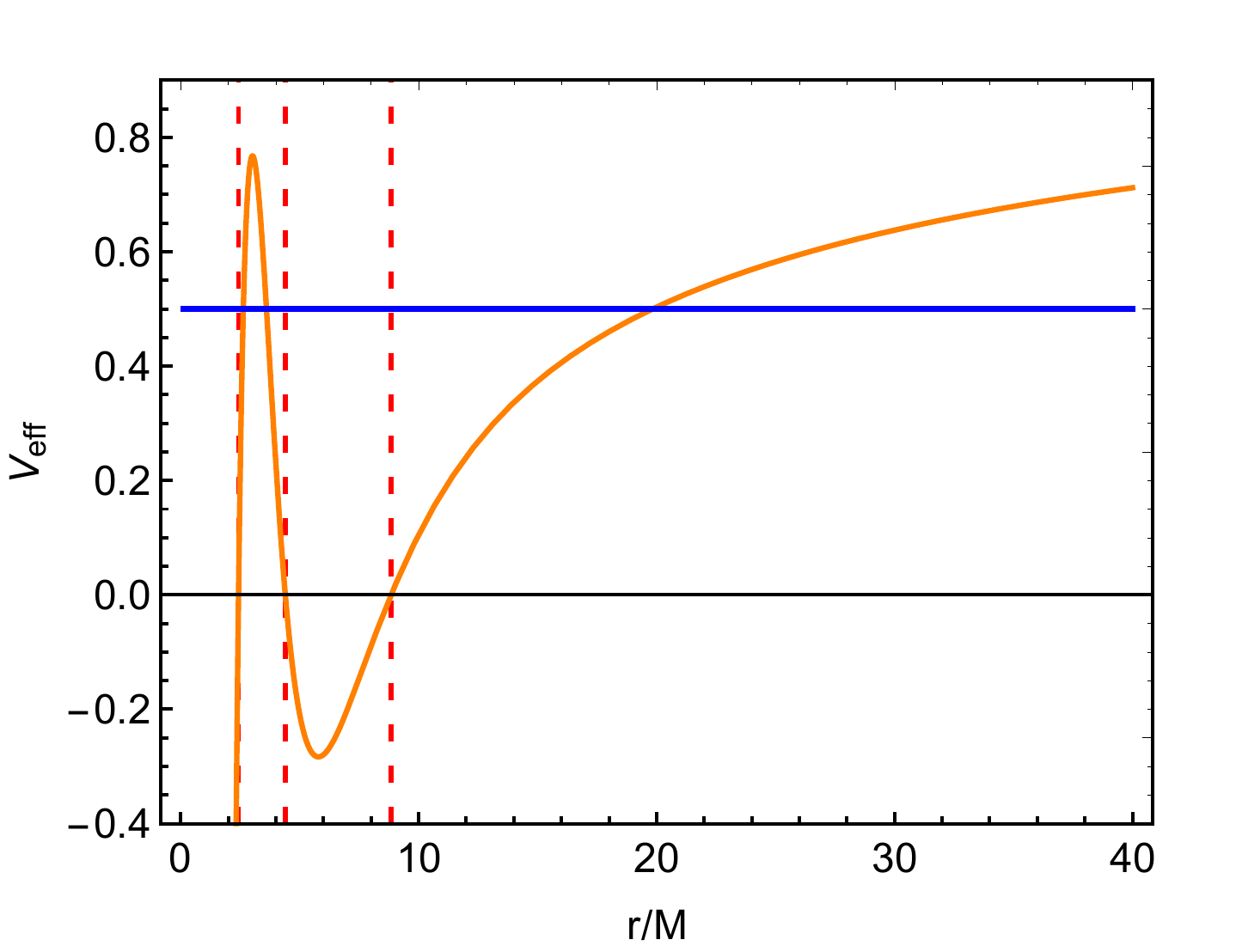}&	\includegraphics[height=4.7cm,keepaspectratio]{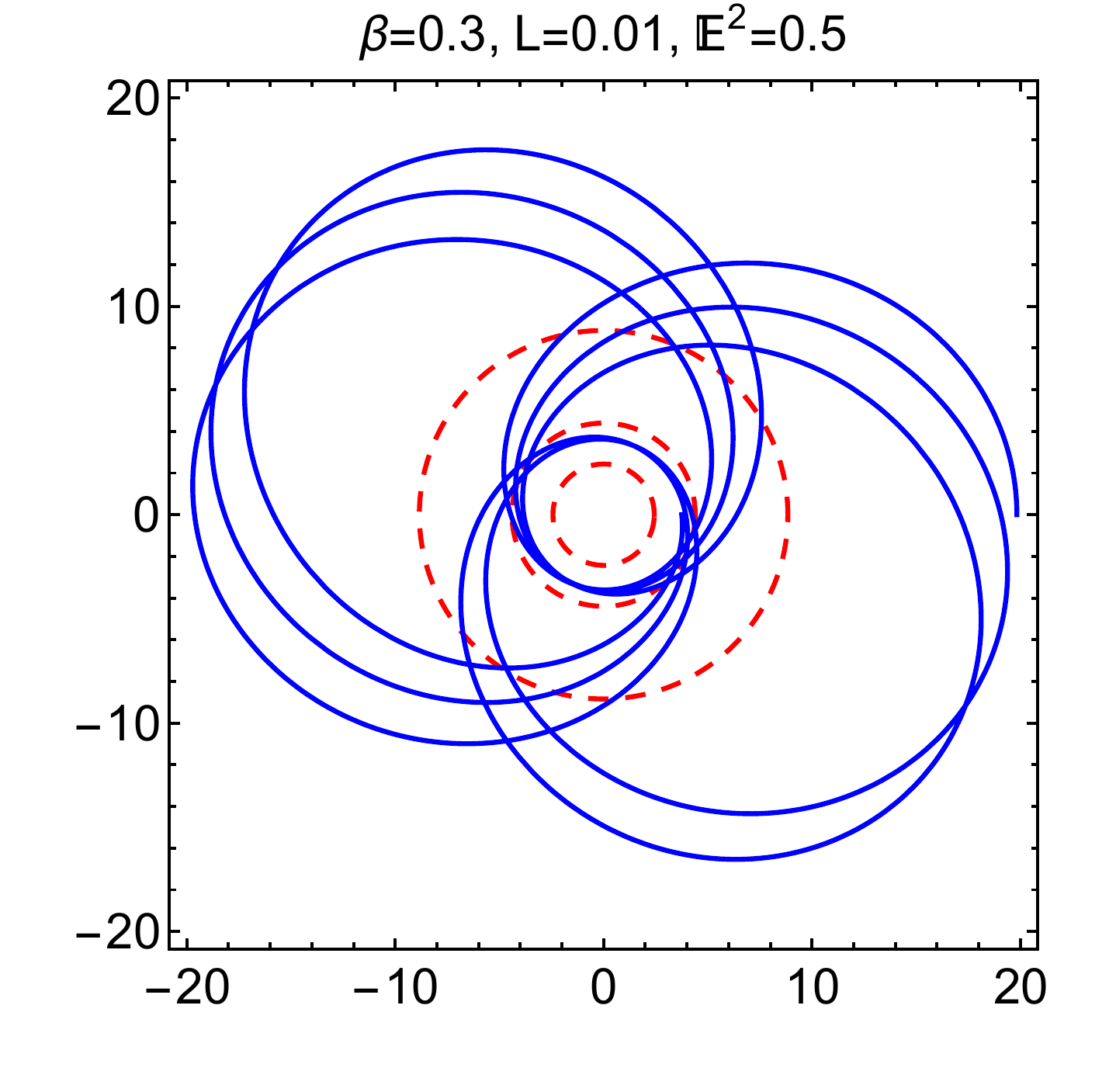}  
	\end{tabular}
	\begin{tabular}{cc}
		\includegraphics[height=4.7cm,keepaspectratio]{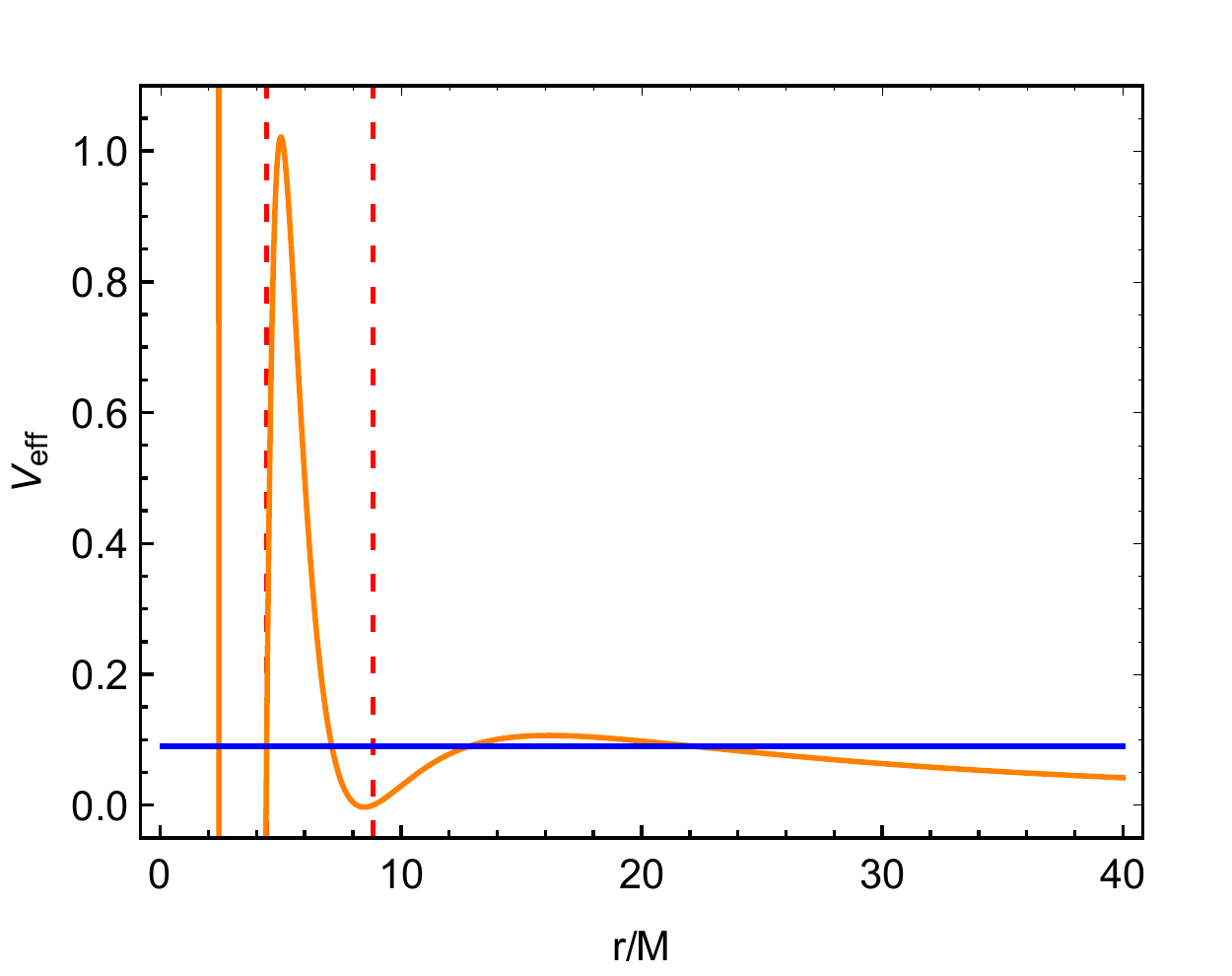}&	\includegraphics[height=4.7cm,keepaspectratio]{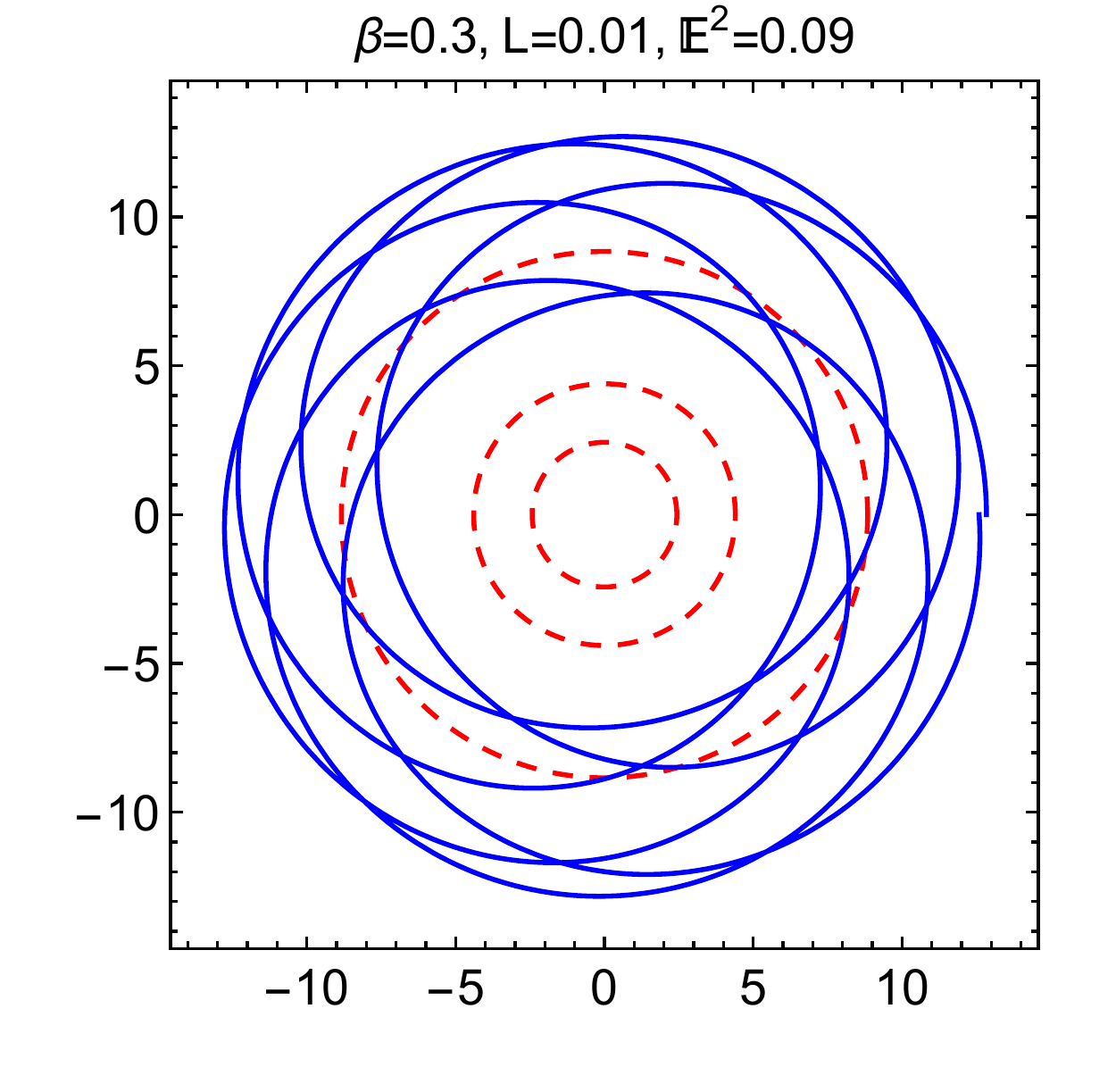} 
	\end{tabular}		
	\caption{The possible bound orbits in three-horizons ENE black hole as pairs of an effective potential scenario and its corresponding possible orbit for massive particles (top) and light particles (bottom).}
	\label{E_orbit_3}
\end{figure}

The maximum number of horizons that can arise in these models are three (Fig. \ref{fQ}). This scenario affects particles behaviour due to the change of worldlines in every horizon. In Fig. \ref{E_orbit_3} the $V_{eff}$ for both timelike and null particles along with their corresponding bound orbits around ENE black holes are shown. The resulting orbits perform similar behaviour to the previous textremal case, in which the particles do not cross the inner horizon $r_1$.

\begin{figure}[!ht]
	\centering
	\begin{tabular}{cc}
		\includegraphics[height=4.7cm,keepaspectratio]{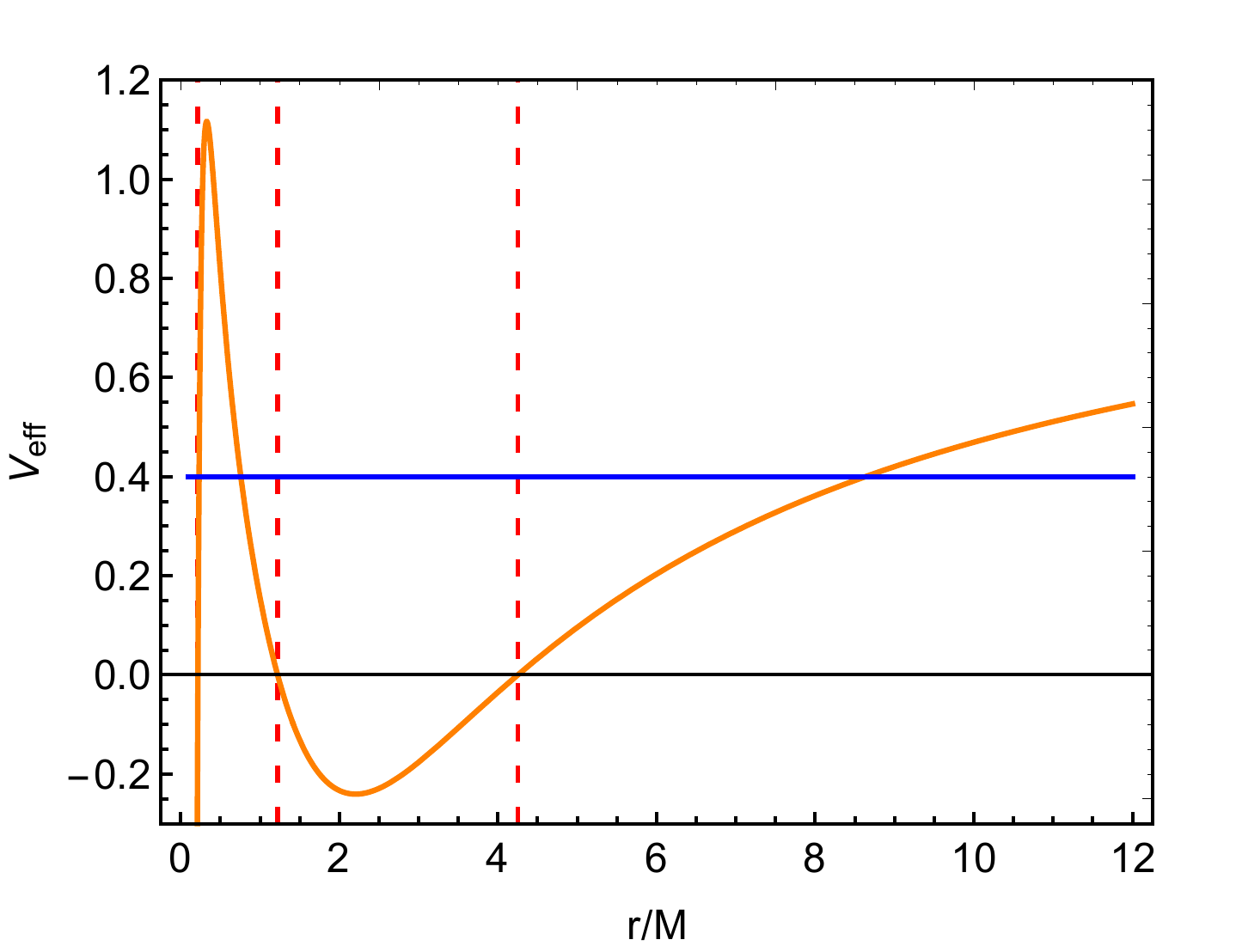}&	\includegraphics[height=4.7cm,keepaspectratio]{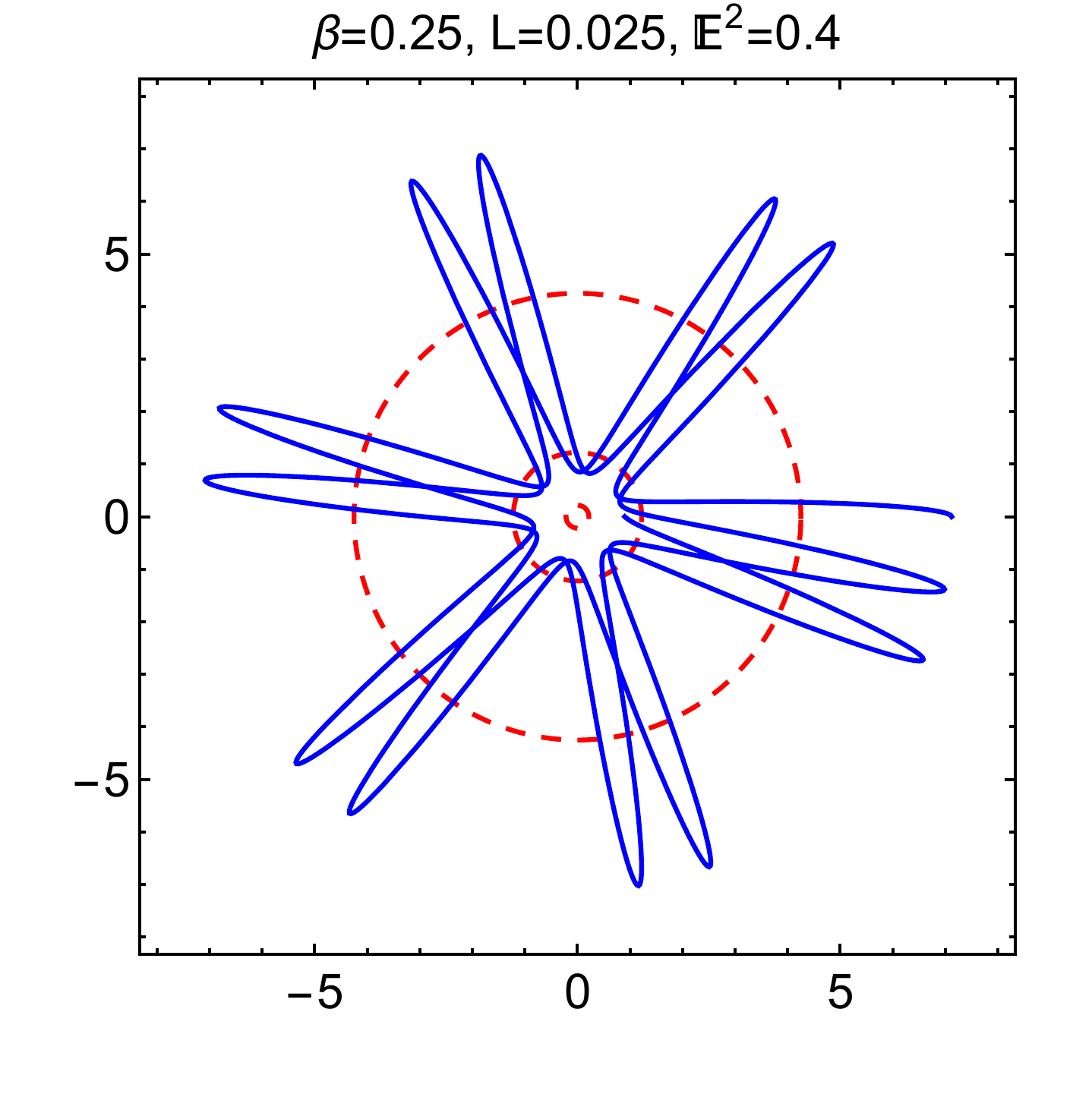}  
	\end{tabular}
	\begin{tabular}{cc}
		\includegraphics[height=4.7cm,keepaspectratio]{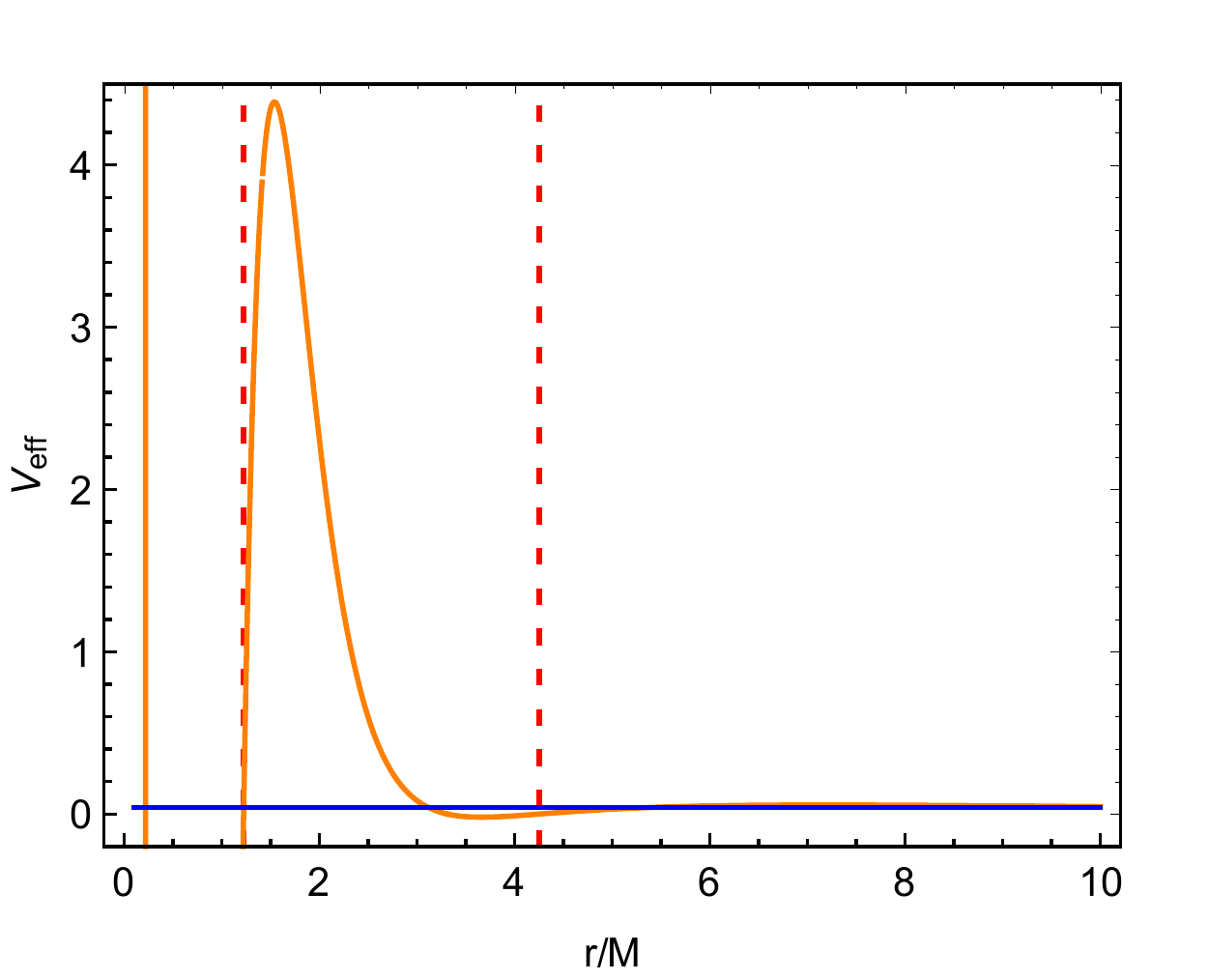}&	\includegraphics[height=4.7cm,keepaspectratio]{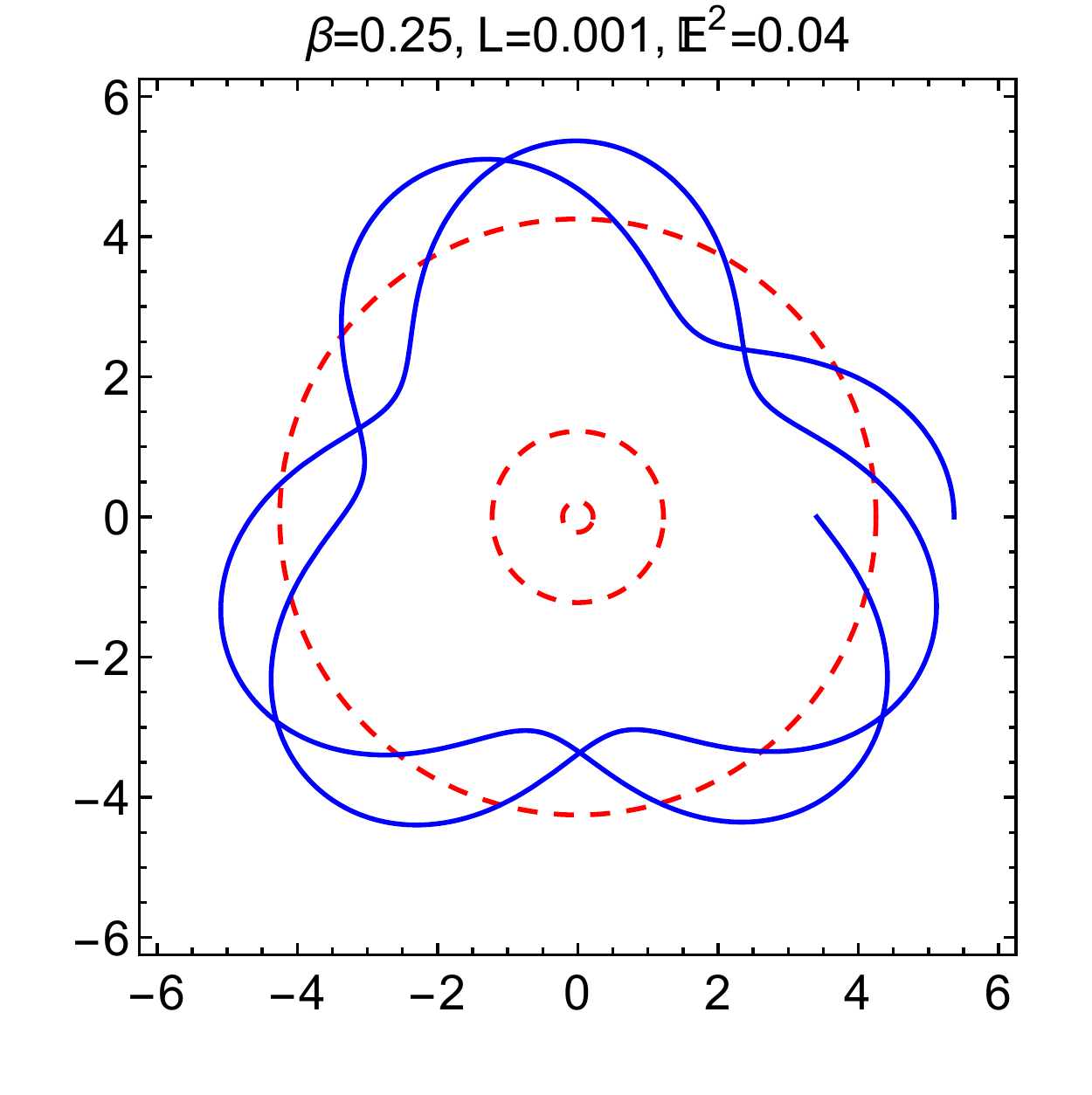} 
	\end{tabular}		
	\caption{The possible bound orbits in three-horizons LNE black hole as pairs of an effective potential scenario and its corresponding possible orbit for massive particles (top) and light particles (bottom).}
	\label{L_orbit_3}
\end{figure}

In the case of three horizons LNE black hole, we discover that the configuration is identical with its two horizons case. Shown in top pair of Fig. \ref{L_orbit_3}, massive particles follow a path with very large radius under a star-polygon-shaped path. Meanwhile, the orbit of photon exhibits a consistent outcome where it is shown that light infiltrate the outer horizon and is repelled out after hitting a particular radius, creating a epitrochoid with certain precession. As in the extremal case, however, the question whether such orbits are physical or not remains open.

\section{Thermodynamics}
\label{t}

From semiclassical perspective, black hole radiates just like a blackbody and has non-decreasing entropy~\cite{Bekenstein:1972tm,Bekenstein:1974ax}. Historically, the thermodynamic properties of vacuum black hole in Anti de Sitter (AdS) spacetime has been studied by Hawking and Page \cite{Hawking:1982dh}. It is shown that the AdS spacetime enables the existence of the minimum temperature to exist. The information about the positivity of heat capacity reveals that the black hole is in stable equilibrium. Charged static version (RN-AdS) have a different behavior in thermodynamical aspects. It has two stationary conditions indicating the phase transition. The thermodynamical properties of nonlinear electrodynamics have also been widely discussed. Thermodynamical aspects of BI-(A)dS black hole was studied by Fernando \cite{Fernando:2003tz}. The corresponding first law mechanics as well as the stability in the grand canonical ensemble were investigated in \cite{Fernando:2006gh} and also by Myung et al in~\cite{Myung:2008eb}. Higher dimensional version of the nonlinear BI black hole and the thermodynamics consequences was extensively discussed in \cite{Dey:2004yt}. In~\cite{Jayawiguna:2018tba} two of us also proved that BI coupled to Eddington-inspired-Born-Infed (EiBI) gravity satisfies the first law and analyzed its entropy. Moreover, in the NLED extension of the low energy string theory, Dehghani obtained the Born-Infeld-dilaton case \cite{Dehghani:2020jcw} where the thermodynamics quantity depends on nonlinear and dilaton parameter. If the nonlinear parameter shifts to infinity, then we have thermodynamics properties in Maxwell-dilaton gravity \cite{Dehghani:2018tcw}. Thermodynamics in different type of nonlinear electrodynamics has been proposed in many extensive literatures \cite{Gonzalez:2009nn}.

Since black holes radiate, the corresponding Hawking temperature are given by
\begin{eqnarray}
T_{H}&=&\frac{f'(r)}{4\pi}\bigg|_{r=r_{+}}, \nonumber \\ &=& \frac{1}{4\pi}
\left\{\begin{array}{cl}
\frac{1}{r_{+}}-2 \beta ^2 r_{+} ~e^{-\frac{Q^2}{4 \beta ^2 r^4_{+}}} \sinh \left(\frac{Q^2}{4 \beta ^2 r^4_{+}}\right) &, \textrm{ENE}\\
\frac{1}{r_{+}} -\beta ^2 r_{+} \ln \left(\frac{Q^2}{2 \beta ^2 r^4_{+}}+1\right) &, \textrm{LNE}
\end{array} \right\}\label{th}.
\end{eqnarray}
\begin{figure}[!ht]
	\centering
	\begin{tabular}{cc}
		\includegraphics[height=6.7cm,keepaspectratio]{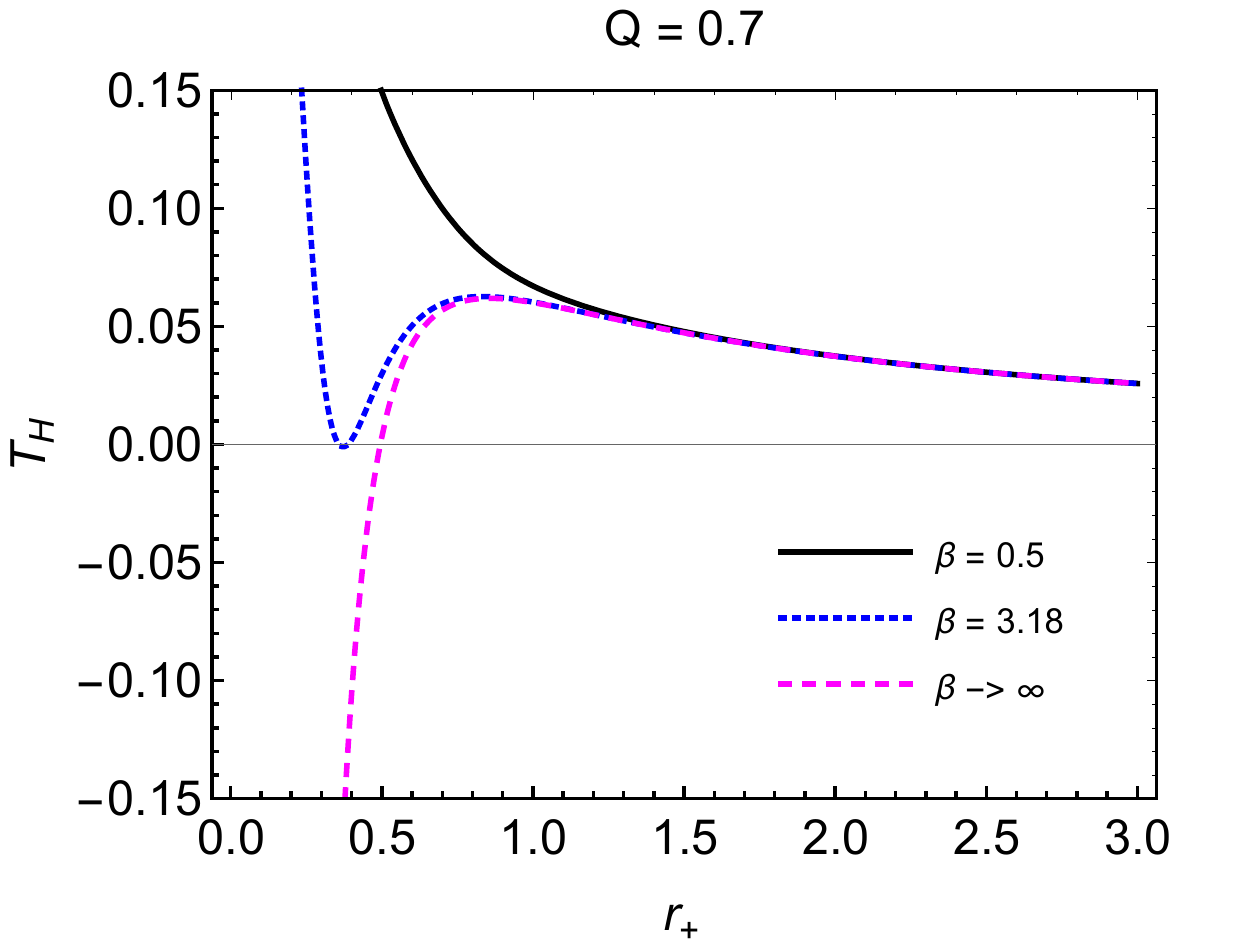} &		\includegraphics[height=6.7cm,keepaspectratio]{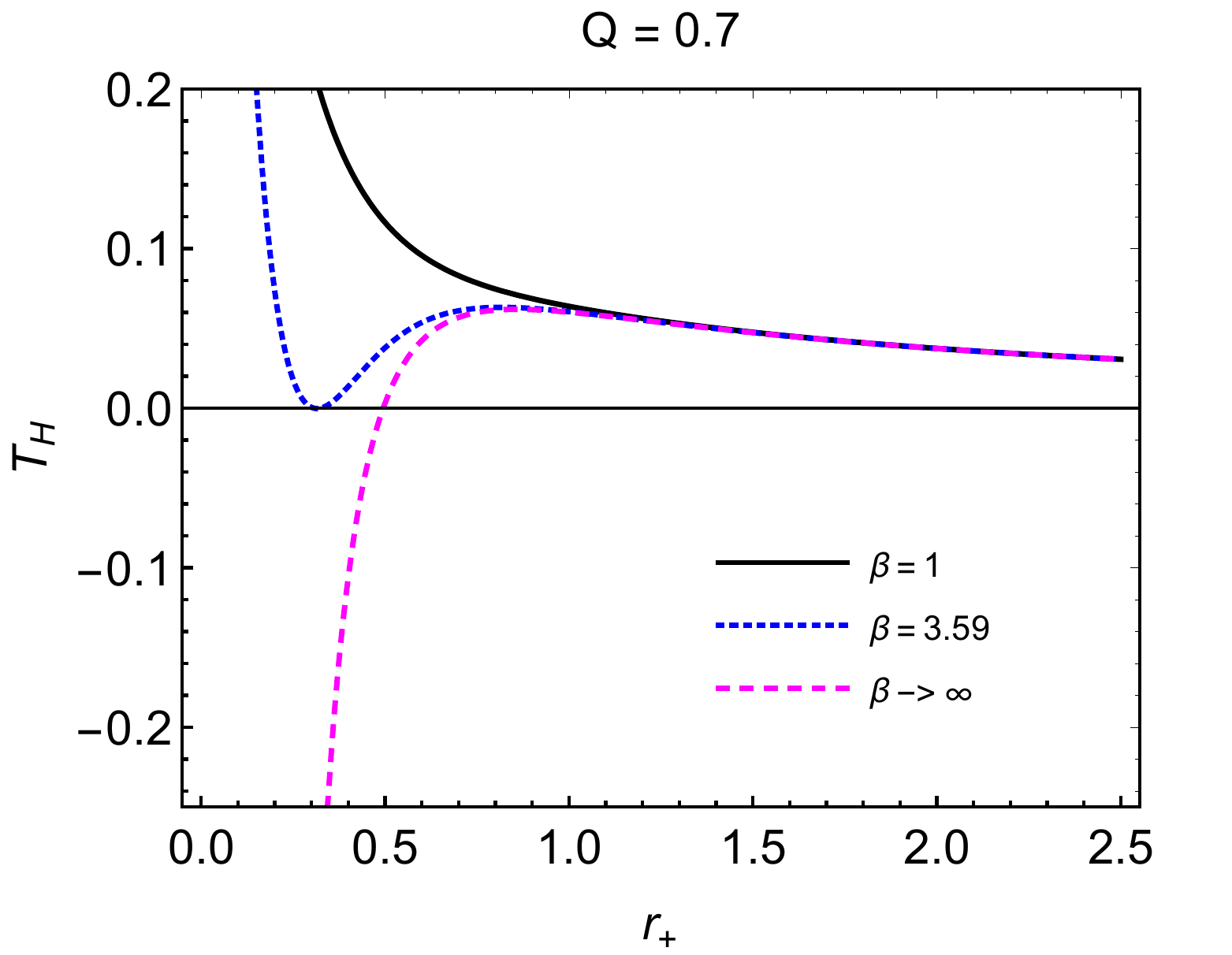} \\ \includegraphics[height=6.7cm,keepaspectratio]{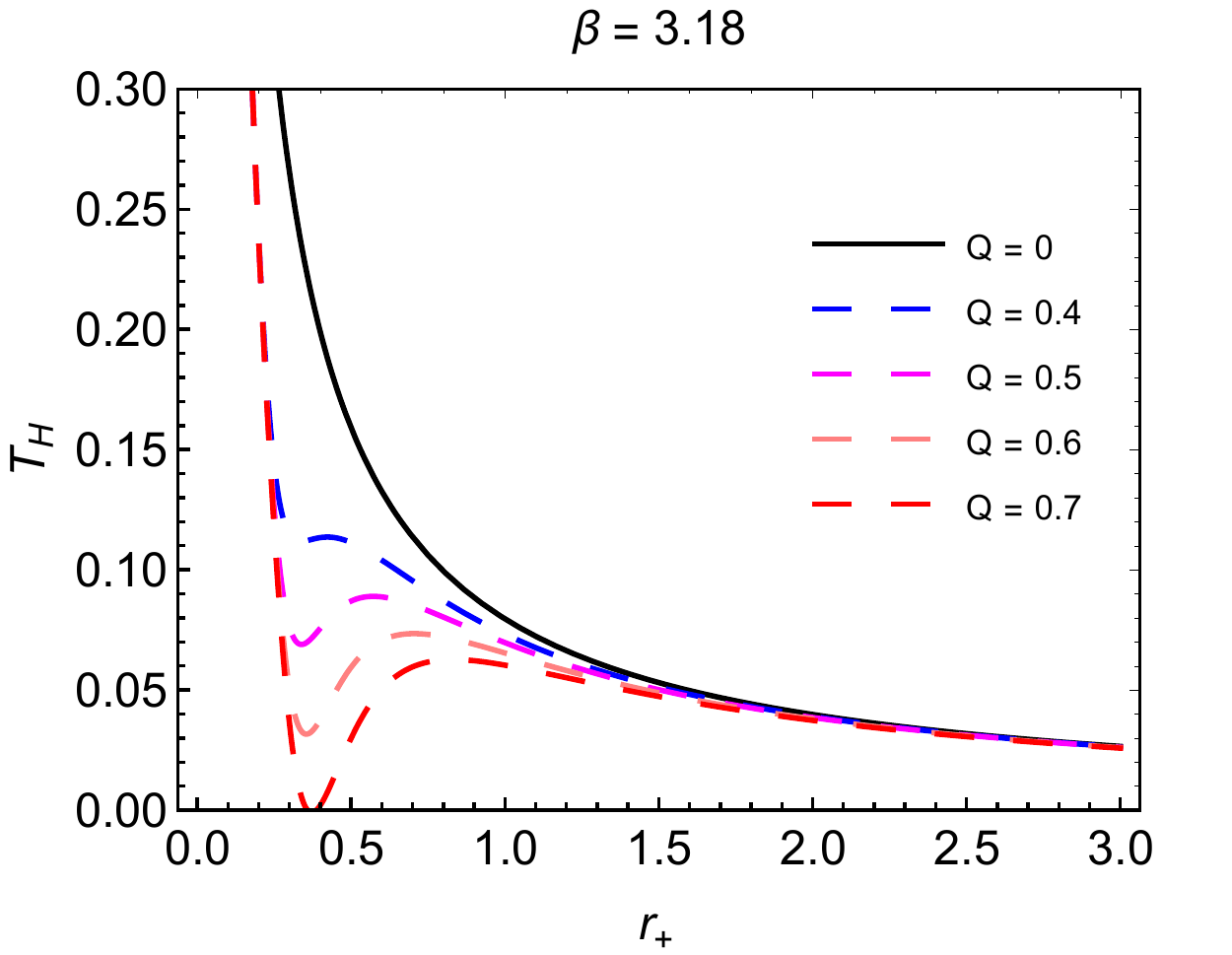} &		\includegraphics[height=6.7cm,keepaspectratio]{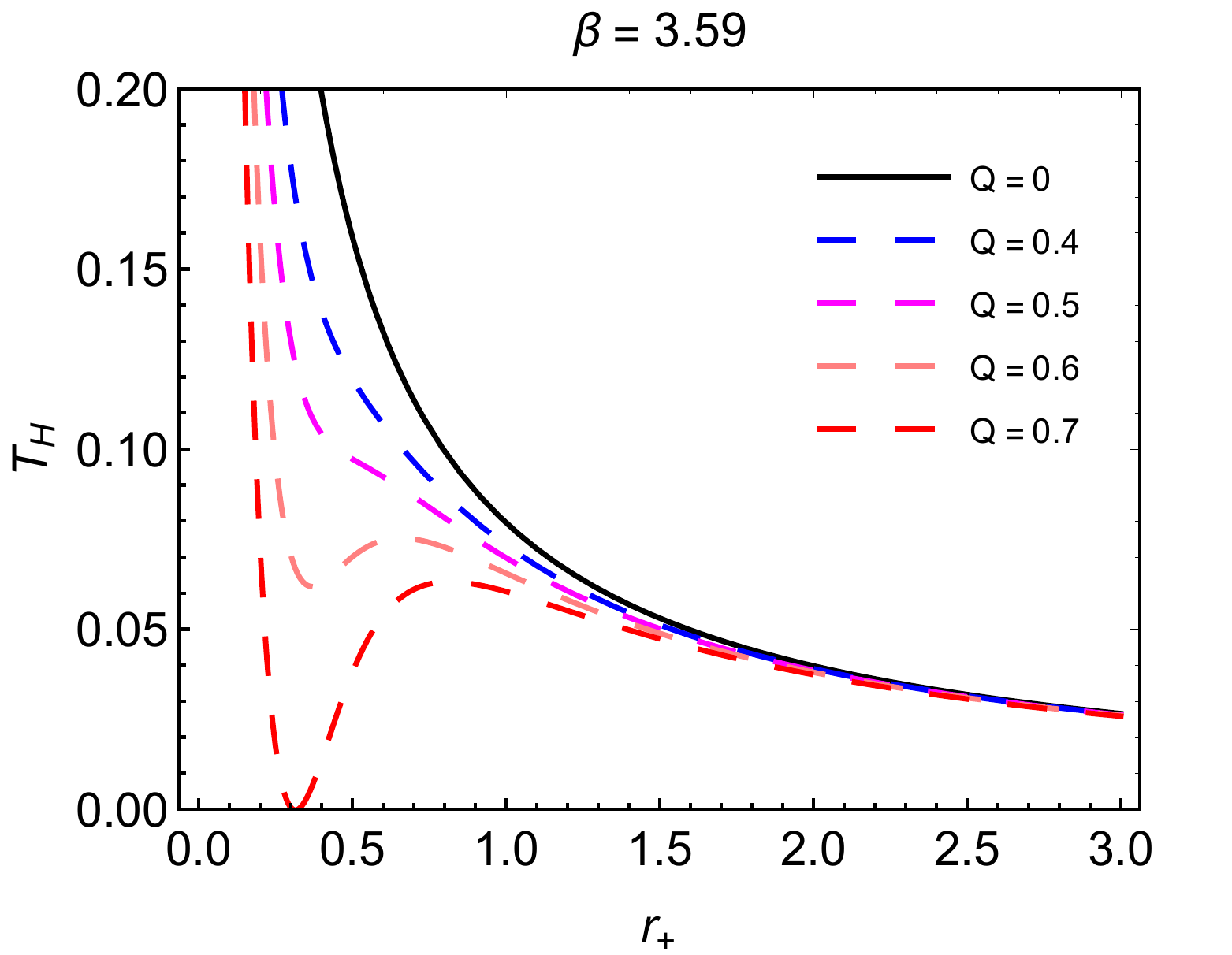} 
	\end{tabular}
	\caption{[Left] Plot of Hawking temperature versus radius in ENE with $Q=0.7$ (with varying $ \beta $) and $ \beta = 3.18 $ (with varying charge). [Right] Plot of Hawking temperature versus radius in LNE with $Q=0.7$ (with varying $ \beta $) and $ \beta = 3.59 $ (with varying charge).}
	\label{ENET}
\end{figure} 
In Fig.~\ref{ENET}, we showed the typical plot of the Hawking temperature versus radius. The temperature in the strong regime $ (\beta<1) $ is Schwarzschild-like. 
The Hawking temperature decreases as the radius increases and becomes colder when it absorbs matter from outside the horizon \cite{Ortin:2015hya}. On the other hand, in the intermediate regime ($1\lesssim\beta\lesssim5$), there exist two local optima, one of which indicates the stability. Particularly, at $ \beta=3.18 $ (for ENE) and $ \beta=3.59 $ (for LNE) there exist $r=r_{eq}$ satisfying $T(r=r_{eq})=0$ and ${dT\over dr}\bigg|_{r=r_{eq}}=0$. At weak coupling limit $ \beta\rightarrow\infty $, the temperature reduces to that of RN-like properties. We also plot the behavior of Hawking temperature with varying charge. From this, we can infer that the local minima tends to disappear when charge $ Q $ gets weaker.

In any case, the black hole can radiate and undergo phase transitions, but its entropy can decrease. Nevertheless, the total entropy never decreases, as suggested by the generalized second-law of black hole thermodynamics \cite{Bekenstein:1974ax}. The area $ A $ and entropy $ S $ of magnetically charged black hole is given by
\begin{eqnarray}
\mathcal{A} &=& \int_{0}^{2\pi} \int_{0}^{\pi} \sqrt{g_{\theta\theta}g_{\phi\phi}} d\theta d\phi = 4\pi r_{+}^2,\\ S &=& \frac{\mathcal{A}}{4} = \pi r^2_{+},
\end{eqnarray}
respectively. Both NLED models satisfy this entropy. Since our solutions are static it possesses a timelike killing vector $ \xi^{a}=\lbrace1,0,0,0\rbrace $. The corresponding magnetic field vector then is defined by 
\begin{equation}
H_{a}= - \frac{1}{2} \epsilon_{abcd} G^{cd} \xi^{b},
\end{equation}
where $ G^{ab}= \mathcal{L}_{F} F^{ab} $ \cite{Rasheed:1997ns}. Then, the magnetic potential for both ENE and LNE are
\begin{eqnarray}
\label{potential}
\Psi(r)=
\left\{\begin{array}{cl}

-\frac{\sqrt{Q\beta}}{2^{11/4}} ~\Gamma \left(\frac{1}{4},\frac{Q^2}{2 r^4 \beta^2}\right) ,\,\, \textrm{for ENE}. \\

\frac{\sqrt{\beta} \sqrt{Q}}{4\sqrt[4]{2}} \bigg[\ln \bigg|\sqrt{\frac{2^{3/4} \sqrt{\beta } \sqrt{Q}r+Q+\sqrt{2} \beta  r^2}{-2^{3/4}\sqrt{\beta } \sqrt{Q} r+Q+\sqrt{2} \beta r^2}}\bigg|+ \tan^{-1}\left(1-\frac{2^{3/4} \sqrt{\beta}r}{\sqrt{Q}}\right)\nonumber \\  - \tan ^{-1}\left(\frac{2^{3/4}\sqrt{\beta } r}{\sqrt{Q}}+1\right)\bigg],\,\, \textrm{for LNE}
\end{array} \right\}.
\end{eqnarray}

From the conserved quantity, we can conclude that the ADM mass now depends on the entropy, charge, and the parameter $\beta$. Formally the expression is given by $ M\equiv M(S,Q,\beta) $. When $ \beta\rightarrow\infty $, all the solutions reduce to the well known RN with $ M(S,Q) $. As we can see that the $ \beta $ parameter plays an important role in obtaining the first-law of nonlinear black hole thermodynamics \cite{Rasheed:1997ns,Gunasekaran:2012dq}. Therefore with the Smarr formula we get
\begin{equation}
\label{smarr}
M=2\left(\frac{\partial M}{\partial S}\right)S+\left(\frac{\partial M}{\partial Q}\right) Q-\left(\frac{\partial M}{\partial \beta}\right)\beta,
\end{equation}
where $ T = \left(\frac{\partial M}{\partial S}\right)_{Q,\beta} $, $ \Psi = \left(\frac{\partial M}{\partial Q}\right)_{S,\beta} $, and $ \mathcal{B} = \left(\frac{\partial M}{\partial \beta}\right)_{S,Q} $. Using the derivative formula, the quantity $ \mathcal{B} $ reads
\begin{eqnarray}
\label{B}
\mathcal{B}=
\left\{\begin{array}{cl}

\frac{ \beta  r^3_{+}}{24} \left[8 e^{-\frac{Q^2}{2 \beta^2 r^4_{+}}}-\sqrt[4]{2} \left(\frac{Q^2}{\beta^2 r^4_{+}}\right)^{3/4} \Gamma \left(\frac{1}{4},\frac{Q^2}{2r^4_{+} \beta ^2}\right)-8\right],\,\, \textrm{for ENE}. \\

-\frac{1}{3} \beta  r^3_{+} \ln\bigg|\frac{Q^2}{2 \beta ^2 r^4_{+}}+1\bigg|+ \frac{Q^{3/2}}{12 \sqrt[4]{2} \sqrt{\beta }}	 \bigg[\ln\bigg|\sqrt{\frac{2^{3/4}\sqrt{\beta } \sqrt{Q} r_{+}+Q+\sqrt{2} \beta r^2_{+}}{-2^{3/4} \sqrt{\beta } \sqrt{Q} r_{+}+Q+\sqrt{2} \beta r^2_{+}}}\bigg| \\ +\tan ^{-1}\left(1-\frac{2^{3/4} \sqrt{\beta } r_{+}}{\sqrt{Q}}\right)-\tan^{-1}\left(\frac{2^{3/4} \sqrt{\beta } r_{+}}{\sqrt{Q}}+1\right)\bigg],\,\, \textrm{for LNE}
\end{array} \right\}.
\end{eqnarray}
Our solution obeys the first-law by subtituting the corresponding variable ($ r=r_{+} $) to the Smarr formula \eqref{smarr}.

The black hole stability can be probed by looking at the specific heat
\begin{equation}
C_{Q}=T_{H}\frac{\partial S}{\partial T_{H}}.
\end{equation}
Inserting temperature and entropy, we get
\begin{eqnarray}
C_{Q,ENE} &=& \frac{2 \pi  r^4_{+} e^{-\frac{Q^2}{4 \beta ^2 r^4_{+}}} \left[2\beta^2 r^2_{+} \sinh\left(\frac{Q^2}{4\beta^2r^4_{+}}\right)+1\right]}{e^{-\frac{Q^2}{4\beta^2r^4_{+}}} \left[2\beta^2 r^4_{+} \sinh\left(\frac{Q^2}{4\beta^2r^4_{+}}\right)-r^2_{+}\right]+2 Q^2},\label{cqene}\\ C_{Q,LNE} &=& \frac{2 \pi  r^2_{+} \left(Q^2+2 \beta ^2 r^4_{+}\right)\left[\beta ^2  r^2_{+} \ln \bigg|\frac{Q^2}{2 \beta ^2 r^4_{+}}+1\bigg|-1\right]}{Q^2 \left(1-4 \beta^2 r^2_{+}\right)+\beta ^2 r^2_{+}\left(Q^2+2 \beta ^2 r^4_{+}\right) \ln\bigg|\frac{Q^2}{2 \beta ^2 r^4_{+}}+1\bigg|+2 \beta ^2 r^4_{+}}.\label{cqlne}\nonumber \\ 
\end{eqnarray}

\begin{figure}[!ht]
	\centering
	\begin{tabular}{cc}
		\includegraphics[height=6.7cm,keepaspectratio]{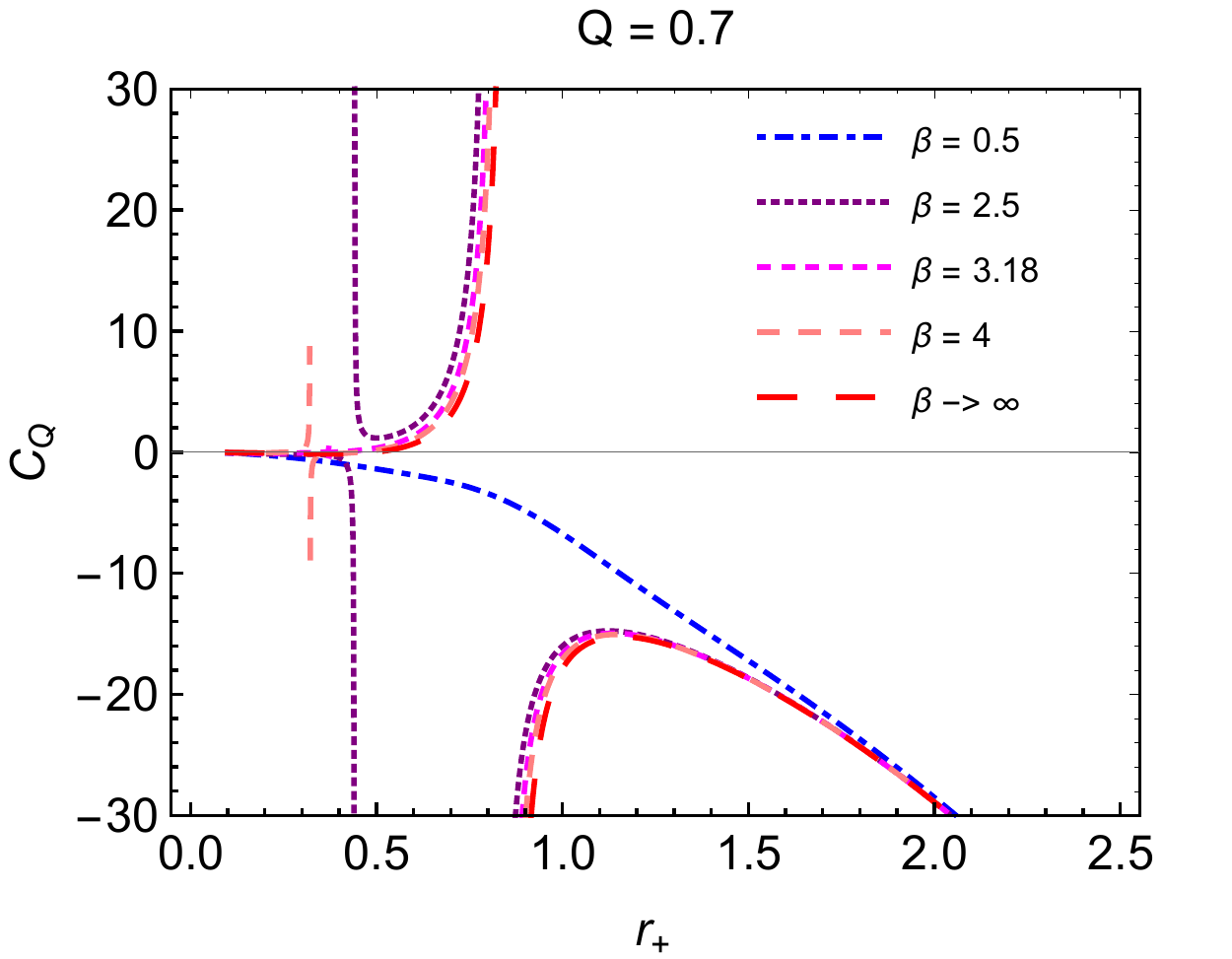} &		\includegraphics[height=6.7cm,keepaspectratio]{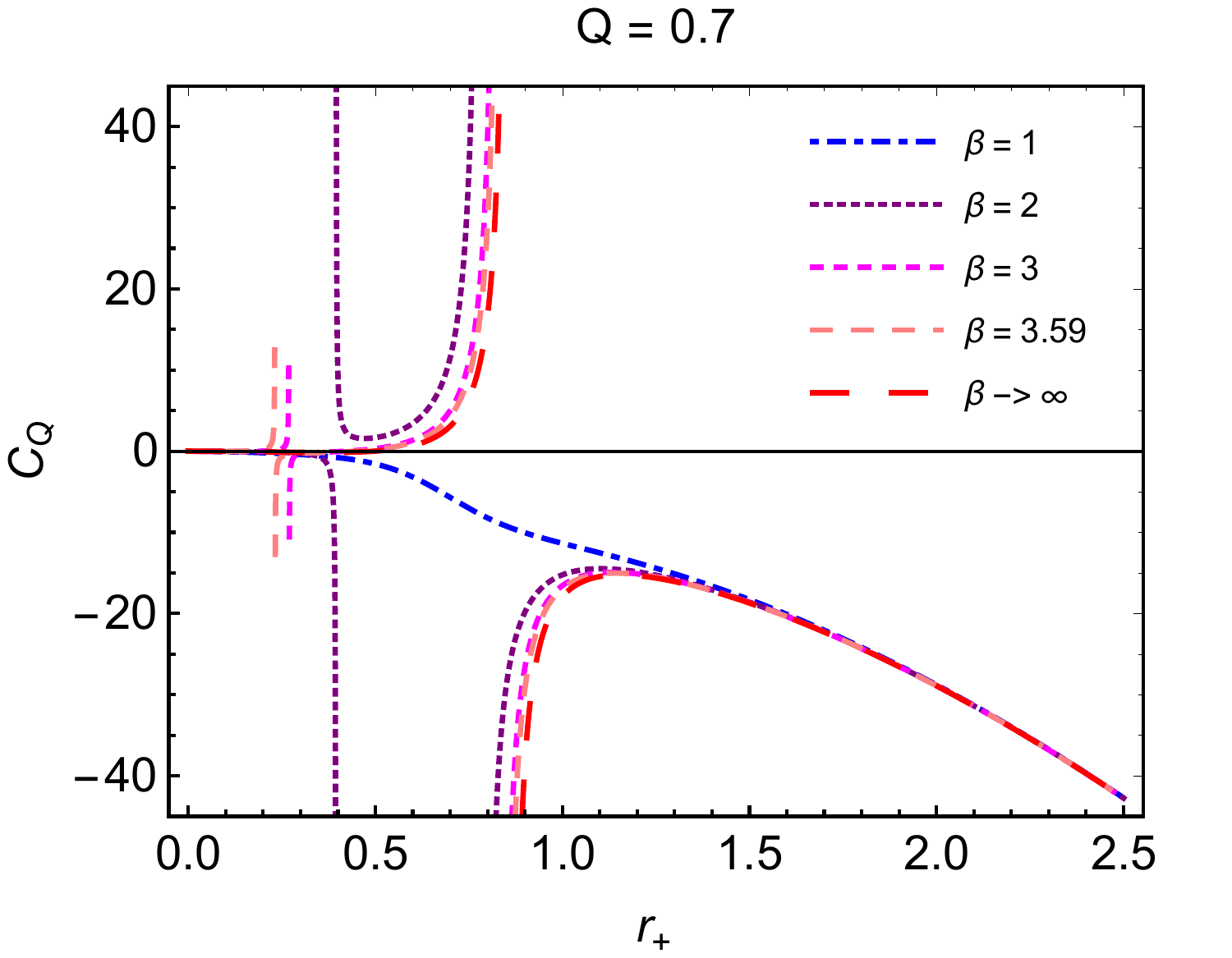}  
	\end{tabular}
	\caption{[Left] Plots of $C_{q}(r_{+})$ in ENE with $Q=0.7$. [Right] Plots of  $C_{q}(r_{+})$ in LNE with $Q=0.7$.}
	\label{LNET}
\end{figure}

In Fig.~\ref{LNET}, the heat capacity for both solutions are plotted with variation of $\beta$. It can be seen that for strongly-coupled NLED the black hole is unstable, in a sense that its rate of absorption is greater than its rate of emission. As $ \beta $ gets larger, the phase transition tends to occur from small to large black hole. On the other hand, we can infer that the nonlinearity of the matter sector enables the existence of stable black holes to have smaller horizons than its RN counterpart.

\section{Conclusion}
\label{con}

QED's recent photon-photon scattering indicates that the vacuum electrodynamics might be nonlinear. Since electrodynamics is believed to be long-range, it thus becomes essential what signatures the NLED leaves on astrophysical scale. It is precisely the motivation behind this work: to study such effects in the most extreme environment; that is in the vicinity of black hole. Our investigation has been quite fruitful. We found exact magnetically-charged solutions of NLED BH whose horizons are smaller than the RN counterpart. They are parametrized by the mass $M$ and charge $Q$, both of which can be identified as the ADM quantities, and the nonlinear parameter $\beta$ whose $\beta\rightarrow\infty$ limit reduces the solutions to RN.

Our main investigation is on the null geodesics, especially the photon's bound orbit. Our solutions can be classified into three-horizon, extremal, and one horizon solutions. As expected, for $\beta>1$ the null orbit is either unbounded (scattering state) or falling into the horizon. This is obvious from the shape of the corresponding $V_{eff}$ who is either not possessing any local minimum outside the horizon or unbounded from below. For $0<\beta<1$, on the other hand, the $V_{eff}$ exhibits the existence of local minimum satisfying $V_{eff}(r_m)|_{r_m>r_+}>0$. As a result, photon is allowed to be in bound (circular as well as non-cirular) state. For the one-horizon case the shape of orbit takes the form of precessed ellipse. For the extremal and three-horizons cases the situation is a bit trickier. The bound orbits exist but they generically cross the outer horizon.  To check the validity of these solutions, we ran the Eddington-Finkelstein diagram analysis, and the results show that in the region between horizons (or between outer and middle horizon) the spacetime structure enables infalling photon to escape back. With this in hand, we then claim that these horizon-crossing photon orbits are physical. For the extremal case the orbit takes the form of epicycloid, while in the three-horizon case it is epitrochoid. In any case, these show that the NLED black holes are much richer in the null orbit phenomenology than the RN. 


Lastly, we study their thermodynamical properties. As with other BH, these ENE and LNE BHs radiate with the Hawking temperature. In strong coupling limit, both solution (ENE and LNE) tends to be Schwarzschild-like. When $ \beta $ gets larger, the corresponding Hawking temperature smoothly transform into RN-like properties. Both solutions do obey the first-law of thermodynamics with the Smarr formula. The non-decreasing entropy signifies that they also obeys the generalized second law. The stability of a black hole can be obtained by extracting the Hawking temperature and the entropy information into specific heat. It is shown that the nonlinearity of the matter sector enables the existence of stable black holes to have smaller horizons than its RN counterpart.
\acknowledgments

We acknowledge grants from Universitas Indonesia through Hibah Riset PUTI Q2 No. NKB-1654/UN2.RST/HKP.05.00/2020 and Hibah Riset PPI Q1 No. NKB-583/UN2. RST/HKP.05.00/2021.

\section*{Data Availability Statement}

Data sharing is not applicable to this article as no data sets were generated or analyzed during the current study.


\appendix
\section{Eddington-Finkelstein Diagram for Extremal Case}
\label{AppA}

\begin{figure}[!h]
	\centering
	\begin{tabular}{cc}
		\includegraphics[height=6.5cm,keepaspectratio]{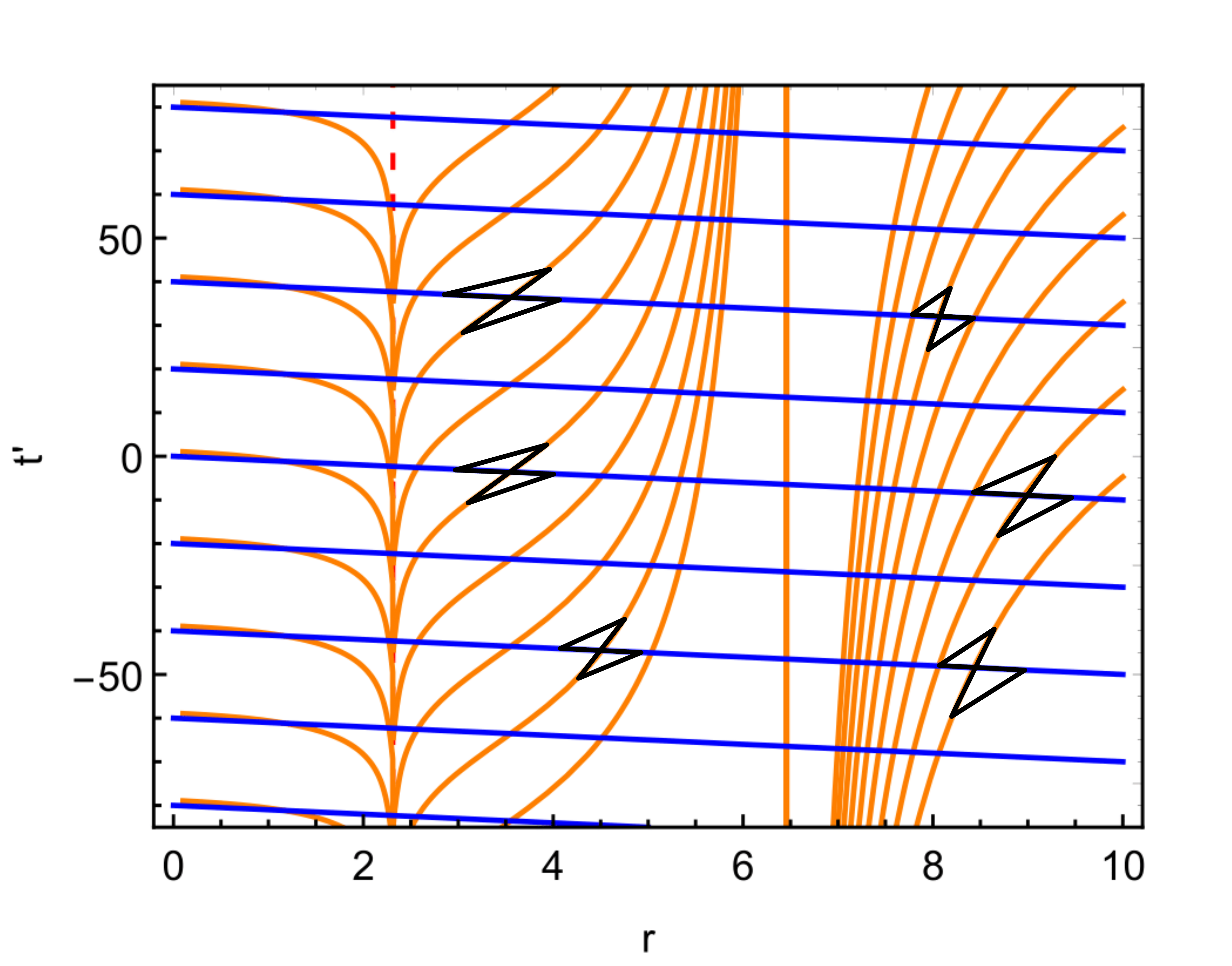} &	\includegraphics[height=6.5cm,keepaspectratio]{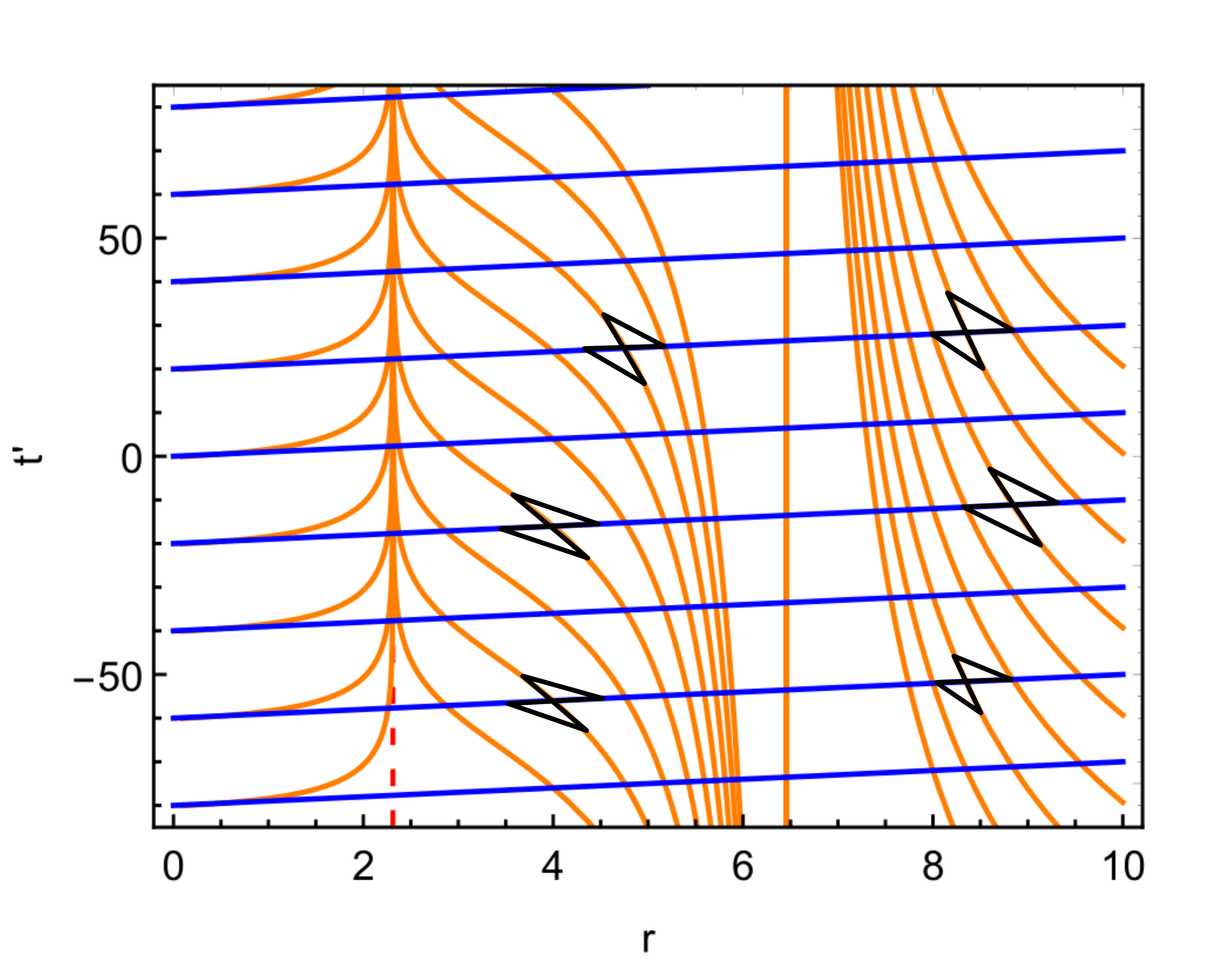}
	\end{tabular}
	\caption{Plots of Eddington-Finkelstein diagram for ingoing photon [left] and outgoing photon [right] of ENE black holes with $Q=10$ and $\beta=0.275$. The horizons can be located at $r_1 = 2.3115$ and $r_2 = 6.4575$. The black lines denote light cones which show the possible direction of light according to the worldlines.}
	\label{EF}
\end{figure}

To investigate the physicality of the extremal case furthermore, we study the null radial geodesic of the spacetime inside the outer horizon $r_2$ in the Eddington-Finkelstein coordinate. We introduce a pair of null coordinate
\begin{equation}
\label{uv}
u\equiv t+r' \,\,\,\, , \,\,\,\, v\equiv t-r' \,\,\,\, \text{where} \,\,\,\, r'=\int f(r)^{-1}dr,
\end{equation}
which give the new temporal coordinate $t'$ as
\begin{eqnarray}
\label{t*}
t'=
\left\{\begin{array}{cl}
u-r,& \textrm{for ingoing photon}\\
v+r,& \textrm{for outgoing photon}
\end{array} \right\}.
\end{eqnarray}
The behaviour of the resulting ingoing and outgoing null coordinates can be seen in Fig. \ref{EF}. Observing the worldlines (orange lines) and the light cones, we see that while the region $r<r_1$ is an inescapable region for light, the space $r_1<r<r_2$ is not. It is possible for a worldline of infalling photon between $r_1<r<r_2$ to avoid $r_1$ and, due to the null radial structure of outgoing photon, redirect its trajectory back across the outer horizon. Thus we can safely conclude that the orbital paths shown in the top and bottom right of Fig.~\ref{E_orbit_ext} are indeed physical.


\section{Eddington-Finkelstein Diagram for Three-Horizon Case}
\label{AppB}

\begin{figure}[!h]
	\centering
	\begin{tabular}{cc}
		\includegraphics[height=6.5cm,keepaspectratio]{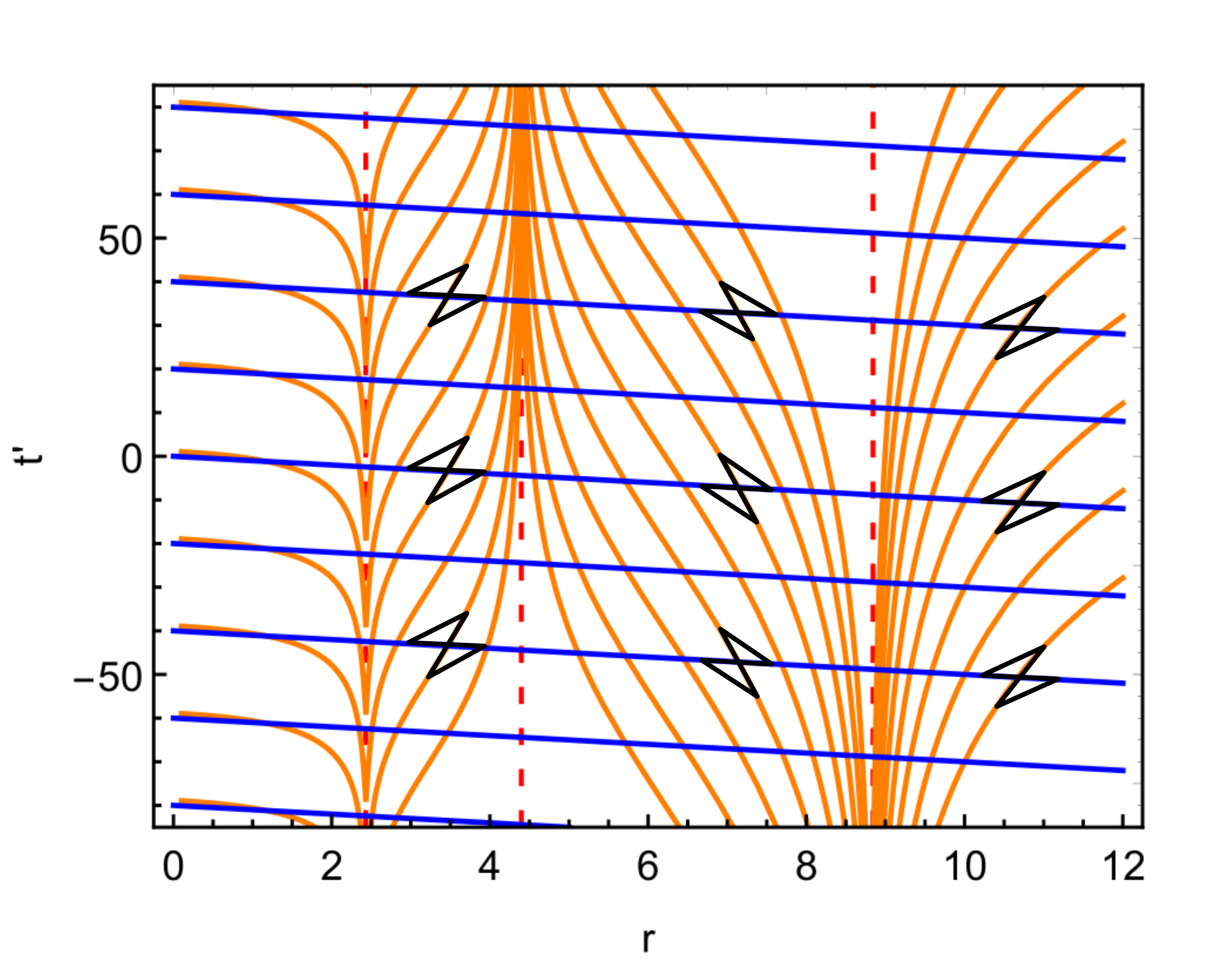} &	\includegraphics[height=6.5cm,keepaspectratio]{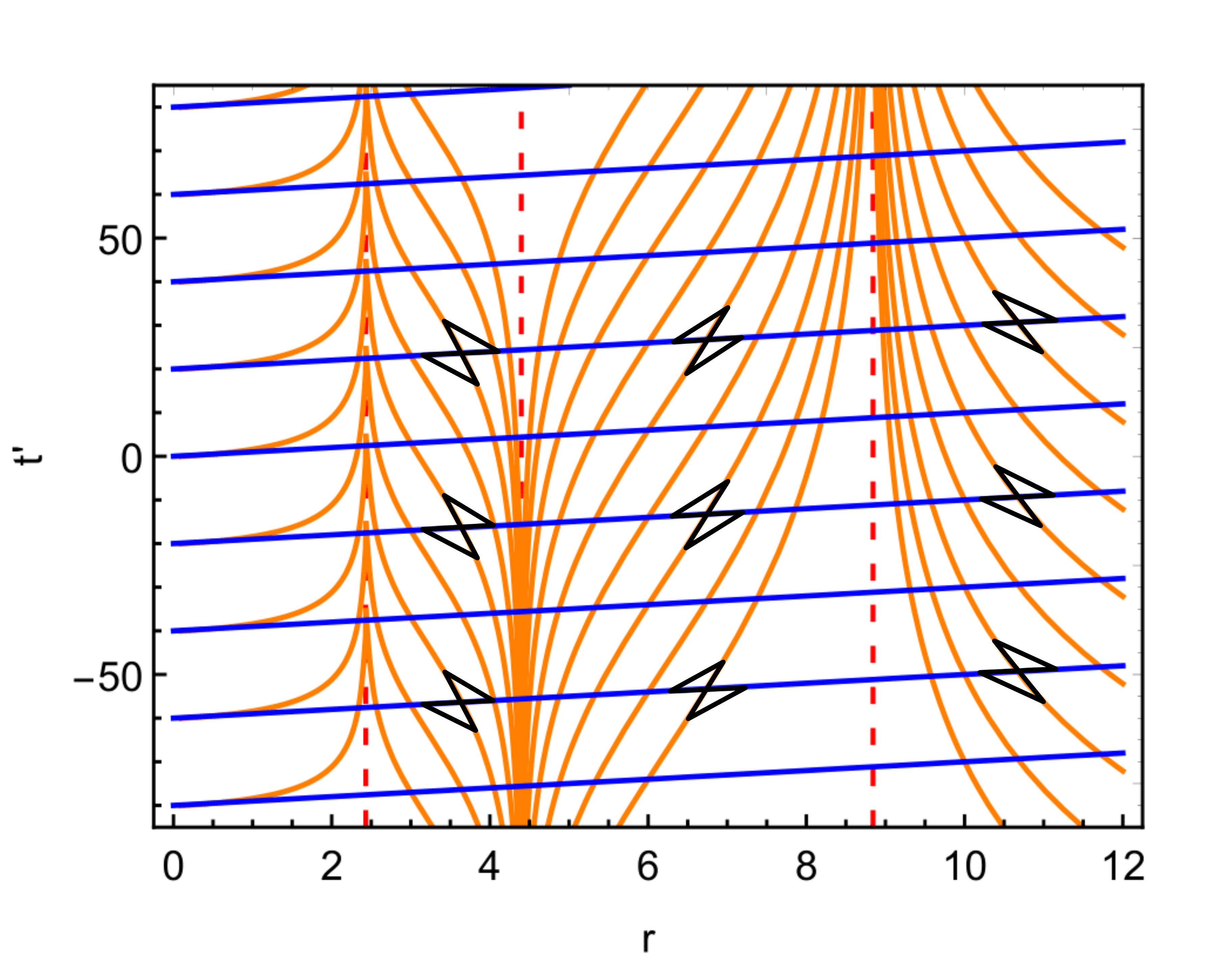}
	\end{tabular}
	\caption{Plots of Eddington-Finkelstein diagram for ingoing photon [left] and outgoing photon [right] of ENE black holes with $Q=10$ and $\beta=0.3$. The horizons can be located at $r_1 = 2.4310$, $r_2 = 4.3965$, and $r3 = 8.8426$. The black lines denote light cones which show the possible direction of light according to the worldlines.}
	\label{EF3}
\end{figure}

The validity of orbits in three-horizon case are also studied by inspecting the Eddington-Finkelsten radial null geodesics, shown in Fig. \ref{EF3}. The three horizons divide the spacetime into 4 regions:
\begin{eqnarray}
\label{4r}
r=
\left\{\begin{array}{cl}
r<r_1,   & \textrm{region 1},\\
r_1<r<r_2,& \textrm{region 2}.\\
r_2<r<r_3,& \textrm{region 3},\\
r>r_3,	& \textrm{region 4}.\\
\end{array} \right\}
\end{eqnarray}
Observing the ingoing diagram (left), once particles enter the horizon they are not allowed to change its direction outward until it reaches region 2. As it switches from ingoing to outgoing (right), we see the wordlines lead the particles outward to region 4. This phenomenon is demonstrated in the orbit of massive particles as shown in top pair of Fig \ref{E_orbit_3}. On the other hand, something quite interesting happens in the null case. It is possible for the null $V_{eff}$ to exhibit a (finite) barrier potential in region 3. As a result, photon cannot pass through to region 2. They are thus bound to orbit the ENE BH only between region 3 and 4. This can be understood as follows. The null radial geodesic described by the Eddington-Finkelstein coordinate is independent on the modification from the effective metric, since its modification only affects the angular part. On the other hand, the null $V_{eff}$ is greatly affected by the effective metric. Thus, it seems that the NLED effective geometry ``modifies" the spacetime structure inside the horizons. For both timelike and null cases he orbital paths are similar to the extremal case, where they both result in a type of epitrochoid orbit.

\end{document}